\documentclass[11pt]{article}

\usepackage[a4paper]{geometry}
\usepackage{authblk}
\usepackage{arydshln}
\usepackage{psfrag}
\usepackage{multirow}
\usepackage{changepage}
\usepackage{lscape}

\usepackage[numbers,sort&compress]{natbib}
\usepackage{url}
\usepackage{tikz}
\usepackage[symbol]{footmisc}
\usepackage{geometry}
\usepackage{setspace}
\usepackage{graphicx}
\usepackage{scalerel}

\usepackage[ruled,vlined]{algorithm2e}

\usepackage{changepage}
\usepackage[british]{babel}
\usepackage{float}
\usepackage{mathrsfs,amsmath}
\usepackage{wrapfig}
\usepackage{cases, mathtools}
\usepackage{amssymb, amsbsy} 
\usepackage[labelfont=bf]{caption}
\usepackage{subcaption}
\usepackage{mathrsfs,amsmath}
\usepackage{listings}
\usepackage{bm}

\usepackage[colorlinks]{hyperref}
\usepackage{xcolor}
\hypersetup{
    colorlinks,
    linkcolor={red},
    citecolor={red},
    urlcolor={red}
}

\definecolor{lightblue}{RGB}{6,69,173}
\definecolor{myRED}{RGB}{149, 37, 51}
\definecolor{myBLUE}{RGB}{45, 115, 184}

\usepackage[colorlinks]{hyperref}
\usepackage{xcolor}
\hypersetup{
    colorlinks,
    linkcolor={lightblue},
    citecolor={lightblue},
    urlcolor={lightblue}
}

\usetikzlibrary{shapes.geometric}
\usepackage{rotating}
\usetikzlibrary{quotes,angles}
\usetikzlibrary{patterns}
\usetikzlibrary{decorations.pathmorphing}
\usetikzlibrary{decorations.markings}
\usetikzlibrary{matrix}
\usetikzlibrary{angles, quotes}
\usetikzlibrary{calc}

\def\XXint#1#2#3{{\setbox0=\hbox{$#1{#2#3}{\int}$}
     \vcenter{\hbox{$#2#3$}}\kern-.5\wd0}}

\usepackage{listings}
\usepackage{color}
 
\definecolor{codegreen}{rgb}{0,0.6,0}
\definecolor{codegray}{rgb}{0.5,0.5,0.5}
\definecolor{codepurple}{rgb}{0.58,0,0.82}
\definecolor{backcolour}{rgb}{0.95,0.95,0.92}
 
\lstdefinestyle{mystyle}{
    backgroundcolor=\color{backcolour},   
    commentstyle=\color{codegreen},
    keywordstyle=\color{magenta},
    numberstyle=\tiny\color{codegray},
    stringstyle=\color{codepurple},
    basicstyle=\footnotesize,
    breakatwhitespace=false,         
    breaklines=true,                 
    captionpos=b,                    
    keepspaces=true,                 
    numbers=left,                    
    numbersep=5pt,                  
    showspaces=false,                
    showstringspaces=false,
    showtabs=false,                  
    tabsize=2
}
 
\definecolor{myGreen}{RGB}{0,128,0}
\definecolor{myPurple}{RGB}{139,0,139}
\definecolor{myRed}{RGB}{162, 20, 47}
\definecolor{myYellow}{RGB}{255, 215, 0}
\definecolor{myGrey}{RGB}{189, 201, 199}

\usepackage{float,yfonts}

  \def\xmax{2.8}

\title{Reconfigurable topological valley-Hall interfaces: Asymptotics of arrays of Dirichlet and Neumann inclusions for multiple scattering in metamaterials}

\author[1,2]{Richard Wiltshaw\thanks{Corresponding author: r.wiltshaw17@imperial.ac.uk, wiltshaw.richard.w82@osaka-u.ac.jp}} 
\author[3]{Henry J. Putley}
\author[4]{Christelle Bou Dagher}
\author[1]{Mehul P. Makwana}
\date{}
\affil[1]{Department of Mathematics,
Imperial College London,
London, SW7 2AZ, United Kingdom}
\affil[2]{Graduate School of Engineering, The University of Osaka, Suita, Osaka 565-0871, Japan}
\affil[3]{School of Physics and Astronomy, University of Birmingham, Birmingham, B15 2TT, United Kingdom}
\affil[4]{School of Physics and Astronomy,
University of Southampton,
Southampton, SO17 1BJ, United Kingdom}

\begin{document}
\maketitle

\begin{abstract}
We study two-dimensional periodic metamaterials in which idealised cylindrical inclusions are modelled by  boundary conditions. In the scalar time-harmonic setting, the background field satisfies the Helmholtz equation, and high-contrast inclusion limits reduce to Dirichlet or Neumann conditions, with direct analogues in dielectric and acoustic media. By switching the condition assigned to selected inclusions, we break point-group symmetries of the primitive cell and thereby lift symmetry-induced degeneracies in the Floquet--Bloch spectrum of hexagonal and square lattices, opening valley-type band gaps with Berry curvature localised near opposite valleys.

To analyse infinite and finite structures within a unified framework, we derive matched-asymptotic point-scatterer approximations for mixed Dirichlet--Neumann arrays. For doubly periodic systems, this yields a finite-dimensional generalised eigenvalue problem for the Floquet--Bloch spectrum; for finite arrays, it yields a generalised Foldy multiple-scattering system. In both hexagonal and square lattices, geometrically identical crystals can realise distinct valley-Hall phases solely through boundary-condition assignment while retaining an overlapping bulk gap. Spatially varying this assignment therefore creates and relocates internal interfaces without altering the underlying geometry, enabling the associated valley-Hall interfacial modes to be repositioned within the same crystal.
\end{abstract}


\section{Introduction}

Topological photonics \cite{lu2014topological,ozawa2019topological,sun2017two,mousavi2015topologically,khanikaev2017two} concerns wave-propagation phenomena controlled by spectral topology in structured electromagnetic media, including topological-insulator phases and the emergence of interface-localised modes that lie in bulk spectral gaps and propagate along material boundaries or internal domains. In infinitely periodic settings, these modes are naturally framed via the Floquet--Bloch spectrum and its symmetry- and topology-driven degeneracies. Moreover, finite topological waveguides behave as Fabry--Pérot resonators \cite{levy2017topological} and can be understood in terms of leaky cavity modes, offering a powerful framework for designing finite topological devices \cite{marti2025fabry,vial2024quasinormal}. A canonical route to valley-Hall guiding in photonic and phononic crystals is to begin with a lattice whose band structure possesses Dirac-type or symmetry-induced degeneracies and then to break inversion \cite{xiao2007valley,makwana2019topological,makwana2020hybrid,marti2025fabry} or reflection symmetry of the primitive cell. This lifts the degeneracy and opens a bulk band gap \cite{xiao2007valley}, thereby engineering valley-Hall-type insulators that preserve time-reversal symmetry (TRS) \cite{ma2016all,makwana2020hybrid, lu2016topological, makwana2018designing, gao2017valley, dong2017valley, yang2018topological, kang2018pseudo, chen_tunable_2018, he2019silicon, lu_observation_2016, ye2017observation, zhang_topological_2018,  wu_direct_2017, jung2018midinfrared, gao2018topologically, zhang2019valley,  xia2018observation, shalaev2018experimental, liu2018tunable}. The resulting gap is accompanied by Berry-curvature localisation \cite{berry1984quantal} (equivalently, a non-zero valley Chern number \cite{ma2016all,ezawa2013topological}) near distinct valleys and by non-trivial valley indices for the adjacent bands \cite{wong2020gapless}. Valley-Hall insulators of this kind have been designed, both theoretically and experimentally, for guiding waves in photonic \cite{ma2016all,makwana2019topological,makwana2020hybrid,ochiai2001dispersion,sakoda2005optical,sakoda1995symmetry,yang2019realization}, acoustic \cite{wiltshaw2020asymptotic,laforge2021acoustic,wiltshaw2023analytical}, elastic \cite{makwana2018geometrically,makwana2018designing,ungureanu2021localizing,makwana2019tunable,mousavi2015topologically}, and water-wave \cite{makwana2020experimental} systems.

When two bulk media with overlapping gapped spectra and opposite valley indices are joined, their interface supports confined modes, often termed zero-line modes (ZLMs) \cite{makwana2018designing}. In practice, valley-Hall interfaces are commonly created by joining two \emph{geometrically} distinct chiral unit cells, for example mirror-related arrangements of inclusions. The central design question addressed here is different: we retain a fixed geometric pattern of inclusions and instead control the \emph{boundary conditions imposed on the inclusions}. This provides an active mechanism by which the bulk phase can be switched locally, so that the location of the interface--and hence the spatial position of the ZLMs--can be moved within the same underlying crystal. 

Topological metamaterials provide a route to structurally robust wave transport, most notably through boundary, interface, or domain-wall modes residing within bulk band gaps. However, for practical devices, a dominant limitation is that the topological phase and microstructure are fixed during fabrication; as a result, the propagation path, operating frequency band, and functionality are difficult to alter post hoc without sacrificing the robustness one seeks. Over roughly the last decade, the field has therefore shifted from static topological insulators and crystals towards reconfigurable topological platforms\cite{ota2020active} capable of switching phases, opening and closing gaps, and dynamically rerouting edge or interface channels across photonic, elastic, and phononic settings. These capabilities have been realised using mechanisms ranging from physically rotating or translating symmetry-broken unit cells \cite{tang2020observations,bahrami2025topological} to electronically programming unit-cell states \cite{darabi2020experimental} or switching material phases for phononic biosensing \cite{okita2026mass}. In photonics and electromagnetic metamaterials, this trend is exemplified by digitally reprogrammable valley-Hall-type platforms \cite{you2021reprogrammable}, in which the topological phase and domain walls are updated through fast electronic switching at the unit-cell level. Related developments include on-chip valley photonic crystals \cite{qi2022electrical}, where electrical control, for example via thermo-optic \cite{cao2019dynamically} or thermo-acoustic \cite{lian2026reversible} tuning, modulates the transmission of kink or edge channels. In parallel, phase-change approaches \cite{wu2018dial,xiu2022topological} exploit large, reversible material contrasts, often without geometric motion, to toggle the existence and spectral position of topological edge states, thereby motivating non-volatile “topological memory” \cite{uemura2024photonic} concepts.

Reconfigurable topological metamaterials have rapidly evolved from demonstrations of static robustness into actively switchable, routable, and programmable wave platforms spanning electromagnetics and photonics, elasticity, and acoustics/phononics, with switching speeds of nanoseconds in digitally reprogrammed metasurfaces \cite{you2021reprogrammable} or thermally driven \cite{cao2019dynamically} soft-matter transitions. Across these platforms, topological switching can be organised into a small set of physically distinct control mechanisms: geometry- or symmetry-based switching; material-property switching, including phase-change and thermoresponsive materials; and active or electronic modulation of effective couplings or boundary conditions. Each approach carries characteristic trade-offs in speed, reversibility, robustness, and manufacturability.

In this manuscript, we develop (i) a mixed Dirichlet--Neumann point-scatterer model for small inclusions, (ii) a Floquet--Bloch eigenvalue formulation for infinite arrays, (iii) a generalised Foldy multiple-scattering formulation for finite collections, and (iv) applications to hexagonal and square lattices that illustrate symmetry breaking, valley-type gaps, ribbon ZLMs, and reconfigurable interface relocation via boundary-condition switching. The semi-analytical solutions are validated against full numerical Finite Element Method (FEM) simulations, and are shown to be accurate.

\section{Model Formulation}
We consider the dynamic electric field propagating in an isotropic dielectric structure.  We consider a two-dimensional scalar time-harmonic wavefield in the plane $\textbf{x}=x \textbf{e}_{x}+y \textbf{e}_{y}$, obtained from a structured medium that is translationally invariant along the cylinder axis $\textbf{e}_{z}$. The medium consists of a homogeneous background in which cylindrical inclusions are embedded. In the cross-section, the inclusions are discs centred at points $\textbf{x} = \textbf{X}_{IJ}$ and of radii $\eta_{i}$, arranged periodically in primitive cells. The dimensionless equation governing the electric field potential $\phi$, in the plane, is given by \cite{zolla2005foundations,guenneau2004comparisons}

\begin{equation}
\left\lbrace \nabla^{2} + \Omega^{2} \right\rbrace \phi (\textbf{x}) = 0. \label{Helmholtz}
\end{equation}

 We enumerate primitive cells by $I=1,\ldots,N$ and inclusions within a cell by $J$ indices; later we will distinguish Dirichlet-type and Neumann-type idealised inclusions. The boundary conditions along the surface between two isotropic dielectrics are presented as \cite{landau1960electrodynamics}
\begin{equation}
\phi \Big|_{r_{IJ}= \eta_{i}} = \phi_{i} \Big|_{r_{IJ}= \eta_{i}}, \quad \quad \epsilon \textbf{n} \cdot \nabla \phi \Big|_{r_{IJ}= \eta_{i}} = \epsilon_{i} \textbf{n} \cdot \nabla \phi_{i} \Big|_{r_{IJ}= \eta_{i}}. \label{DilectricContCond}
\end{equation}

In \eqref{DilectricContCond}, $\epsilon$ denotes the (piecewise constant) dielectric \emph{permittivity}, $\textbf{n}$ is the unit normal at the inclusion boundary, and subscript $i$ denotes quantities within the inclusions. We focus on two singular-contrast idealisations for the inclusion material: $\epsilon_{i}\to 0$ and $\epsilon_{i}\to \infty$. In these limits, the continuity conditions in \eqref{DilectricContCond} reduce to Neumann or Dirichlet constraints on the exterior field at the inclusion boundary

\begin{align}
\frac{\partial \phi}{\partial r_{IJ}} \Big|_{r_{IJ}= \eta_{i}}  & = 0 \quad \mbox{as $\epsilon_{i} \to 0$},  \label{Neumann}\\
\phi \Big|_{r_{IJ}= \eta_{i}}  & = 0 \quad \mbox{as $\epsilon_{i} \to \infty$}. \label{Dirichlet}
\end{align}

Although we phrase the discussion for photonic crystals, the same scalar Helmholtz model with Dirichlet/Neumann inclusions is also the standard idealisation of sound-soft/sound-hard scatterers in acoustics; the subsequent spectral and multiple-scattering framework therefore apply identically to the scalar phononic analogue.

\section{The mixed Dirichlet-Neumann Inclusion problem}

We consider small inclusions whose radii satisfy $\eta_{i}\ll 1$ in the non-dimensionalisation implicit in \eqref{Helmholtz}. In this regime, Dirichlet inclusions are asymptotically represented by monopoles \cite{schnitzer2017bloch}, whereas Neumann inclusions require a combination of monopoles and dipoles \cite{wiltshaw2020asymptotic} - both of which can be viewed as a singular perturbation to the wavefield approaching the centre of the inclusions. This is most naturally derived via the method of matched asymptotic expansions~\cite{schnitzer2017bloch,wiltshaw2020asymptotic}: an inner solution is matched to an outer solution of the Helmholtz equation, and the singular asymptotics naturally cancel during the matching procedure and enable the appropriate boundary conditions to be asymptotically satisfied throughout the procedure. For mixtures of Dirichlet and Neumann inclusions, the singular asymptotics exists tending towards the centre of each inclusion, and is therefore inherited from the corresponding pure problems (refer to \cite{schnitzer2017bloch,wiltshaw2020asymptotic} for the derivations of the singular asymptotics).

To support both of the computational settings used later, we treat:
(i) an infinite periodic arrangement (leading to a Floquet--Bloch eigenvalue problem for band structure); and
(ii) a finite collection of inclusions (leading to a multiple-scattering problem with radiation conditions).
Both follow from the same mixed point-scatterer approximation.

\subsection{Singular perturbations to approximate small Dirichlet and Neumann inclusions}
Assuming each inclusion radius is small and inclusions remain well separated within the primitive unit cell, the mixed boundary-value problem governed by \eqref{Helmholtz} with Dirichlet conditions on a subset of inclusions and Neumann conditions on the remainder can be reduced to a Helmholtz equation with distributional source terms supported at the inclusion centres. Dirichlet inclusions contribute monopole sources; Neumann inclusions contribute a monopole and a dipole term - as follows

\begin{equation}
\begin{split}
\Large\lbrace \nabla^{2} + \Omega^{2} \Large\rbrace \phi = 4i \left\lbrace \sum_{I = 1}^{N} \sum_{J = 1}^{P} a_{IJ} \delta(\textbf{x} - \textbf{X}_{IJ}) + \color{white} \right\rbrace \\ + \color{white} \left\lbrace  \color{black} \sum_{I = 1}^{N} \sum_{K = 1}^{Q} \eta_{IK}^{2} \left[ b_{IK} \delta(\textbf{x} - \textbf{X}_{IK})  - \textbf{c}_{IK} \cdot \nabla \delta(\textbf{x} - \textbf{X}_{IK}) \right] \right\rbrace 
\end{split} \quad .  \label{PFSmyDrude}
 \end{equation}

In \eqref{PFSmyDrude}, the indices $(I,J)$ enumerate Dirichlet inclusions and $(I,K)$ enumerate Neumann inclusions within the $I$th primitive cells. Each primitive cell contains $P$ Dirichlet and $Q$ Neumann inclusions respectively. The coefficients $a_{IJ}$ (Dirichlet) and $(b_{IK},\textbf{c}_{IK})$ (Neumann) are determined by asymptotically matching the inner and outer solutions so that the required boundary conditions are approximately satisfied. The representation \eqref{PFSmyDrude} is interpreted in the distributional sense in the outer domain (the plane with the inclusion centres removed) and is the starting point for the periodic and finite formulations that follow.

\section{An eigenvalue problem for Floquet--Bloch waves} \label{EigenSection}

\begin{figure}[h]
\centering
\centering
\begin{tikzpicture}[scale=0.4, transform shape,cross/.style={path picture={ 
  \draw[black]
(path picture bounding box.south east) -- (path picture bounding box.north west) (path picture bounding box.south west) -- (path picture bounding box.north east);
}}]

\begin{scope}[xshift=-37cm, yshift=69cm]
		\node[regular polygon, regular polygon sides=4,draw, inner sep=5.4cm,rotate=0,line width=0.0mm, white,
           path picture={
               \node[rotate=0] at (-0.425,-0.325){
                   \includegraphics[scale=1.25]{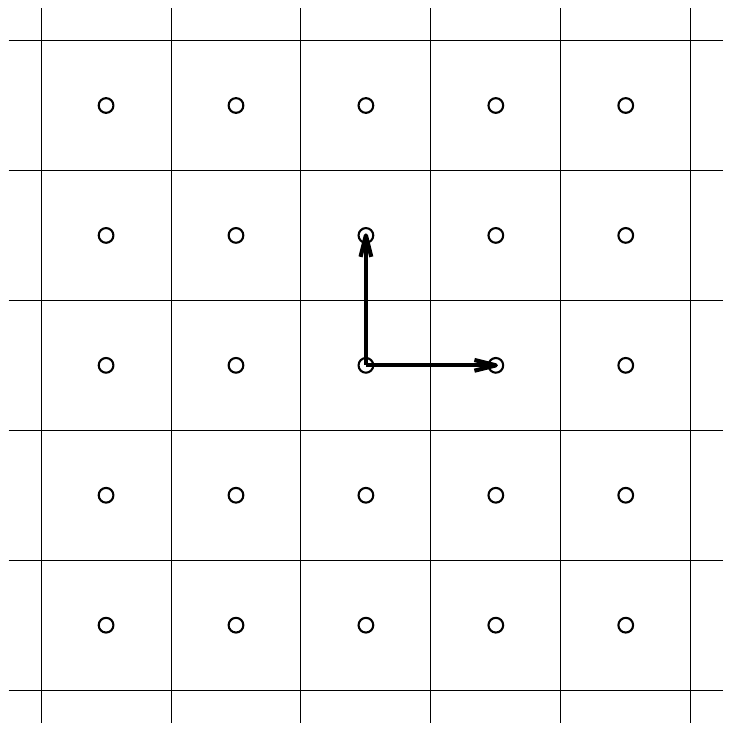}
               };
           }]{};
           		\node[below, scale=3,black] at (3.4,0.25) {$\displaystyle  \boldsymbol{\alpha}_{1}$}; 
				\node[below, scale=3,black] at (0.25,4.25) {$\displaystyle  \boldsymbol{\alpha}_{2}$}; 
				\node[below, scale=4,] at (8.5,0.6) {$\displaystyle  \ldots$}; 
				\node[below, scale=4,] at (-8.5,0.6) {$\displaystyle  \ldots$}; 
				\node[below, scale=4,] at (0.1,10.1) {$\displaystyle  \vdots$}; 
				\node[below, scale=4,] at (0.1,-6.9) {$\displaystyle  \vdots$}; 
           		\node[below, scale=2.5, black] at (-8,9) {$\displaystyle (a)$};     
\end{scope}

\begin{scope}[xshift=-15cm, yshift=69cm]
		\node[regular polygon, regular polygon sides=4,draw, inner sep=5.4cm,rotate=0,line width=0.0mm, white,
           path picture={
               \node[rotate=0] at (-0.675,-0.5){
                   \includegraphics[scale=2.0]{Figs/simplestSkem.pdf}
               };
           }]{};
           		\node[below, scale=3,black] at (5.5,0.0) {$\displaystyle  \boldsymbol{\beta}_{1}$}; 
				\node[below, scale=3,black] at (0.25,6.5) {$\displaystyle  \boldsymbol{\beta}_{2}$}; 
				\node[below, scale=4,] at (8.5,0.6) {$\displaystyle  \ldots$}; 
				\node[below, scale=4,] at (-8.5,0.6) {$\displaystyle  \ldots$}; 
				\node[below, scale=4,] at (0.1,10.1) {$\displaystyle  \vdots$}; 
				\node[below, scale=4,] at (0.1,-6.9) {$\displaystyle  \vdots$}; 
           		\node[below, scale=2.5, black] at (-8,9) {$\displaystyle (b)$};     
\end{scope}

\begin{scope}[xshift=-37cm, yshift=46cm]
		\node[regular polygon, regular polygon sides=4,draw, inner sep=5.4cm,rotate=0,line width=0.0mm, white,
           path picture={
               \node[rotate=0] at (-0.425,-0.325){
                   \includegraphics[scale=1.25]{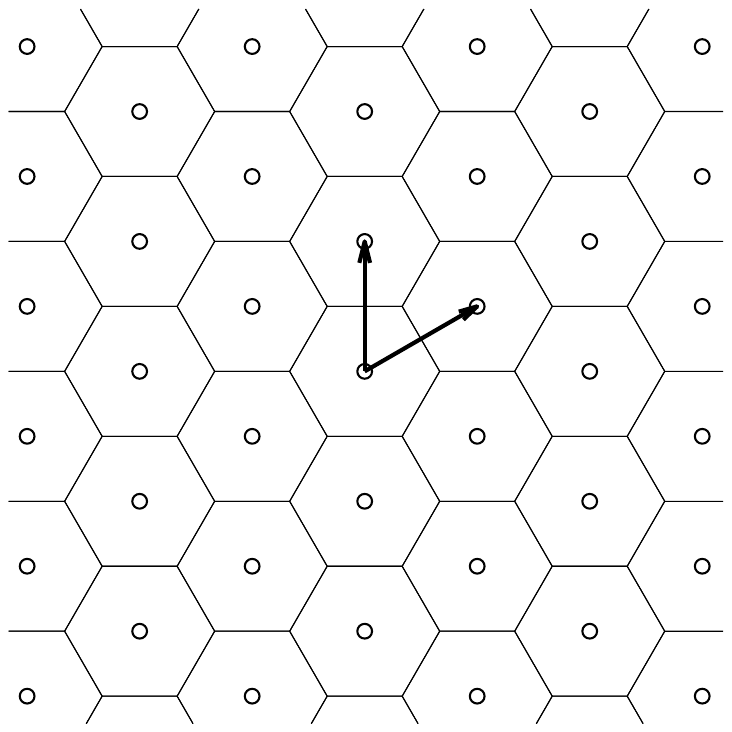}
               };
           }]{};
           		\node[below, scale=3,black] at (2.6,2.75) {$\displaystyle  \boldsymbol{\alpha}_{1}$}; 
				\node[below, scale=3,black] at (0.25,4.25) {$\displaystyle  \boldsymbol{\alpha}_{2}$}; 
				\node[below, scale=4,] at (8.5,0.45) {$\displaystyle  \ldots$}; 
				\node[below, scale=4,] at (-8.5,0.45) {$\displaystyle  \ldots$}; 
				\node[below, scale=4,] at (0.0,10.4) {$\displaystyle  \vdots$}; 
				\node[below, scale=4,] at (0.0,-6.6) {$\displaystyle  \vdots$}; 
          		\node[below, scale=2.5, black] at (-8,10)  {$\displaystyle (c)$};     
\end{scope}  

 \begin{scope}[xshift=-15cm, yshift=46cm]
		\node[regular polygon, regular polygon sides=4,draw, inner sep=5.4cm,rotate=0,line width=0.0mm, white,
           path picture={
               \node[rotate=-30] at (-0.675,-0.0){
                   \includegraphics[scale=2.0]{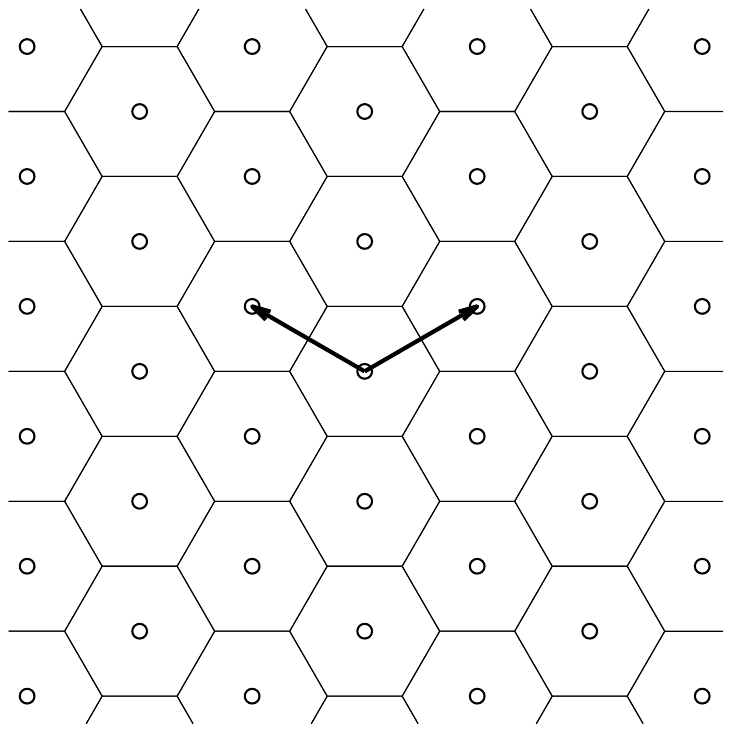}
               };
           }]{};
           
          	    \node[below, scale=3,black] at (5.2,0.0) {$\displaystyle  \boldsymbol{\beta}_{1}$}; 
				\node[below, scale=3,black] at (-2.0,6.2) {$\displaystyle  \boldsymbol{\beta}_{2}$}; 
				\node[below, scale=4,] at (8.5,0.45) {$\displaystyle  \ldots$}; 
				\node[below, scale=4,] at (-8.5,0.45) {$\displaystyle  \ldots$}; 
				\node[below, scale=4,] at (0.1,10.4) {$\displaystyle  \vdots$}; 
				\node[below, scale=4,] at (0.1,-6.6) {$\displaystyle  \vdots$}; 
           		\node[below, scale=2.5, black] at (-8,10)  {$\displaystyle (d)$};     
\end{scope}

\begin{scope}[xshift=-33cm, yshift=35cm,scale=1.75, transform shape]
\draw[thick,-latex] (-7 , 0 ) -- (-6 , 0);
\draw[thick,-latex] (-7 , 0 ) -- (-7 , 1);
\draw[thick,cross,fill=white] (-7,0) circle (0.15);
\node[right, scale=1.75] at (-6,0) {$\displaystyle \textbf{e}_{x}$};
\node[above, scale=1.75] at (-7,1) {$\displaystyle \textbf{e}_{y}$};
\node[left, scale=1.75]at (-7.15,0) {$\displaystyle \textbf{e}_{z}$};
\end{scope}

\end{tikzpicture}
\caption{Two-dimensional square (a) and hexagonal (c) Bravais lattices in physical space, generated by $\boldsymbol{\alpha}_{1}$ and $\boldsymbol{\alpha}_{2}$. Panels (b) and (d) show the corresponding reciprocal lattices, generated by $\boldsymbol{\beta}_{1}$ and $\boldsymbol{\beta}_{2}$. In all panels, Bravais lattice points are indicated by $\circ$.}
\label{fig:PhysRecipIntro}
\end{figure}

We now consider an infinite doubly periodic arrangement of inclusions for the mixed point-scatterer formulation in the $xy$-plane, as illustrated in Fig.~\ref{fig:PhysRecipIntro}. In \eqref{PFSmyDrude} we take $N\to\infty$ and partition free space into primitive cells translated by the two-dimensional Bravais lattice
\begin{equation}
\textbf{R} = n \boldsymbol{\alpha}_{1} + m \boldsymbol{\alpha}_{2},
\qquad n,m \in \mathbb{Z}.
\label{IntroBigR}
\end{equation}
The reciprocal basis vectors $\boldsymbol{\beta}_{1}$ and $\boldsymbol{\beta}_{2}$ are defined by
\begin{equation}
\boldsymbol{\alpha}_{i} \cdot \boldsymbol{\beta}_{j} = 2 \pi \delta_{ij},
\qquad i,j = 1,2,
\label{Intro:physRecipRelate}
\end{equation}
and the corresponding reciprocal Bravais lattice is
\begin{equation}
\textbf{G} = n \boldsymbol{\beta}_{1} + m \boldsymbol{\beta}_{2},
\qquad n,m \in \mathbb{Z}.
\label{IntroBigG}
\end{equation}

Because the structure is periodic, it is sufficient to work with a reference cell, which we label by $I=1$; the source coefficients in all other cells are then determined by the Bloch phase. Let $\mathcal{A}$ denote the area of a primitive cell. For waves propagating in an infinite periodic medium, Bloch's theorem \cite{kittel1996introduction,brillouin1953wave} implies that the field may be written as
\begin{equation}
\phi(\textbf{x}) = \Phi(\textbf{x}) \exp(i \boldsymbol{\kappa} \cdot \textbf{x}) .
\label{BlochDecompIntro}
\end{equation}
Here $\Phi(\textbf{x}+\textbf{R})=\Phi(\textbf{x})$ and $\boldsymbol{\kappa}$ is the Bloch wavevector. Once $\Phi$ is known in a single primitive cell, the field throughout the lattice follows from the Bloch phase factor. It is therefore natural to expand the periodic part in reciprocal-lattice harmonics:
\begin{equation}
\Phi(\textbf{x}) = \sum_{\textbf{G}} \Phi_{\textbf{G}} \exp(i \textbf{G} \cdot \textbf{x}),
\label{IntroFourierSeriesRep}
\end{equation}
where the sum is taken over the reciprocal lattice \eqref{IntroBigG}. Writing
\[
\textbf{K}_{\textbf{G}}=\boldsymbol{\kappa}+\textbf{G},
\]
and substituting \eqref{BlochDecompIntro} and \eqref{IntroFourierSeriesRep} into \eqref{PFSmyDrude}, then multiplying by $\exp[-i(\boldsymbol{\kappa}+\textbf{G}')\cdot \textbf{x}]$, integrating over the reference cell, and applying orthogonality, yields
\begin{equation}
\mathcal{A} \left\{ \textbf{K}_{\textbf{G}} \cdot \textbf{K}_{\textbf{G}} - \Omega^{2} \right\} \Phi_{\textbf{G}}
=
\sum_{J=1}^{P} \frac{4 a_{1J} \exp(-i \textbf{K}_{\textbf{G}} \cdot \textbf{X}_{1J})}{i}
+
\sum_{K=1}^{Q} \frac{4 \eta_{1K}^{2} \left[ b_{1K} - i \textbf{c}_{1K} \cdot \textbf{K}_{\textbf{G}} \right] \exp(-i \textbf{K}_{\textbf{G}} \cdot \textbf{X}_{1K})}{i}.
\label{EIGEN1}
\end{equation}
Hence the field admits the series representation
\begin{equation}
\phi(\textbf{x})
=
\frac{4}{i \mathcal{A}}
\sum_{\textbf{G}}
\left\{
\sum_{J=1}^{P}
\frac{ a_{1J} \exp(i \textbf{K}_{\textbf{G}} \cdot \textbf{r}_{1J})}
{\textbf{K}_{\textbf{G}} \cdot \textbf{K}_{\textbf{G}} - \Omega^{2}}
+
\sum_{K=1}^{Q}
\frac{ \eta_{1K}^{2} \left[ b_{1K} - i \textbf{c}_{1K} \cdot \textbf{K}_{\textbf{G}} \right] \exp(i \textbf{K}_{\textbf{G}} \cdot \textbf{r}_{1K})}
{\textbf{K}_{\textbf{G}} \cdot \textbf{K}_{\textbf{G}} - \Omega^{2}}
\right\},
\label{DM}
\end{equation}
where
\[
\textbf{r}_{1J}=\textbf{x}-\textbf{X}_{1J}, \qquad
\textbf{r}_{1K}=\textbf{x}-\textbf{X}_{1K}.
\]
We denote $r_{1J}=|\textbf{r}_{1J}|$ and $r_{1K}=|\textbf{r}_{1K}|$. As is standard for lattice sums of this type, the representation \eqref{DM} is only conditionally convergent. Moreover, it becomes singular as the observation point approaches an inclusion centre: the field diverges as $r_{1J}\to 0$ at a Dirichlet inclusion and as $r_{1K}\to 0$ at a Neumann inclusion. Resolving these singular contributions is therefore the crucial step in obtaining a practical finite-dimensional eigenvalue problem.

\subsection{Double sum asymptotics of the inner limit of the outer solution}

The conditional convergence of the Bloch series creates the main computational difficulty: in order to evaluate the field near an inclusion centre, one must isolate the singular contribution from the regular part of the sum. To this end, we introduce a reciprocal-space cut-off \(R\) and decompose the series into truncated and residual parts:
\begin{equation}
\begin{split}
\lim_{r_{1J} \to 0} \phi
&=
\lim_{r_{1J} \to 0}
\left\{
\overbrace{\sum_{|\textbf{G}| < R} \Phi_{\textbf{G}} \exp(i \textbf{K}_{\textbf{G}} \cdot \textbf{x})}^{\phi_{\mathrm{tr}}}
+
\overbrace{\sum_{|\textbf{G}| > R} \Phi_{\textbf{G}} \exp(i \textbf{K}_{\textbf{G}} \cdot \textbf{x})}^{\phi_{\mathrm{res}}}
\right\}, \\
\lim_{r_{1K} \to 0} \phi
&=
\lim_{r_{1K} \to 0}
\left\{
\overbrace{\sum_{|\textbf{G}| < R} \Phi_{\textbf{G}} \exp(i \textbf{K}_{\textbf{G}} \cdot \textbf{x})}^{\phi_{\mathrm{tr}}}
+
\overbrace{\sum_{|\textbf{G}| > R} \Phi_{\textbf{G}} \exp(i \textbf{K}_{\textbf{G}} \cdot \textbf{x})}^{\phi_{\mathrm{res}}}
\right\}.
\end{split}
\label{HelmCondConvSer}
\end{equation}
Here, \(\phi_{\mathrm{tr}}\) denotes the truncated sum and \(\phi_{\mathrm{res}}\) the remainder beyond the cut-off. For an absolutely convergent series, the residual term vanishes as \(R \to \infty\), so truncation introduces negligible error. In the present problem, however, the series diverges as \(r_{1J} \to 0\) and \(r_{1K} \to 0\), and the residual part therefore contains the singular asymptotics required for the matched asymptotic analysis.

For a large but finite cut-off satisfying \(R \gg 1\), \(R r_{1J} \ll 1\), and \(R r_{1K} \ll 1\), the singular asymptotics of \(\phi_{\mathrm{res}}\) can be extracted analytically using the Euler--Maclaurin formula \cite{olver1997asymptotics}. For the Dirichlet case, \cite{schnitzer2017bloch} gives
\begin{equation}
\phi_{\mathrm{res}} \sim \frac{2 a_{1J}}{i \pi} \left( \log \frac{2}{r_{1J} R} - \gamma_{E} \right) + \ldots,
\qquad \text{as } r_{1J} \to 0 \text{ and } R \to \infty.
\label{DirOuter}
\end{equation}
Here \(\gamma_{E}\) denotes the Euler--Mascheroni constant. Similarly, from the Neumann analysis in \cite{wiltshaw2020asymptotic}, one obtains
\begin{equation}
\begin{split}
& \phi_{\mathrm{res}} \sim \eta_{1K}^{2} \frac{\exp \left( i  \boldsymbol{\kappa} \cdot \textbf{r}_{1K} \right)}{i \pi} \Big\lbrace \frac{2 \textbf{c}_{1K} \cdot \textbf{e}_{r_{1K}}}{r_{1K}} J_{0}(Rr_{1K}) + Ji_{0} (Rr_{1K}) \Big[ i r_{1K} (i \textbf{c}_{1K} \Omega^{2} + 2 b_{1K} \boldsymbol{\kappa}) \cdot \textbf{e}_{r_{1K}} - 2  b_{1K}  \Big] - \\
& \quad - \frac{J_{1}(R'r)}{R'r} \Big[ 2 i \Big( \textbf{c}_{1K} \cdot \boldsymbol{\kappa}\cos(2 \varphi_{1K} - 2 \theta_{\kappa}) - (\textbf{c}_{1K} \times \boldsymbol{\kappa})\cdot \textbf{e}_{z} \sin(2 \varphi_{1K} - 2 \theta_{\kappa}) \Big) + \\
&\quad \quad \quad  \quad \quad \quad \quad \quad \quad  \quad \quad \quad \quad \quad \quad  \quad \quad \quad  + i r_{1K}(i \textbf{c}_{1K} \Omega^{2} + 2 b_{1K} \boldsymbol{\kappa}) \cdot \textbf{e}_{r_{1K}} \Big] - \\
& \quad -  2 \kappa \Big[ \textbf{c}_{1K} \cdot \boldsymbol{\kappa} \cos( 3 \theta_{\kappa} - 3 \varphi_{1K}) + (\textbf{c}_{1K} \times \boldsymbol{\kappa})\cdot \textbf{e}_{z} \sin(3 \theta_{\kappa} - 3 \varphi_{1K})  \Big] \frac{J_{2} (R'r_{1K})}{R'^{2} r_{1K}}  \Big\rbrace + \ldots \, , \quad \\
&\quad \quad \quad  \quad \quad \quad \quad \quad \quad  \quad \quad \quad \quad \quad \quad \quad \quad  \quad \quad \quad \quad \quad \quad  \quad \quad \quad \quad \quad \quad  \quad \quad \quad \mbox{as $r_{1K} \to 0$ and $R \to \infty$.}
\end{split} \label{NAP::G>Rcont}
\end{equation}
Here $\theta_{\kappa}$ and $\varphi_{1K}$ are the arguments of $\boldsymbol{\kappa}$ and $\textbf{x}-\textbf{X}_{1K}$, respectively. In addition, $\textbf{e}_{r_{1K}}$ denotes the radial base vector centred on the $1K$th Neumann inclusion, \(Ji_{0}\) denotes Van der Pol's zero-order Bessel-integral function, whereas \(J_{0}\), \(J_{1}\), and \(J_{2}\) denote Bessel functions. Their small-argument behaviour is standard \cite{abramowitz1964handbook}; in particular, Humbert \cite{humbert1933bessel} showed that
\begin{equation}
Ji_{0}(x) = Ci(x) - \log 2
= \log \frac{x}{2} + \gamma_{E} - \frac{x^{2}}{4} + \mathcal{O}(x^{4}),
\qquad \text{as } x \to 0,
\label{eqn::Ji0small}
\end{equation}
where \(Ci(x)\) is the cosine-integral function.

\subsection{The outer limit of the inner problem}

The matching conditions require the outer limit of the corresponding inner solutions near the Dirichlet and Neumann inclusions. These are inherited from the pure Dirichlet and pure Neumann problems. For a Dirichlet inclusion, \cite{schnitzer2017bloch} gives
\begin{equation}
\lim_{r_{1J} \to 0} \phi = \frac{2 a_{1J}}{i \pi} \log \frac{\eta_{1J}}{r_{1J}}.
\label{DirInner}
\end{equation}
For a Neumann inclusion, we \cite{wiltshaw2020asymptotic} obtained
\begin{equation}
\begin{split}
\lim_{r_{1K} \to 0} \phi \sim
&\frac{4 i}{\pi} \frac{b_{1K}}{\Omega^{2}}
\left[ 1 - \frac{\Omega^{2} r_{1K}^{2}}{4} \right]
+ \frac{4}{i \pi} \frac{\textbf{c}_{1K}\cdot \textbf{e}_{r_{1K}}}{\Omega}
\left( \frac{r_{1K}\Omega}{2} - \frac{\Omega^{3} r_{1K}^{3}}{16} \right) \\
&+ \eta_{1K}^{2}
\left\{
\frac{2 i b_{1K}}{\pi}
\left[ \log \frac{r_{1K}}{\eta_{1K}} + \frac{3}{4} \right]
+ \frac{\textbf{c}_{1K} \cdot \textbf{e}_{r_{1K}}}{i \pi}
\left(
\frac{2}{r_{1K}} - \Omega^{2} r_{1K}
\left[ \log \frac{r_{1K}}{\eta_{1K}} - \frac{7}{4} \right]
\right)
\right\}.
\end{split}
\label{NAP::HelmholtzInnerLim}
\end{equation}

\subsection{A generalised eigenvalue problem from the method of matched asymptotic expansions}

Assume \(\eta_{1J} \ll 1\) and \(\eta_{1K} \ll 1\). In the vicinity of a given inclusion, the local singular behaviour is asymptotically dominated by that inclusion alone. Since the Fourier series \eqref{DM} diverges as \(r_{1J} \to 0\) and \(r_{1K} \to 0\), we introduce a large but finite reciprocal-space cut-off satisfying \(R \gg 1\), \(R r_{1J} \ll 1\), and \(R r_{1K} \ll 1\). Equations \eqref{DirOuter} and \eqref{NAP::G>Rcont}  quantify the error introduced by truncating the conditionally convergent series.

Combining \eqref{HelmCondConvSer}, \eqref{DirOuter}, and \eqref{DirInner} yields the Dirichlet matching condition
\begin{align}
\sum_{|\textbf{G}| < R} \Phi_{\textbf{G}} \exp ( i \textbf{K}_{\textbf{G}} \cdot \textbf{X}_{1J} )
- \frac{2 a_{1J}}{\pi i}
\left\{
\log \frac{\eta_{1J} R}{2} + \gamma_{E}
\right\}
= 0,
\qquad \text{as } r_{1J} \to 0.
\label{HelmholtzEigeAsympD}
\end{align}
Similarly, combining \eqref{HelmCondConvSer}, \eqref{NAP::G>Rcont}, and \eqref{NAP::HelmholtzInnerLim} gives
 \begin{equation}
 \begin{split}
\sum_{|\textbf{G}| < R} \Phi_{\textbf{G}} \exp ( i \textbf{K}_{\textbf{G}} \cdot \textbf{X}_{1K} )  - \frac{4ib_{1K}}{\pi \Omega^{2}} - \eta_{1K}^{2} \left\lbrace \frac{2b_{1K}i}{\pi} \left[ \log \frac{2}{\eta_{1K} R} + \frac{3}{4} - \gamma_{E} \right]  +  \frac{\boldsymbol{\kappa} \cdot \textbf{c}_{1K}}{\pi} \right\rbrace = 0, \\
\quad \quad \quad \quad \quad \quad \mbox{as $r_{1K} \to 0$}.
\end{split} \label{HelmholtzEigeAsympN}
 \end{equation}

A Neumann inclusion introduces three unknown coefficients, namely \(b_{1K}\) and the two independent components of \(\textbf{c}_{1K}\). Two additional relations are therefore needed to close the system. To obtain these, and to preserve the Hermitian structure (Floquet-Bloch dispersion relation should be purely real) of the resulting generalised eigenvalue problem, we also match the gradient of the field near each Neumann inclusion:
\begin{equation}
\nabla \phi_{\mathrm{tr}} = \nabla \phi - \nabla \phi_{\mathrm{res}}
\qquad \text{as } r_{1K} \to 0.
\end{equation}
It is convenient to evaluate the gradients in local polar coordinates centred on each inclusion and then project onto the global Cartesian basis. This gives the \(\textbf{e}_{x}\)- and \(\textbf{e}_{y}\)-components as
\begin{equation}
\begin{split}
\textbf{e}_{x}: \quad  \sum_{G < R'} i K_{1 \, \textbf{G}}  \Phi_{\textbf{G}} \exp(i \textbf{K}_{\textbf{G}} \cdot \textbf{X}_{1K}) & = \quad \quad \quad \quad \quad \quad \quad \quad \quad \quad \quad \quad \\
= \Big[ \frac{2}{i \pi} + i \frac{\eta_{1K}^{2}}{\pi} \Omega^{2} \Big( \log \frac{2}{\eta_{1K} R} & - \frac{5}{4} - \gamma_{E} \Big)  + \frac{\eta_{1K}^{2}}{2i \pi } R^{2} \Big] \textbf{e}_{x} \cdot \textbf{c}_{1k} + \\
+ \eta_{1K}^{2} \frac{b_{1K}}{\pi} \kappa_{1}  & -  \frac{\eta_{1K}^{2} i}{4 \pi} \Big[  \kappa_{1}  \textbf{c}_{1K} \cdot \boldsymbol{\kappa} - \kappa_{2} (\textbf{c}_{1K} \times \boldsymbol{\kappa} ) \cdot \textbf{e}_{z} \Big]  + \mathcal{O}(r_{1K}) \quad \mbox{as $r_{1K} \to 0$}.
\end{split} \label{HelmholtzGradExEigeAsymp} 
\end{equation}

\begin{equation}
\begin{split}
\textbf{e}_{y}: \quad \sum_{G < R'} i K_{2 \, \textbf{G}}  \Phi_{\textbf{G}} \exp(i \textbf{K}_{\textbf{G}} \cdot \textbf{X}_{1K}) & = \quad \quad \quad \quad \quad \quad \quad \quad \quad \quad \quad \quad \\
= \Big[ \frac{2}{i \pi} + i \frac{\eta_{1K}^{2}}{\pi} \Omega^{2} \Big( \log \frac{2}{\eta_{1K} R} & - \frac{5}{4} - \gamma_{E} \Big)+ 
\frac{\eta_{1K}^{2}}{2i \pi } R^{2} \Big] \textbf{e}_{y} \cdot \textbf{c}_{1K} + \\
+\eta_{1K}^{2} \frac{b_{1K}}{\pi} \kappa_{2} 
& -  \frac{\eta_{1K}^{2} i}{4 \pi} \Big[ \kappa_{2} \textbf{c}_{1K} \cdot \boldsymbol{\kappa} + \kappa_{1} (\textbf{c}_{1K} \times \boldsymbol{\kappa} ) \cdot \textbf{e}_{z} \Big] + \mathcal{O}(r_{1K}) \quad \mbox{as $r_{1K} \to 0$}.
\end{split} \label{HelmholtzGradEyEigeAsymp} 
\end{equation}

Together, \eqref{EIGEN1}, \eqref{HelmholtzEigeAsympD}, the Neumann relation \eqref{HelmholtzEigeAsympN} after multiplication by \(\Omega^{2}\), and the gradient conditions \eqref{HelmholtzGradExEigeAsymp}--\eqref{HelmholtzGradEyEigeAsymp} can be assembled into the generalised eigenvalue problem
\begin{equation}
\left\lbrace \mathfrak{A}(\boldsymbol{\kappa}) - \Omega^{2} \mathfrak{B}(\boldsymbol{\kappa}) \right\rbrace \boldsymbol{\Phi} = \textbf{0}.
\label{HelmholtzHardScheme_1}
\end{equation}
If we assume $M$ reciprocal lattice points lie within \(|\textbf{G}| < R\), then \(\mathfrak{A}\) and \(\mathfrak{B}\) are matrices of dimensions $(M+P+3Q) \times (M+P+3Q)$ which depend on the Bloch wavevector \(\boldsymbol{\kappa}\). The eigenvalue is \(\Omega^{2}\), and the eigenvector \(\boldsymbol{\Phi}\) contains the truncated Fourier coefficients \(\Phi_{\textbf{G}}\) together with the source coefficients \(a_{1J}\), \(b_{1K}\), \(\textbf{e}_{x}\cdot\textbf{c}_{1K}\), and \(\textbf{e}_{y}\cdot\textbf{c}_{1K}\).

\section{The multiple scattering problem utilising generalised Foldy's method} \label{MSTmannn}

We consider scattering from a finite number of inclusions. Here we take $N$ to be finite in \eqref{PFSmyDrude} and replace the double sums over $I,J$ and $I,K$ by single sums over $j$ and $k$, respectively. The scattered field is then
\begin{equation}
\phi_{\mathrm{sc}}
=
\sum_{j=1}^{p} a_{j} H_{0}^{(1)}(\Omega r_{j})
+
\sum_{k=1}^{q} \eta_{k}^{2}
\left\{
b_{k} H_{0}^{(1)}(\Omega r_{k})
+
\mathbf{c}_{k} \cdot \mathbf{e}_{r_{k}} \, \Omega H_{1}^{(1)}(\Omega r_{k})
\right\},
\label{Green}
\end{equation}
where $p$ and $q$ are the total numbers of Dirichlet and Neumann inclusions, respectively, $r_{j} = |\mathbf{x} - \mathbf{X}_{j}|$, and $\mathbf{e}_{r_{j}}$ is the radial unit vector in polar coordinates centred on the scatterer at $\mathbf{x}=\mathbf{X}_{j}$ (and similarly for subscript $k$). For a derivation of \eqref{Green}, see Section 1.3 of \cite{Wiltshaw2022Thesis}. We take the incident field to be a plane wave arriving from infinity, an angle $\theta_{inc}$ centred upon $\textbf{X}_{\mathrm{inc}}$. The incident field $\phi_{\mathrm{inc}}$ is
\begin{equation}
\phi_{\mathrm{inc}}
=
A_{\mathrm{inc}}
\exp\!\left[- i \Omega r_{\mathrm{inc}} \cos(\psi)\right],
\end{equation}
where $r_{\mathrm{inc}} = |\mathbf{x} - \mathbf{X}_{\mathrm{inc}}|$ and $\psi = \theta - \theta_{\mathrm{inc}}$. The total field is therefore
\begin{equation}
\phi = \phi_{\mathrm{sc}} + \phi_{\mathrm{inc}},
\label{mscSOLNtotal}
\end{equation}
where the unknown coefficients $a_{j}$, $b_{k}$, and $\mathbf{c}_{k}$, i.e.\ the scattering coefficients, must be determined so that the boundary conditions \eqref{Neumann} and \eqref{Dirichlet} are satisfied asymptotically. This is achieved by using the methods of matched asymptotic expansions, in which the appropriate conditions are applied in the limits $r_{j} \to 0$ and $r_{k} \to 0$ for every scatterer in the field.

Typically, Foldy's method \cite{foldy1945multiple,martin2006multiple} determines the scattering coefficients heuristically \cite{schnitzer2017bloch} by assuming that the scattering coefficient of the $m$th scatterer is proportional to its ``external field'', i.e.\ the total field with the contribution from the $m$th scatterer removed. As $r_{m} \to 0$, the external field remains finite, allowing one to determine the constant of proportionality, and hence the scattering coefficient. However, the same result follows directly from matched asymptotics: the singularities in the inner limit of the outer solution (i.e.\ $\lim_{r_{m} \to 0} \phi$ from \eqref{mscSOLNtotal}) match those in the outer limit of the inner solution (i.e.\ $\phi$ from \eqref{DirInner} or \eqref{NAP::HelmholtzInnerLim}, as appropriate).

Consequently, the singular terms cancel, leaving the following expression for the $m$th Dirichlet inclusion \cite{schnitzer2017bloch}
\begin{equation}
\begin{split}
- \left[ \frac{2 i}{\pi} \left( \log \frac{\Omega \eta_{m}}{2} + \gamma_{E} \right) +1 \right] a_{m} - \sum_{\substack{j=1 \\ j \neq m}}^{p} a_{j} H_{0}^{(1)}(\Omega r_{mj} ) - \quad \quad \quad \quad \quad \quad \quad \quad \quad \quad \quad \quad \quad \quad \quad \quad \\ 
- \sum_{\substack{k =1}}^{q} \eta_{k}^{2} \Big\lbrace b_{k} H^{(1)}_{0}(\Omega r_{mk}) + \textbf{c}_{k} \cdot \textbf{e}_{r \, mk} \Omega H_{1}^{(1)} (\Omega r_{mk} ) \Big\rbrace =  \phi_{\mathrm{inc}} \Big|_{\textbf{x} = \textbf{X}_{m}},
\end{split} \label{DirNeuScatt}
\end{equation} similarly, for the $n$th Neumann inclusion \cite{wiltshaw2020asymptotic},
\begin{equation}
\begin{split}
b_{n} \left\lbrace \frac{4i}{\pi \Omega^{2}} - \eta^{2}_{n} \left[ 1 - \frac{2i}{\pi} \left( \log \frac{2}{\epsilon \Omega} + \frac{3}{4} - \gamma_{E}  \right) \right] \right\rbrace - \sum_{\substack{j=1 }}^{p} a_{j} H_{0}^{(1)}(\Omega r_{nj} ) - & \\
- \sum_{\substack{k=1 \\ k \neq n}}^{q} \eta_{j}^{2} \Big\lbrace b_{j} H^{(1)}_{0}(\Omega r_{nj}) + \textbf{c}_{j} \cdot \textbf{e}_{r \, nj} \Omega & H_{1}^{(1)} (\Omega r_{nj} ) \Big\rbrace = \phi_{\mathrm{inc}} \Big|_{\textbf{x} = \textbf{X}_{n}}.
\end{split} \label{NeuDirScatt}
\end{equation} Again, since each Neumann inclusion introduces three unknowns into the system, we also impose matching conditions on the gradient of the wavefield as it approaches the $n$th Neumann inclusion:
\begin{equation}
\begin{split}
\textbf{c}_{n} \left\lbrace \frac{2}{i \pi} + \frac{\eta_{n}^{2} i \Omega^{2}}{\pi} \left( \log \frac{2}{\eta_{n} \Omega} - \frac{5}{4} - \gamma_{E} \right) - \frac{\eta_{n}^{2} \Omega^{2}}{2}  \right\rbrace  +  \sum_{\substack{j=1 }}^{p} \left\lbrace a_{j} \Omega H_{1}^{(1)}(\Omega r_{nj} ) \right\rbrace \textbf{e}_{r \, nj}  - &  \\ 
- \sum_{\substack{k=1 \\ k \neq n}}^{q} \eta_{k}^{2} \Big\lbrace \Big( - b_{k} \Omega H^{(1)}_{1}(\Omega r_{nk}) + \frac{\textbf{c}_{k} \cdot \textbf{e}_{r \,nk} }{2} \Omega^{2} \Big[ H^{(1)}_{0} (\Omega r_{nk}) - H^{(1)}_{2} (\Omega r_{nk}) \Big] \Big)  \textbf{e}_{r \, nk} + & \\
+ \Big( \frac{\textbf{c}_{k} \cdot \textbf{e}_{\varphi \, nk}}{r_{nk}}  \Omega H_{1}^{(1)} (\Omega r_{nk}) \Big) \textbf{e}_{\varphi \, nk} \Big\rbrace =  \nabla \phi_{\mathrm{inc}}\Big|_{\textbf{x} = \textbf{X}_{n}}, &
\end{split} \label{gradNeuDirScatt}
\end{equation}

Equations \eqref{DirNeuScatt}, \eqref{NeuDirScatt}, together with the $\mathbf{e}_{x}$- and $\mathbf{e}_{y}$-components of \eqref{gradNeuDirScatt}, evaluated for every $m = 1, \ldots, p$ and $n = 1, \ldots, q$, form a system of $(p+3q)$ equations for $(p+3q)$ unknowns, which can be written compactly as
\begin{equation}
\mathcal{G} \boldsymbol{\Lambda} = \boldsymbol{\Lambda}_{\mathrm{inc}}.
\label{scatteringSoln}
\end{equation}
Here $\mathcal{G}$ is a $(p+3q)\times(p+3q)$ matrix, while $\boldsymbol{\Lambda}$ and $\boldsymbol{\Lambda}_{\mathrm{inc}}$ are $(p+3q)\times 1$ vectors, with

\begin{equation*}
\boldsymbol{\Lambda}^{T} = \left[ a_{1}, \ldots, a_{p}, b_{1}, \ldots, b_{q},  \textbf{e}_{x} \cdot \textbf{c}_{1}, \ldots, \textbf{e}_{x} \cdot \textbf{c}_{q} , \textbf{e}_{y} \cdot \textbf{c}_{1}, \ldots, \textbf{e}_{y} \cdot \textbf{c}_{q}   \right], \quad \mbox{and}
\end{equation*}

$\boldsymbol{\Lambda}^{T}_{\mathrm{inc}} = \Big[ \phi_{\mathrm{inc}}  \Big|_{\textbf{x} = \textbf{X}_{1} }, \ldots, \phi_{\mathrm{inc}}  \Big|_{\textbf{x} = \textbf{X}_{p} }, $ $ \phi_{\mathrm{inc}}  \Big|_{\textbf{x} = \textbf{X}_{1} } , \ldots, \phi_{\mathrm{inc}}  \Big|_{\textbf{x} = \textbf{X}_{q} }, \textbf{e}_{x} \cdot \nabla \phi_{\mathrm{inc}}  \Big|_{\textbf{x} = \textbf{X}_{1}} , \ldots,  \textbf{e}_{x} \cdot \nabla \phi_{\mathrm{inc}}  \Big|_{\textbf{x} = \textbf{X}_{q} }, \textbf{e}_{y} \cdot \nabla \phi_{\mathrm{inc}}  \Big|_{\textbf{x} = \textbf{X}_{1} }, \ldots,  \textbf{e}_{y} \cdot \nabla \phi_{\mathrm{inc}}  \Big|_{\textbf{x} = \textbf{X}_{q} }   \Big]$

Provided $\mathcal{G}$ is invertible, the unknown source terms approximating the Dirichlet and Neumann inclusions are then determined from \eqref{scatteringSoln}.

\section{Topologically Non-Trivial Band Gaps in hexagonal and square lattices} \label{Sec:TOPO}
We apply the Floquet--Bloch formulation of Section \ref{EigenSection} to two canonical planar lattices: hexagonal and square. In each case we begin with a symmetry configuration that enforces (or permits) a spectral degeneracy in the band structure. We then break an appropriate reflection symmetry by switching the boundary condition on a subset of inclusions, thereby lifting the degeneracy and opening a bulk band gap. The gapped bands exhibit Berry-curvature localisation near valley points, and mirror-related perturbations yield chiral pairs with opposite valley signatures. Adjoining these chiral bulk phases yields an interface supporting ZLMs whose one dimensional dispersion relation lies within the two dimensional bulk gap. Finally, we demonstrate the central reconfigurable mechanism: by changing only the boundary-condition assignment (with geometry fixed), the location of the interface -- and hence the spatial position of the ZLM -- can be moved within the crystal.

\subsection{Actively creating topological insulators in two dimensional lattices}
Our aim is to provide an \emph{active} mechanism for producing and relocating valley-Hall interfaces in two-dimensional crystals. The starting point is a periodic lattice whose primitive cell enforces a symmetry-induced degeneracy in the Floquet--Bloch band structure (Dirac-type in hexagonal lattices; accidental but symmetry-protected in square lattices).   A controlled perturbation that breaks an appropriate reflection or inversion symmetry lifts this degeneracy and opens a bulk band gap. The formation of topologically non-trivial bulk band gaps, arising from symmetry broken degeneracies is well understood from group theory \cite{atkins2011molecular,dresselhaus2008group} and topological concepts; these form the basis of Valleytronics\cite{ma2016all,makwana2020hybrid, lu2016topological, makwana2018designing, gao2017valley, dong2017valley, yang2018topological, kang2018pseudo, chen_tunable_2018, he2019silicon, lu_observation_2016, ye2017observation, zhang_topological_2018,  wu_direct_2017, jung2018midinfrared, gao2018topologically, zhang2019valley,  xia2018observation, shalaev2018experimental, liu2018tunable}, a popular approach used to design topological insulators by exploiting the QVHE.   Provided the perturbation produces two mirror-related chiral unit cells whose valley indices have opposite sign, adjoining the corresponding bulk media yields an interface supporting ZLMs within the overlapping band gap.

The key feature exploited here is that the symmetry-breaking perturbation is realised not by moving inclusions, but by \emph{switching the condition applied on selected inclusion boundaries}. In the idealised mathematical model, this corresponds to switching between Dirichlet and Neumann constraints, which is a particularly clean way to generate chiral pairs while keeping the geometry fixed. In practice, the same concept may be implemented by inclusions whose effective electromagnetic response can be tuned  - for example, via reconfigurable elements~\cite{tang2020observations,putley2022tunable}, photo-responsive materials~\cite{gliozzi2020tunable}, or optically pumped conductive oxides (e.g. indium-tin-oxide (ITO)~\cite{tirole2022saturable}) - so that the effective boundary response of an inclusion changes under external control.

Once a pair of bulk phases is available within a \emph{common} frequency window with an overlapping gapped spectrum, the boundary-condition assignment can be made spatially dependent across the lattice. This creates internal domain interfaces whose location can be translated by reassigning inclusion conditions, thereby repositioning the interfacial and the associated ZLMs \emph{within the same fixed array of inclusions}. The remainder of the manuscript formalises this mechanism via asymptotic point-scatterer models and applies it to hexagonal~\cite{sakoda2004optical} and square~\cite{heine_group_nodate}  lattices.

\subsection{Hexagonal arrays}

By applying basic group-theoretical arguments \cite{dresselhaus_group_2008,atkins2011molecular} in periodic media \cite{heine_group_nodate,sakoda2004optical}, we can engineer symmetry-protected degeneracies and exploit them for robust waveguiding. Let $G_{\boldsymbol{\kappa}}$ denote the point-group symmetry \cite{makwana2018geometrically} of the hexagonal lattice in reciprocal space at Bloch wavevector $\boldsymbol{\kappa}$. A convenient basis for the irreducible representations of $G_{\boldsymbol{\kappa}}$ is the Floquet--Bloch eigenfunctions at $\boldsymbol{\kappa}$ \cite{sakoda2004optical}.

\begin{figure}[H]
\centering
\hspace*{-2cm}
\begin{tikzpicture}[scale=0.3, transform shape]

\begin{scope}[xshift=22.0cm, yshift=48.5cm]
		\node[regular polygon, regular polygon sides=4,draw, inner sep=7.0cm,rotate=0,line width=0.0mm, black, opacity=0.0,
           path picture={
               \node[rotate=0,opacity=1.0] at (-1,1){
                   \includegraphics[scale=1.25]{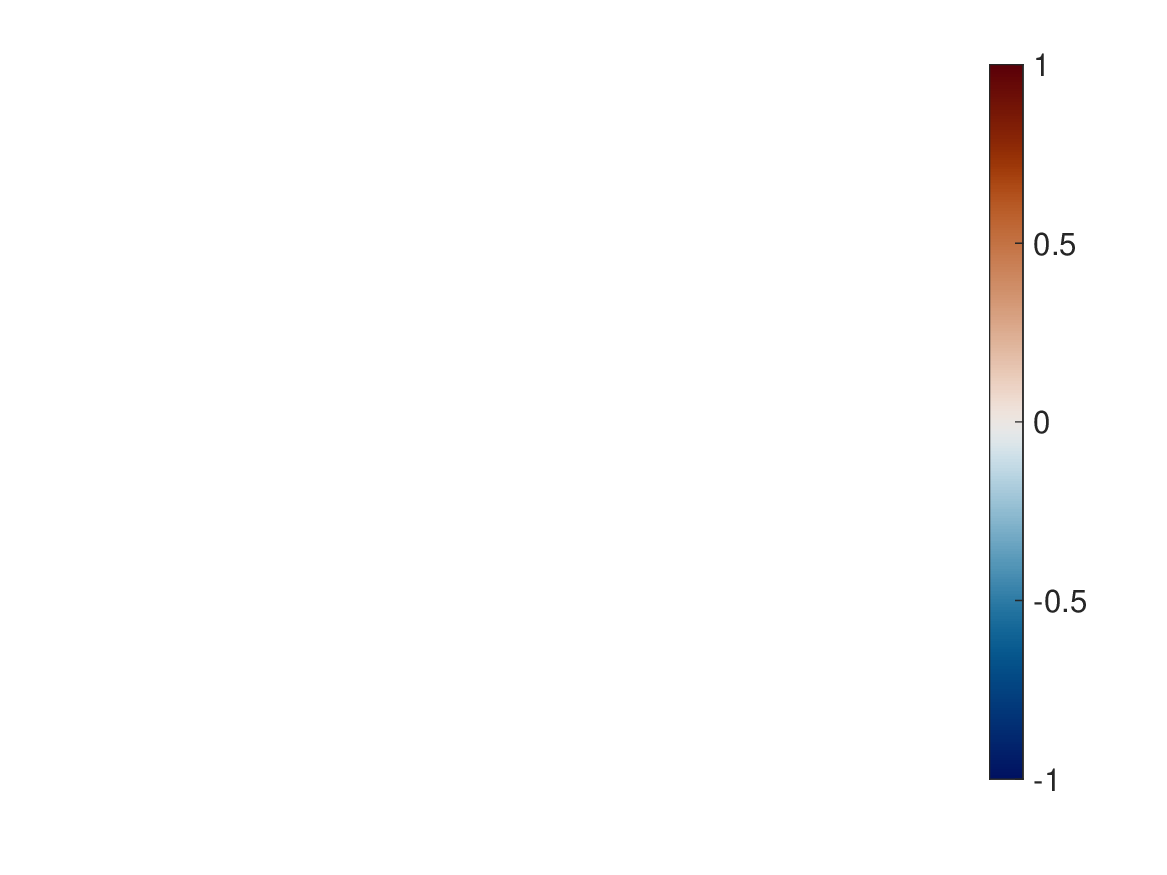}
               };
           }]{};
\end{scope}

\begin{scope}[xshift=17cm, yshift=56cm,scale=0.45]
		\node[regular polygon, regular polygon sides=6,draw, inner sep=6.25cm,rotate=0,line width=0.0mm, black, opacity=0.0,
           path picture={
               \node[rotate=0,opacity=1.0] at (-0.45,-0.35){
                   \includegraphics[scale=1.4]{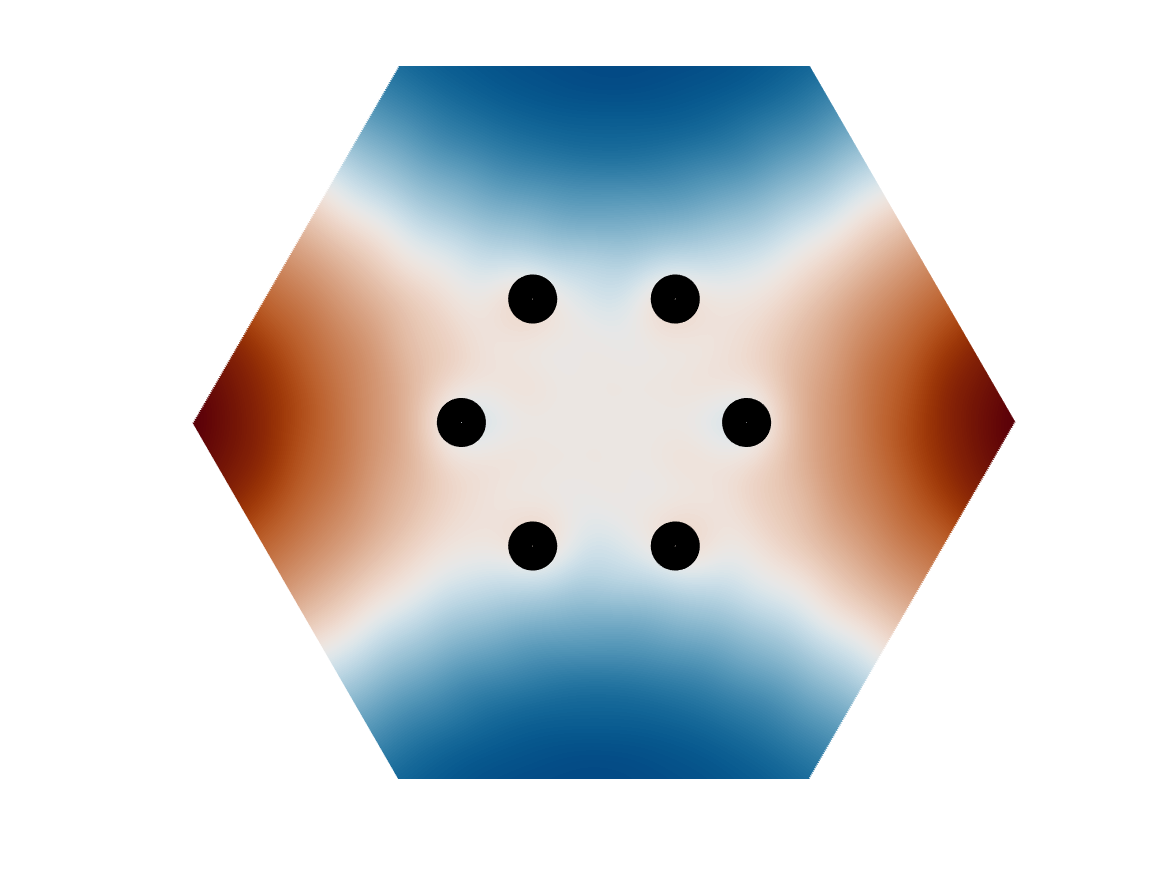}
               };
           }](polygon){};
		\node[below, scale=5.5, black] at (-9.0,6.75) {$\displaystyle (b)$};           
		\foreach \x in {1,4}{
        \draw [black,dashed, shorten <=-0.1cm,shorten >=-0.1cm,line width=0.5mm,](polygon.center) -- (polygon.side \x);}
       
\end{scope}  
\begin{scope}[xshift=17cm, yshift=48.05cm,scale=0.45]
		\node[regular polygon, regular polygon sides=6,draw, inner sep=6.25cm,rotate=0,line width=0.0mm, black, opacity=0.0,
           path picture={
               \node[rotate=0,opacity=1.0] at (-0.45,-0.35){
                   \includegraphics[scale=1.4]{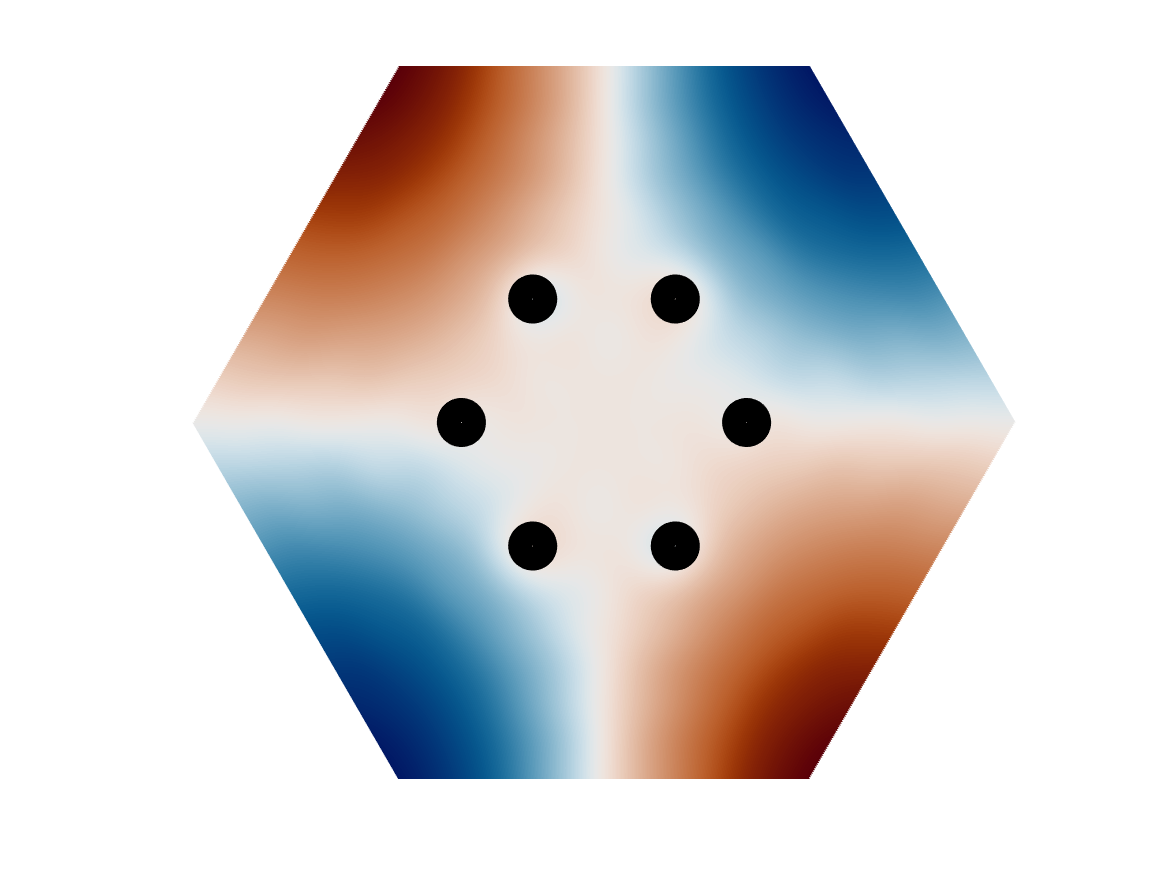}
               };
           }](polygon){};
		\node[below, scale=5.5, black] at (-9.0,6.75) {$\displaystyle (c)$};       
		\foreach \x in {1,4}{
        \draw [black,dashed, shorten <=-0.1cm,shorten >=-0.1cm,line width=0.5mm,](polygon.center) -- (polygon.side \x);}
           
\end{scope}  

\begin{scope}[xshift=23.9cm, yshift=52.025cm,scale=0.45]
		\node[regular polygon, regular polygon sides=6,draw, inner sep=6.25cm,rotate=0,line width=0.0mm, black, opacity=0.0,
           path picture={
               \node[rotate=0,opacity=1.0] at (-0.45,-0.35){
                   \includegraphics[scale=1.4]{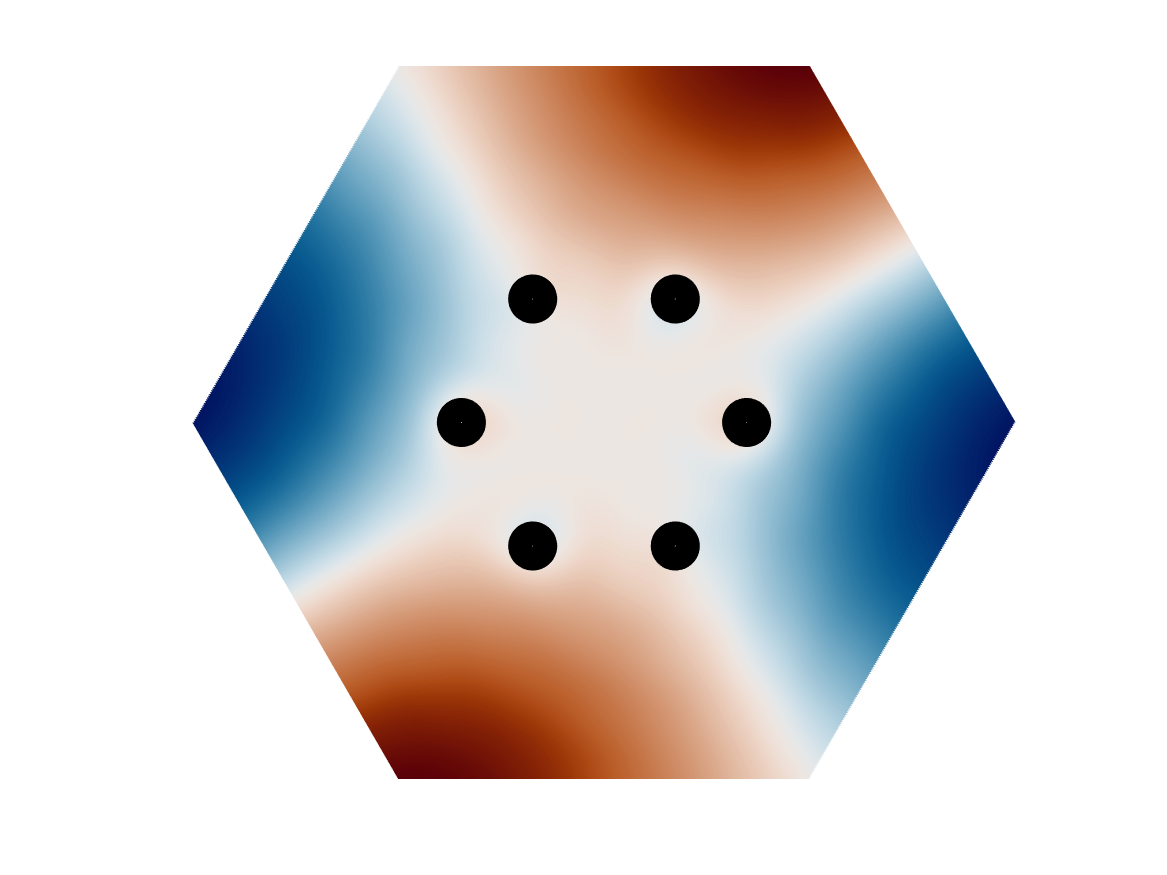}
               };
           }](polygon){};
		\node[above,left, scale=5.5, black] at (9.75,6.75) {$\displaystyle(d)$};           
		\foreach \x in {3,6}{
        \draw [black,dashed, shorten <=-0.1cm,shorten >=-0.1cm,line width=0.5mm,](polygon.center) -- (polygon.side \x);}
         
\end{scope}  

\begin{scope}[xshift=23.9cm, yshift=44.075cm,scale=0.45]
		\node[regular polygon, regular polygon sides=6,draw, inner sep=6.25cm,rotate=0,line width=0.0mm, black, opacity=0.0,
           path picture={
               \node[rotate=0,opacity=1.0] at (-0.45,-0.35){
                   \includegraphics[scale=1.4]{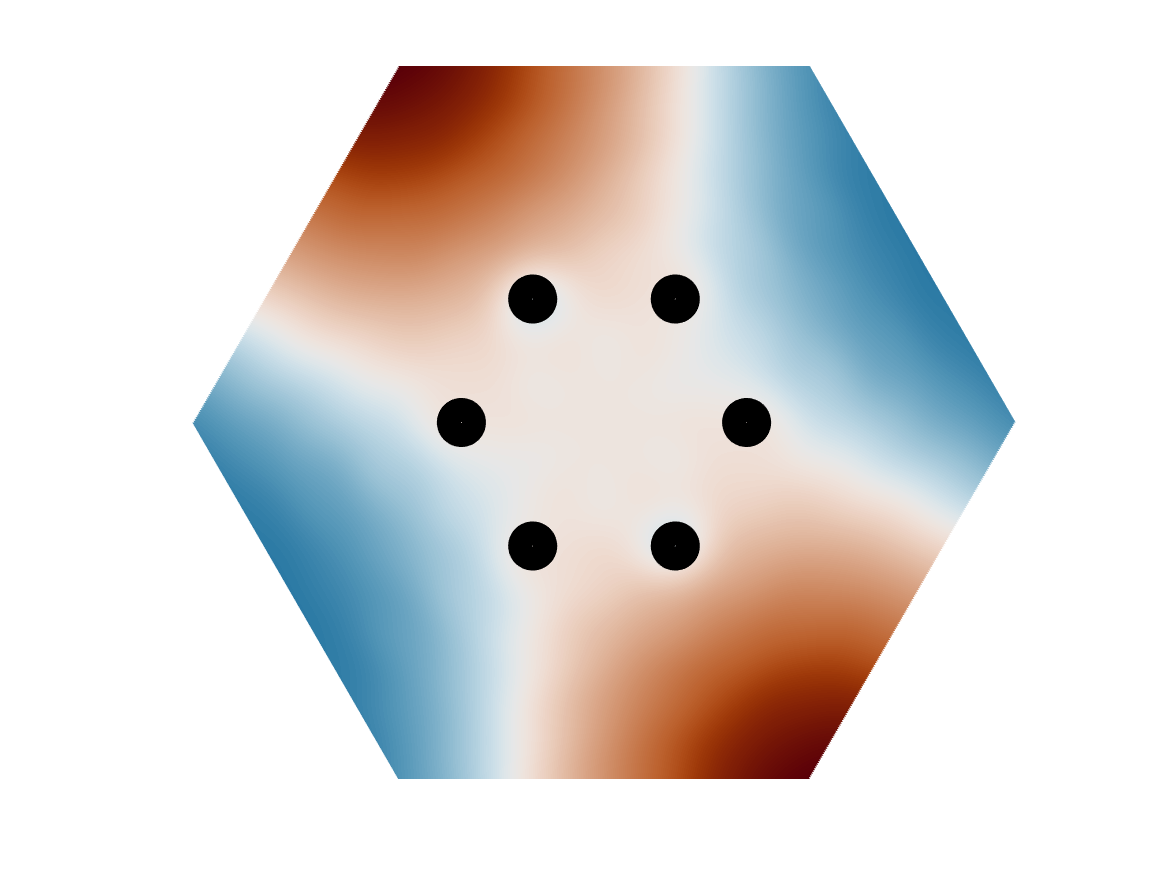}
               };
           }](polygon){};
		\node[above,left, scale=5.5, black] at (9.75,6.75) {$\displaystyle(e)$};    
		\foreach \x in {3,6}{
        \draw [black,dashed, shorten <=-0.1cm,shorten >=-0.1cm,line width=0.5mm,](polygon.center) -- (polygon.side \x);}
                
\end{scope}

\begin{scope}[xshift=-0.5cm, yshift=49.0cm]
		\node[regular polygon, regular polygon sides=4,draw, inner sep=9.0cm,rotate=0,line width=0.0mm, white,
           path picture={
               \node[rotate=0,opacity=1.0] at (2.5,1){
                   \includegraphics[scale=1.5]{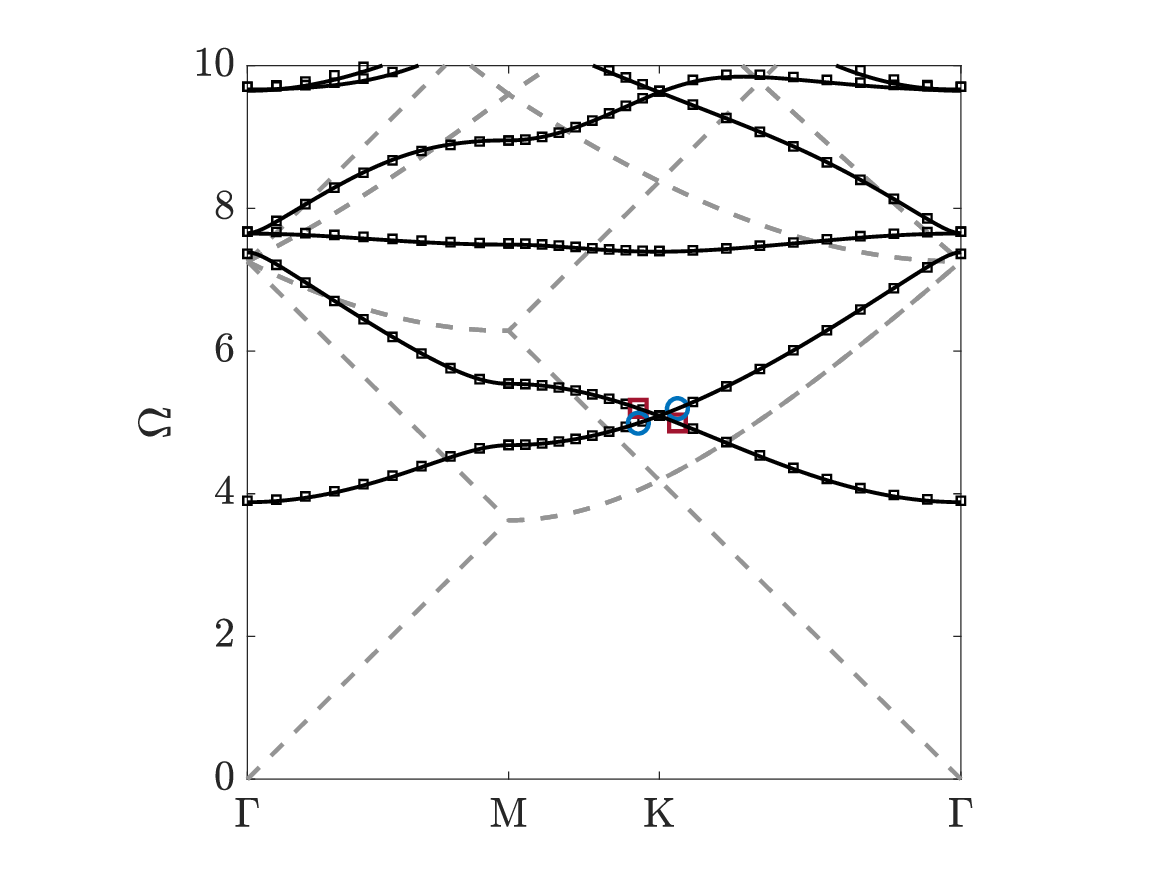}
               };
           }]{};
		\node[below, scale=2.5, black] at (2.4,-9) {$\displaystyle (a)$};    
\end{scope}

\begin{scope}[xshift=-2.5cm, yshift=50.0cm,scale=1.5]
	\node[regular polygon, regular polygon sides=6, draw, inner sep=0.5*6.28*10.0 pt,rotate=90,fill = white] at (0 pt,-6.28*10.0*1.5 pt) {};
	\draw[line width=0.5mm,gray,-] (0pt,-6.28*10.0*1.5 pt) -- (0+ 0.5*6.28*10*1.41 pt,-6.28*10.0*1.5 pt);
	\draw[line width=0.5mm,gray,-] (0+ 0.5*6.28*10*1.41 pt,-6.28*10.0*1.5 pt) -- (0+ 0.5*6.28*10*1.41  pt, 0.577*0.5*6.28*10*1.41 -6.28*10.0*1.5 pt);
	\draw[line width=0.5mm,gray,-] (0+ 0.5*6.28*10*1.41  pt, 0.577*0.5*6.28*10*1.41 -6.28*10.0*1.5 pt) -- (0pt,-6.28*10.0*1.5 pt);
	\node[below,left,scale=1.75] at (0pt,-6.28*10.0*1.5 pt) {$\displaystyle  \Gamma$}; 
	\node[below,right,scale=1.75] at (0+ 0.5*6.28*10*1.41 pt,-6.28*10.0*1.5 pt) {$\displaystyle  M$};
	\node[above,right,scale=1.75] at (0+ 0.5*6.28*10*1.41  pt, 0.577*0.5*6.28*10*1.41 -6.28*10.0*1.5 pt) {$\displaystyle  K$};
		\node[above,right,scale=1.75] at (0- 0.9*6.28*10*1.41  pt, 0.577*0.5*6.28*10*1.41 -6.28*10.0*1.5-3.4*10.0*1.5 pt) {$\displaystyle  K'$};
		\node[regular polygon, circle, draw, inner sep=1.25pt,rotate=0,line width=0.5mm,shading=fill,outer color=gray,gray] at (0pt,-6.28*10.0*1.5 pt)  {};
\node[regular polygon, circle, draw, inner sep=1.25pt,rotate=0,line width=0.5mm,shading=fill,outer color=gray,gray] at (0+ 0.5*6.28*10*1.41 pt,-6.28*10.0*1.5 pt)  {};
\node[regular polygon, circle, draw, inner sep=1.25pt,rotate=0,line width=0.5mm,shading=fill,outer color=gray,gray] at (0+ 0.5*6.28*10*1.41  pt, 0.577*0.5*6.28*10*1.41 -6.28*10.0*1.5 pt)  {};
\end{scope}

\end{tikzpicture}
\vspace*{-1cm}

\caption{Hexagonal primitive cell with six Dirichlet inclusions (Dirichlet inclusions are shown as black filled inclusions). Parameters are $\boldsymbol{\alpha}_{1} = \cos(\frac{\pi}{6}) \textbf{e}_{x} +  \sin(\frac{\pi}{6})\textbf{e}_{y}$, $\boldsymbol{\alpha}_{2} = -\cos(\frac{\pi}{6}) \textbf{e}_{x} +  \sin(\frac{\pi}{6})\textbf{e}_{y}$, $\textbf{X}_{1J} = 0.2( \textbf{e}_{x} \cos \frac{(5-J)\pi}{3} + \textbf{e}_{y} \sin \frac{(5-J)\pi}{3})$, and $\eta_{1J} = 0.03$ for $J = 1, \ldots ,6$. Panel (a) shows the Floquet--Bloch dispersion along $\Gamma$--$M$--$K$--$\Gamma$, the black lines are eigenvalues from \eqref{HelmholtzHardScheme_1} and the black squares are FEM computed results; the four marked points - $\color{myRED}\boldsymbol{\square}$ correspond to (b) \& (e) and $\color{myBLUE}\boldsymbol{\bigcirc}$ to (c) \& (d) - lie on the two branches adjacent to the degeneracy at $K$. Panels (b),(c) correspond to the two marked points on the $M$--$K$ segment  and panels (d),(e) to the two marked points on the $K$--$\Gamma$ segment; they (b)-(e) show the normalised real part of the corresponding Bloch eigenmodes. The dotted line indicates the reflection symmetry line used for parity classification. The unperturbed configuration retains three $\sigma_{v}$ reflection symmetries. The code required to compute all panels of this figure is given in the ancillary files, refer to section \ref{CodeAv}.}

\label{fig:HexDirac}
\end{figure}

Using compatibility relations \cite{sakoda2004optical} and referring to Figure~\ref{fig:HexDirac}, we classify the symmetries at $\Gamma$ and $K$ as $\{G_{\Gamma}, G_{K}\} = \{C_{6v}, C_{3v}\}$. Near the dispersionless crossing at $K$, the eigenstates transform according to the two-dimensional irreducible representation $E$ of $C_{3v}$, with basis functions $x^{2}-y^{2}$ and $xy$. The corresponding degenerate modes may therefore be identified as $E$-type states, namely $x^{2}-y^{2}$ $(\color{myRED}\boldsymbol{\square})$ and $xy$ $(\color{myBLUE}\boldsymbol{\bigcirc})$. Along the high-symmetry lines $\Gamma-K$ and $K-M$, the two adjacent branches, tracked by $\color{myRED}\boldsymbol{\square}\color{myRED}\boldsymbol{\square}$ and $\color{myBLUE}\boldsymbol{\bigcirc}\color{myBLUE}\boldsymbol{\bigcirc}$, have definite parity: one branch is purely even and the other purely odd under $\sigma_{v}$. The presence of $\sigma_{v}$ therefore prevents these modes from hybridising, yielding symmetry-protected degeneracies at $K$ and $K'$. The relevant excerpt of the $C_{3v}$ character table is given in Table~\ref{table:C3VCharacter}.
\vspace*{-0.25cm}
\begin{table}[h]
\centering
\begin{tabular}{c|c c c|c}
\cline{2-4}
& \multicolumn{3}{ c| }{Classes} \\
\hline
\multicolumn{1}{ ||c|  }{IRs} & $E$ & $2C_{3}$ & $3\sigma_{v}$ & \multicolumn{1}{ |c||  }{Appropriate basis functions} \\
\hline\hline
\multicolumn{1}{ ||c|  }{$A_{2}$} & $+1$ & $+1$ & $-1$ & \multicolumn{1}{ |c||  }{$\left\lbrace R_{z} \right\rbrace$} \\
\multicolumn{1}{ ||c|  }{$E$} & $+2$ & $-1$ & $0$ & \multicolumn{1}{ |c||  }{$\left\lbrace x^{2} - y^{2}, \quad xy \right\rbrace$} \\
\hline
\end{tabular}
\vspace*{-0.25cm}
\caption{Excerpt of the $C_{3v}$ character table; only quadratic and rotational basis functions spanning the irreducible representations (IRs) are required.}
\label{table:C3VCharacter}
\end{table}

\vspace*{-0.5cm}

\begin{figure}[H]
\centering
\hspace*{0.5cm}
\begin{tikzpicture}[scale=0.3, transform shape]

\begin{scope}[xshift=28.5cm, yshift=50.5cm]
		\node[regular polygon, regular polygon sides=4,draw, inner sep=7.0cm,rotate=0,line width=0.0mm, white,
           path picture={
               \node[rotate=0] at (-1,1){
                   \includegraphics[scale=1.25]{Figs/ColourBarEigenModes.eps}
               };
           }]{};
\end{scope}  

\begin{scope}[xshift=15cm, yshift=50.5cm]
		\node[regular polygon, regular polygon sides=4,draw, inner sep=7.0cm,rotate=0,line width=0.0mm, white,
           path picture={
               \node[rotate=0] at (1,1){
                   \includegraphics[scale=1.25]{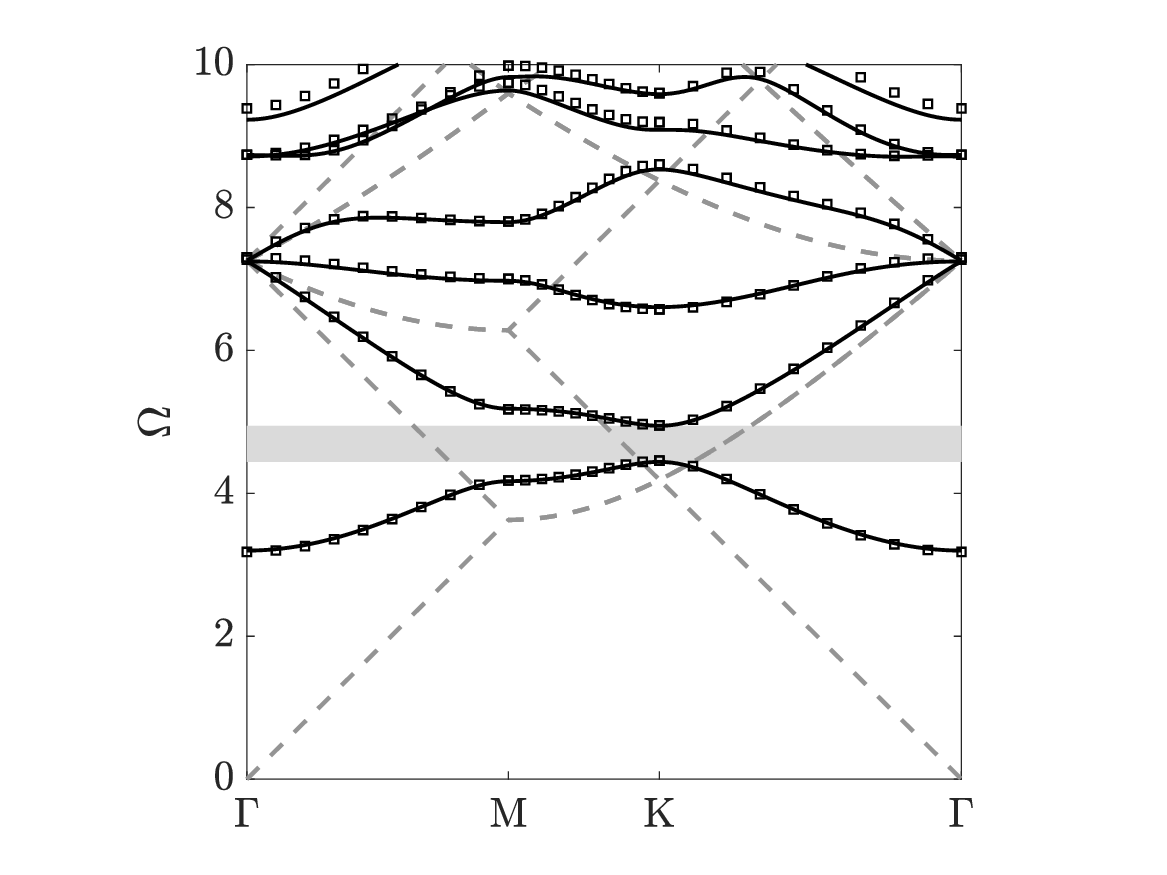}
               };
           }]{};
		\node[below, scale=2.5, black] at (1,-7.5) {$\displaystyle (c)$};           
\end{scope}  

\begin{scope}[xshift=4.0cm, yshift=56cm,scale=0.4]
		\node[regular polygon, regular polygon sides=4,draw, inner sep=6.5cm,rotate=0,line width=0.0mm, white,
           path picture={
               \node[rotate=0] at (0.2,-0.25){
                   \includegraphics[scale=1.4]{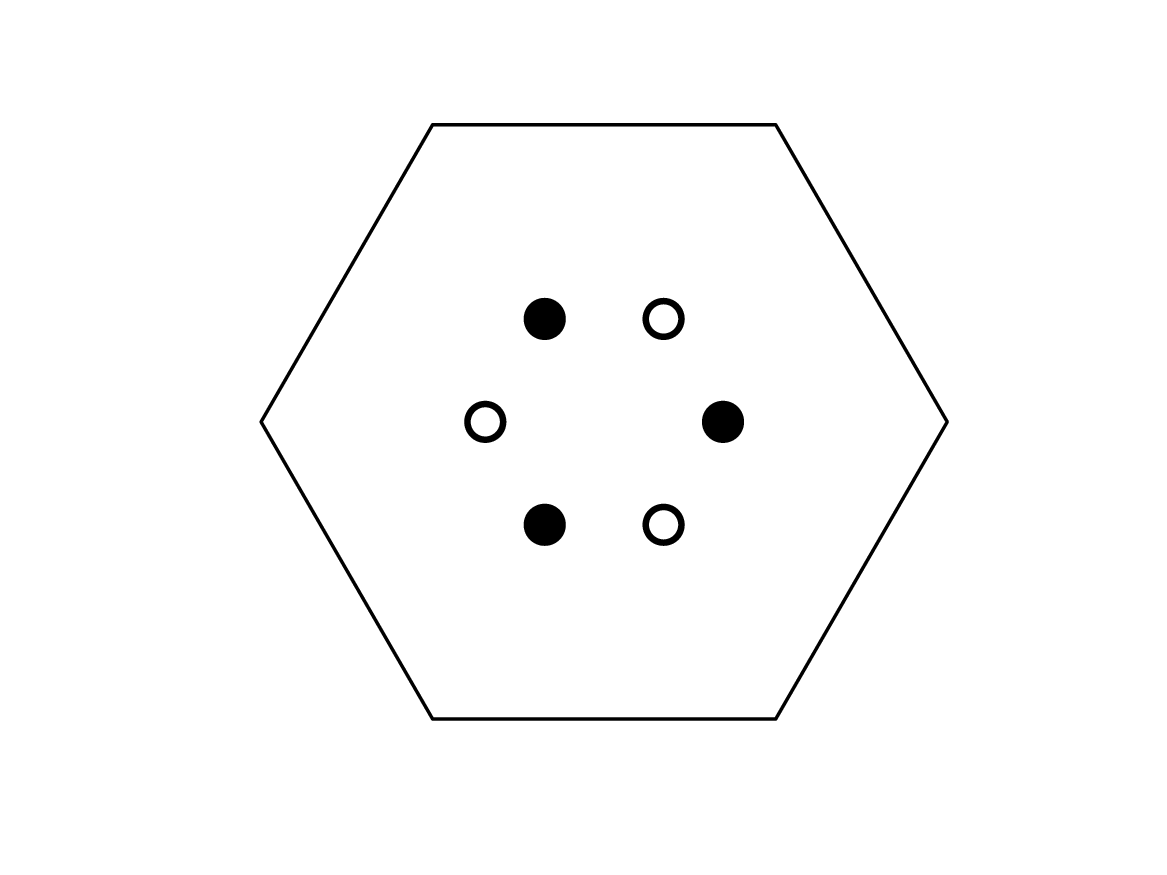}
               };
           }]{};
		\node[below, scale=6.3, black] at (-0.25,-7.5) {$\quad \displaystyle (a)$};             
\end{scope}

\begin{scope}[xshift=4.0cm, yshift=47.25cm,scale=0.4]
	\node[regular polygon, regular polygon sides=4,draw, inner sep=6.5cm,rotate=0,line width=0.0mm, white,
           path picture={
               \node[rotate=0] at (0.2,-0.25){
                   \includegraphics[scale=1.4]{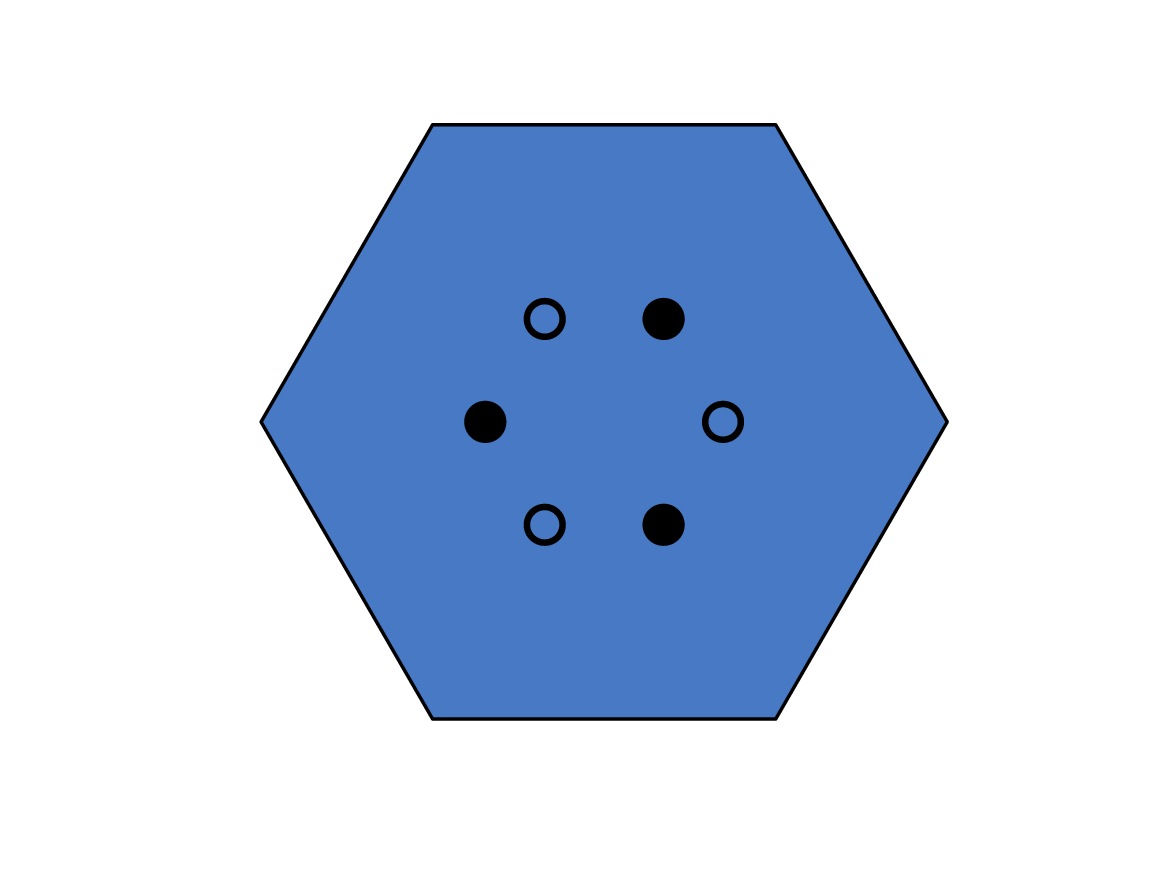}
               };
           }]{};
		\node[below, scale=6.3, black] at (-0.25,-7.5) {$\quad \displaystyle (b)$};           
\end{scope}  

\begin{scope}[xshift=30cm, yshift=58cm,scale=0.6]
		\node[regular polygon, regular polygon sides=4,draw, inner sep=6.5cm,rotate=0,line width=0.0mm, white,
           path picture={
               \node[rotate=0] at (-0.35,-0.25){
                   \includegraphics[scale=1.4]{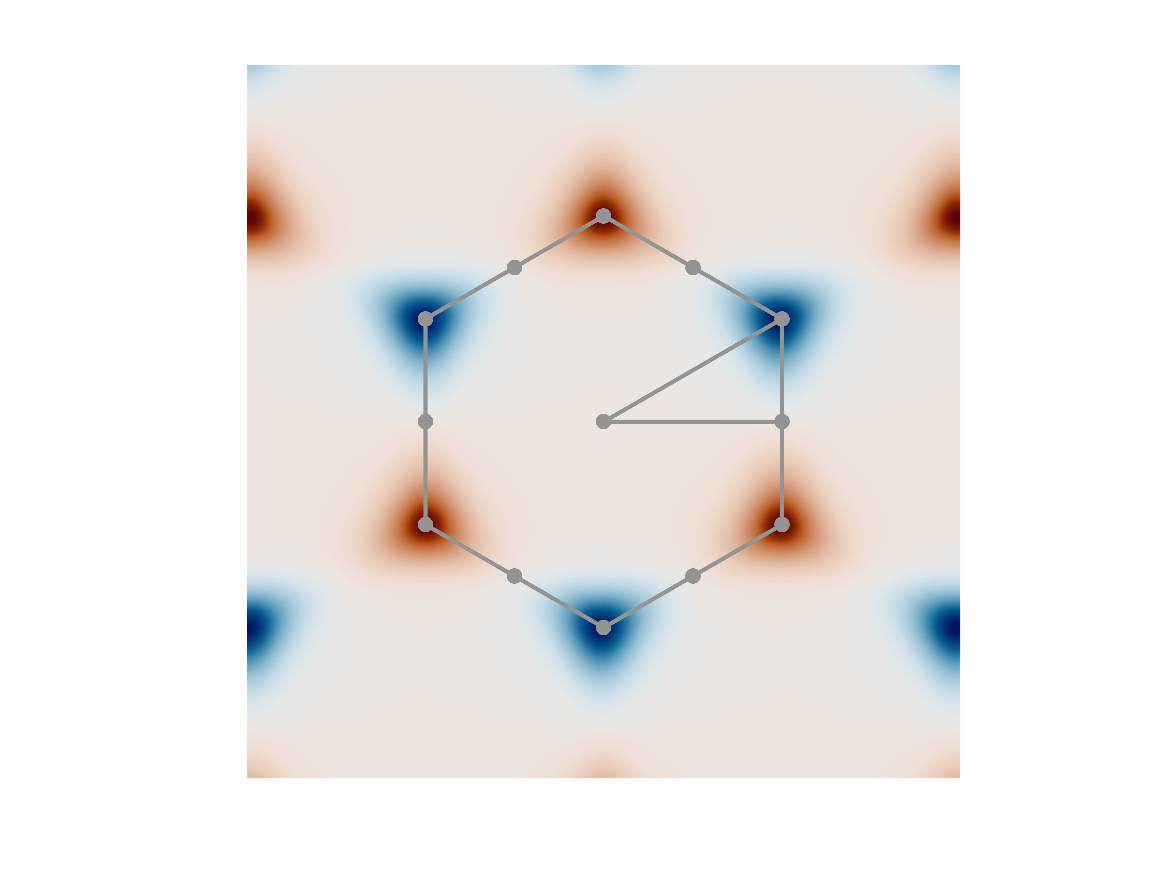}
               };
           }]{};
           \node[below,left,scale=4] at (0.0,-0.3) {$\displaystyle  \Gamma$}; 
			\node[below,right,scale=4] at (4.2,-0.3) {$\displaystyle  M$};
			\node[above,right,scale=4] at (4.2, 2.7) {$\displaystyle K$};
		    \node[above,left,scale=4] at (-4.2, 2.7) {$\displaystyle K$};
	        \node[below,scale=4] at (0, -4.8) {$\displaystyle \, K$};
		    \node[above,right,scale=4] at (4.2, -2.7) {$\displaystyle K'$};
		    \node[above,left,scale=4] at (-4.2, -2.7) {$\displaystyle K'$};
	       \node[above,scale=4] at (0, 5.0) {$\displaystyle \, \, \, K'$};
		   \node[below, scale=4.2, black] at (7,8.75) {$\displaystyle (d)$};              
\end{scope}

\begin{scope}[xshift=30cm, yshift=46.25cm,scale=0.6]
	\node[regular polygon, regular polygon sides=4,draw, inner sep=6.5cm,rotate=0,line width=0.0mm, white,
           path picture={
               \node[rotate=0] at (-0.35,-0.25){
                   \includegraphics[scale=1.4]{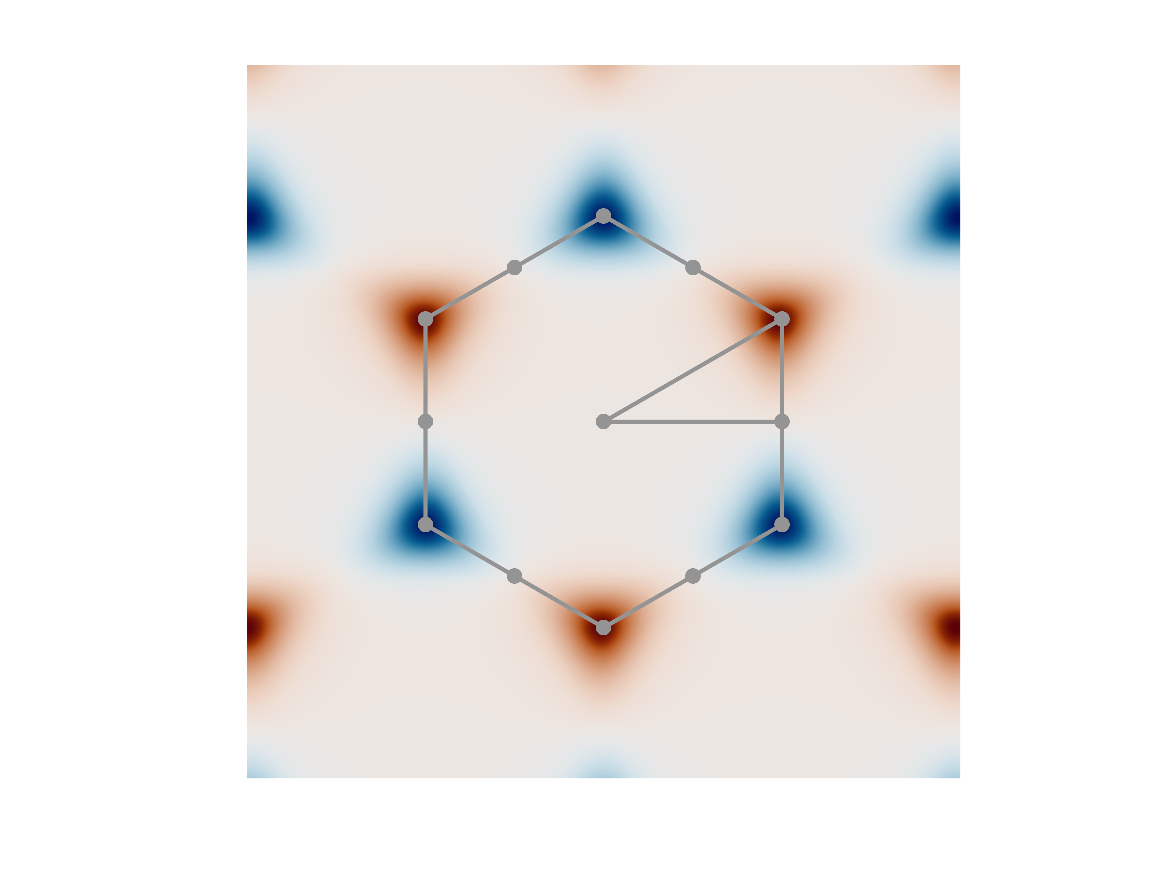}
               };
           }]{};
          \node[below,left,scale=4] at (0.0,-0.3) {$\displaystyle  \Gamma$}; 
			\node[below,right,scale=4] at (4.2,-0.3) {$\displaystyle  M$};
			\node[above,right,scale=4] at (4.2, 2.7) {$\displaystyle K$};
		    \node[above,left,scale=4] at (-4.2, 2.7) {$\displaystyle K$};
	        \node[below,scale=4] at (0, -4.8) {$\displaystyle \, K$};
		    \node[above,right,scale=4] at (4.2, -2.7) {$\displaystyle K'$};
		    \node[above,left,scale=4] at (-4.2, -2.7) {$\displaystyle K'$};
	       \node[above,scale=4] at (0, 5.0) {$\displaystyle \, \, \, K'$};
		\node[below, scale=4.2, black] at (7,8.75) {$\displaystyle (e)$};            
\end{scope}

\end{tikzpicture}

\caption{Perturbing the structure in Figure~\ref{fig:HexDirac} by making every other inclusion Neumann (depicted as  white or blue filled inclusions) creates the mirror-related chiral unit cells in (a) and (b). Panel (c) shows the Floquet--Bloch dispersion relation for cell (a), with the opened bulk band gap shaded in grey- again the black lines are eigenvalues from \eqref{HelmholtzHardScheme_1} and the black squares are FEM computed results. Panels (d) and (e) show the Berry curvature over the Brillouin zone for the bands immediately below and above the gap, respectively, exhibiting valley localisation of opposite sign near $K$ and $K'$. The code required to compute panel (c) of this figure is given in the ancillary files, refer to section \ref{CodeAv}.}

\label{fig:HexValley}
\end{figure}

We then break the relevant reflection symmetries by switching every other inclusion from Dirichlet (black inclusions) to Neumann (white inclusions), producing the mirror-related chiral unit cells shown in Figure~\ref{fig:HexValley}(a) and (b). This perturbation allows the previously symmetry-isolated $E$ modes to hybridise, so the two branches are no longer purely even or odd and band repulsion opens the bulk band gap shown in Figure~\ref{fig:HexValley}(c). The resulting gap is of valley-Hall type: the valley states, related by time-reversal symmetry, carry equal and opposite angular momentum, while the Berry curvature of the bands immediately below and above the gap is localised near $K$ and $K'$ with opposite sign, as shown in Figure~\ref{fig:HexValley}(d) and (e) \cite{xiao2010berry,palmer2022revealing}. The mirror-related chiral pair therefore realises opposite valley signatures within the same frequency window, confirming that the shaded region in Figure~\ref{fig:HexValley}(c) is a topologically non-trivial bulk gap. All the details concerning the computation of the Berry curvature along a single band are given in \cite{ungureanu2021localizing}.

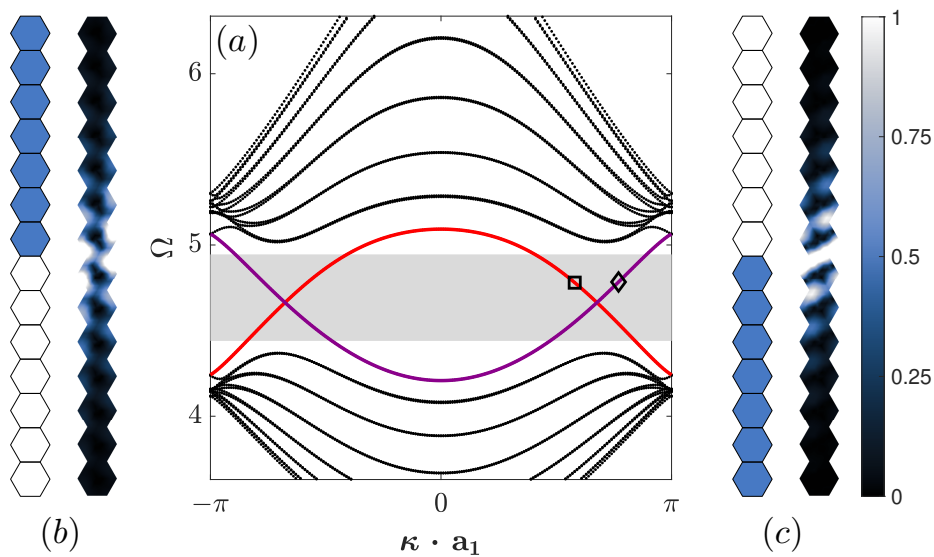
\begin{figure}[H]
\centering
\hspace*{+2cm}
\begin{tikzpicture}[scale=0.3, transform shape]

\begin{scope}[xshift=16.5cm, yshift=-4cm,scale=1.4]
		\node[regular polygon, regular polygon sides=4,draw, inner sep=7cm,rotate=0,line width=0.0mm, white,
           path picture={
               \node[rotate=0] at (-9,0){
                   \includegraphics[scale=1.25]{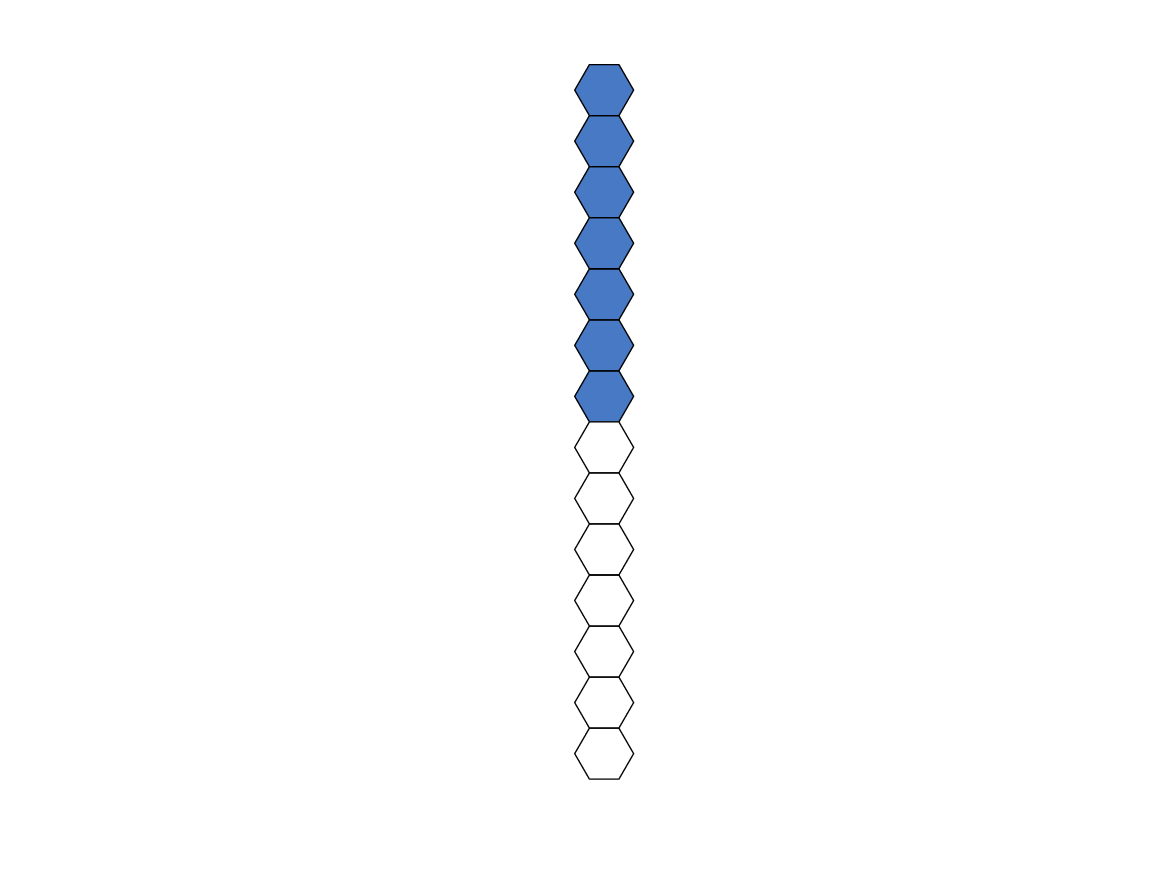}
               };
           }]{};
\end{scope}  

\begin{scope}[xshift=19.5cm, yshift=-4cm,scale=1.4]
		\node[regular polygon, regular polygon sides=4,draw, inner sep=7cm,rotate=0,line width=0.0mm, white,
           path picture={
               \node[rotate=0] at (-9,0){
                   \includegraphics[scale=1.25]{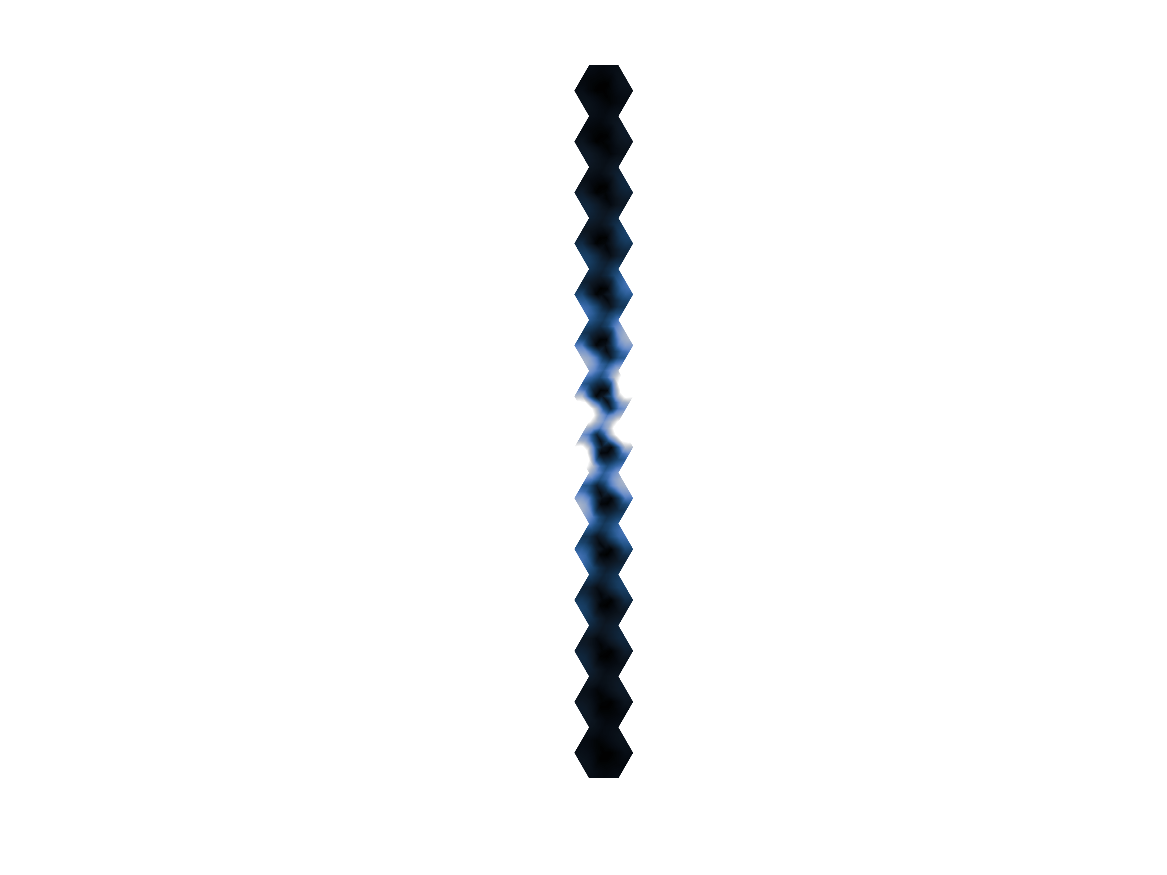}
               };
           }]{};
				\node[below, scale=2.86, black] at (-9.75,-7.5) {$\displaystyle (b)$};           
\end{scope}  

\begin{scope}[xshift=23.5cm, yshift=-4cm,scale=1.4]
		\node[regular polygon, regular polygon sides=4,draw, inner sep=7cm,rotate=0,line width=0.0mm, white,
           path picture={
               \node[rotate=0] at (-1.0,0){
                   \includegraphics[scale=1.25]{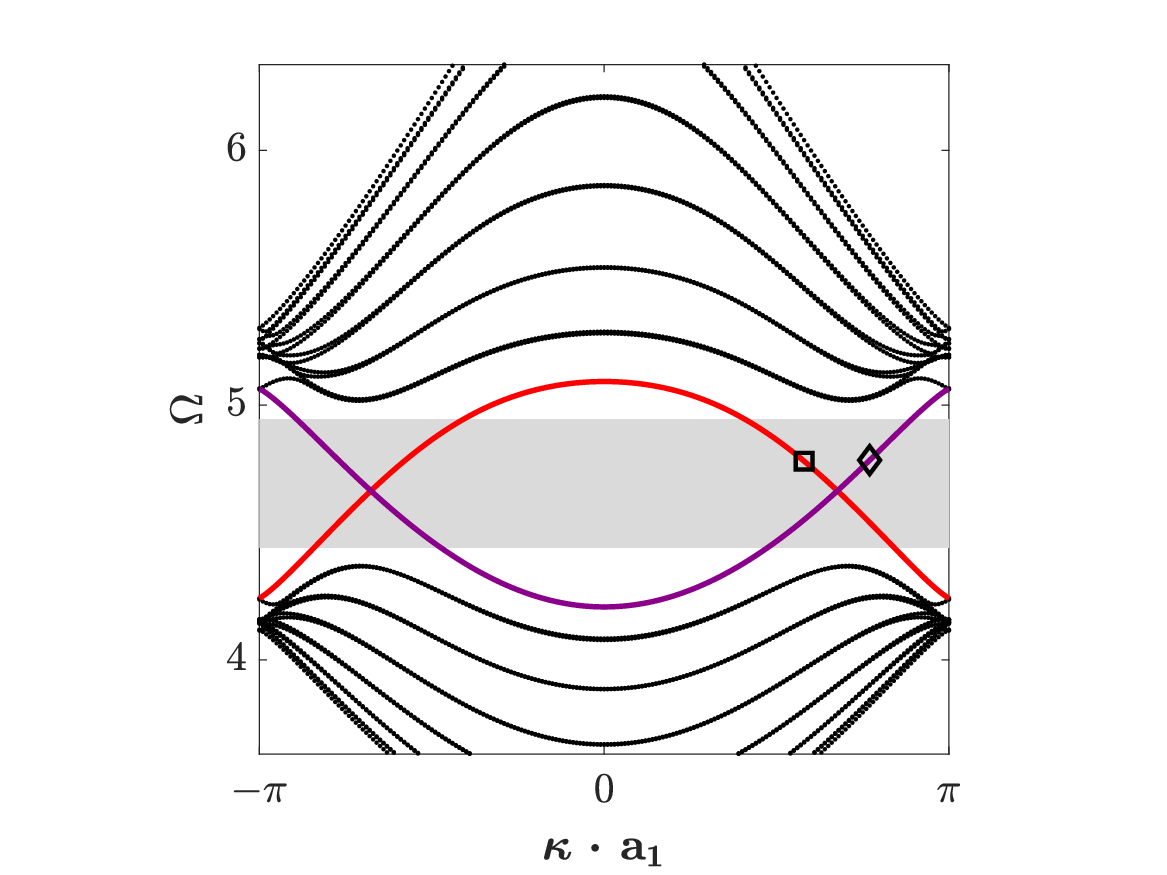}
               };
           }]{};
		\node[below, scale=2.86, black] at (-7,8.0) {$\displaystyle (a)$};           
\end{scope}  

\begin{scope}[xshift=48.5cm, yshift=-4cm,scale=1.4]
		\node[regular polygon, regular polygon sides=4,draw, inner sep=7cm,rotate=0,line width=0.0mm, white,
           path picture={
               \node[rotate=0] at (-9,0){
                   \includegraphics[scale=1.25]{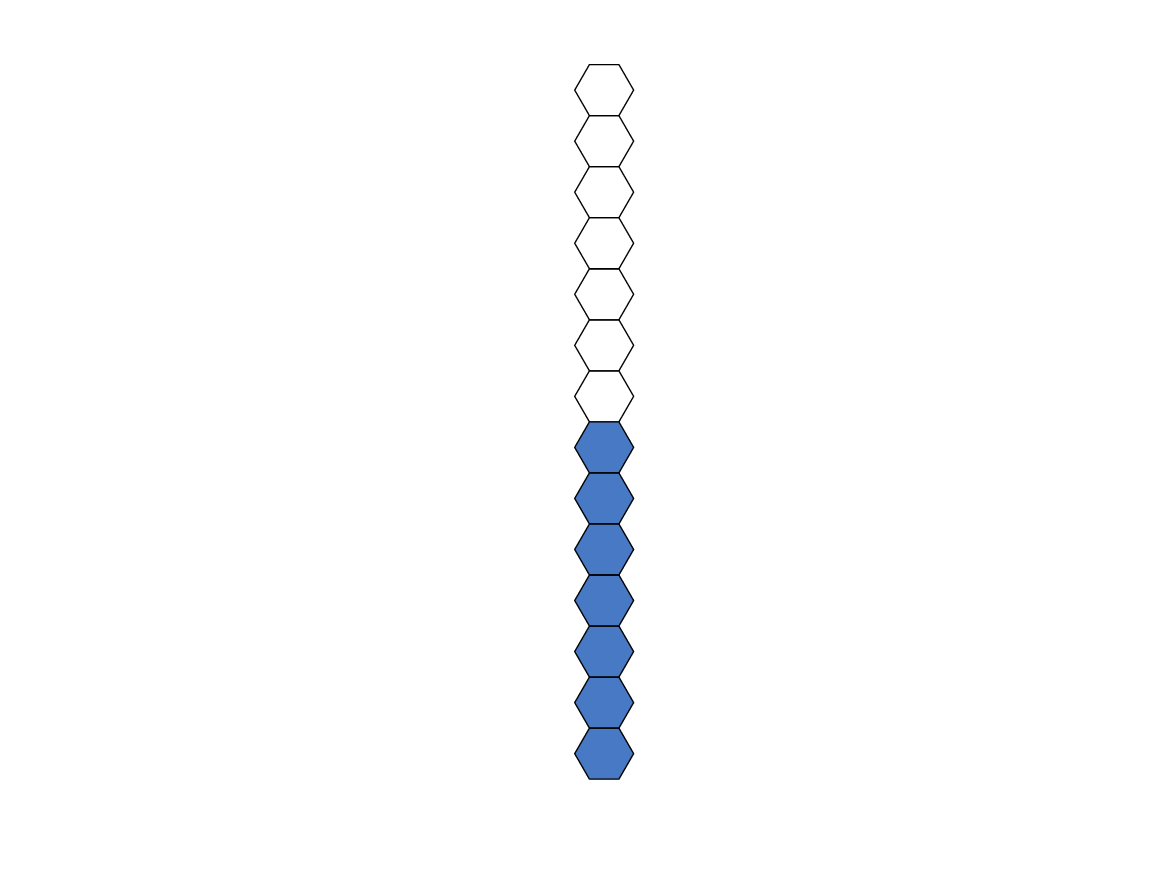}
               };
           }]{};
\end{scope}

\begin{scope}[xshift=51.5cm, yshift=-4cm,scale=1.4]
		\node[regular polygon, regular polygon sides=4,draw, inner sep=7cm,rotate=0,line width=0.0mm, white,
           path picture={
               \node[rotate=0] at (-9,0){
                   \includegraphics[scale=1.25]{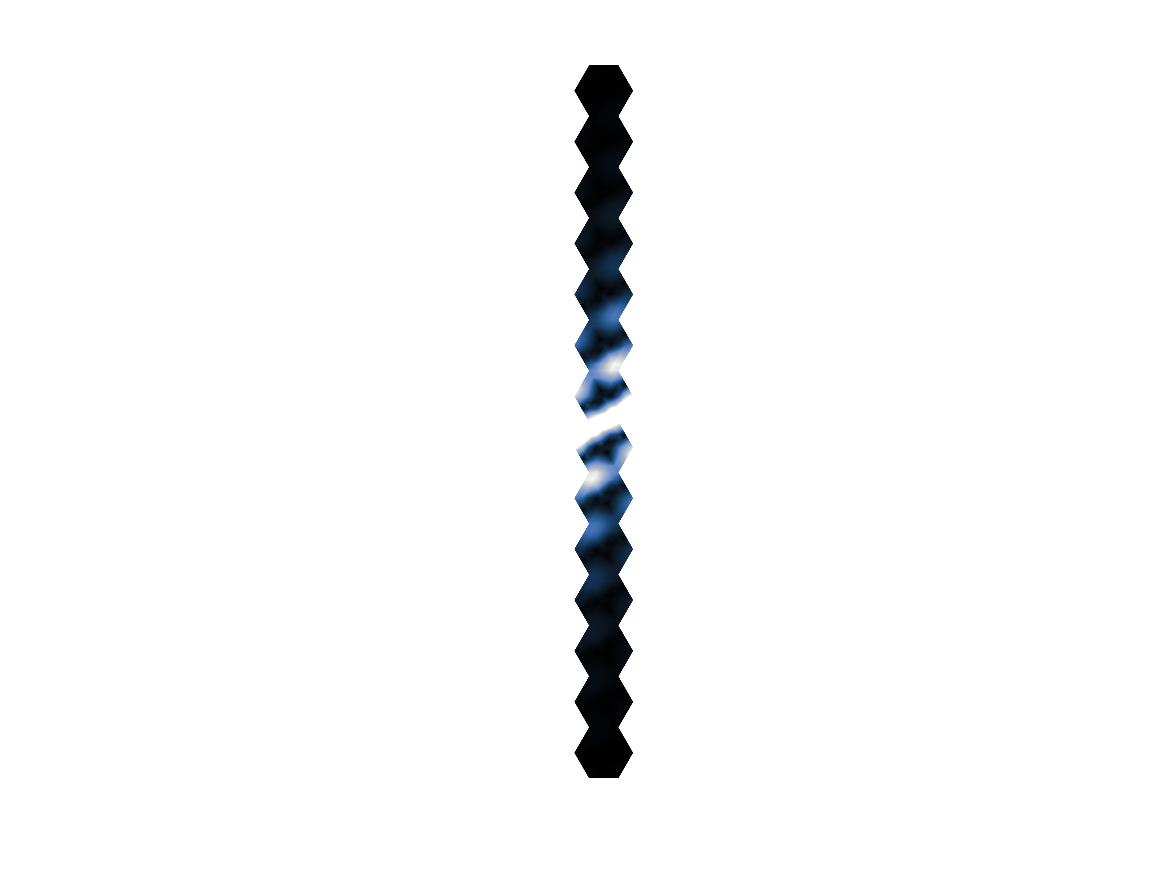}
               };
           }]{};
						\node[below, scale=2.86, black] at (-9.75,-7.5) {$\displaystyle (c)$}; 
\end{scope}  

\begin{scope}[xshift=54.5cm, yshift=-4cm,scale=1.4]
		\node[regular polygon, regular polygon sides=4,draw, inner sep=7cm,rotate=0,line width=0.0mm, white,
           path picture={
               \node[rotate=0] at (-18.0,0){
                   \includegraphics[scale=1.25]{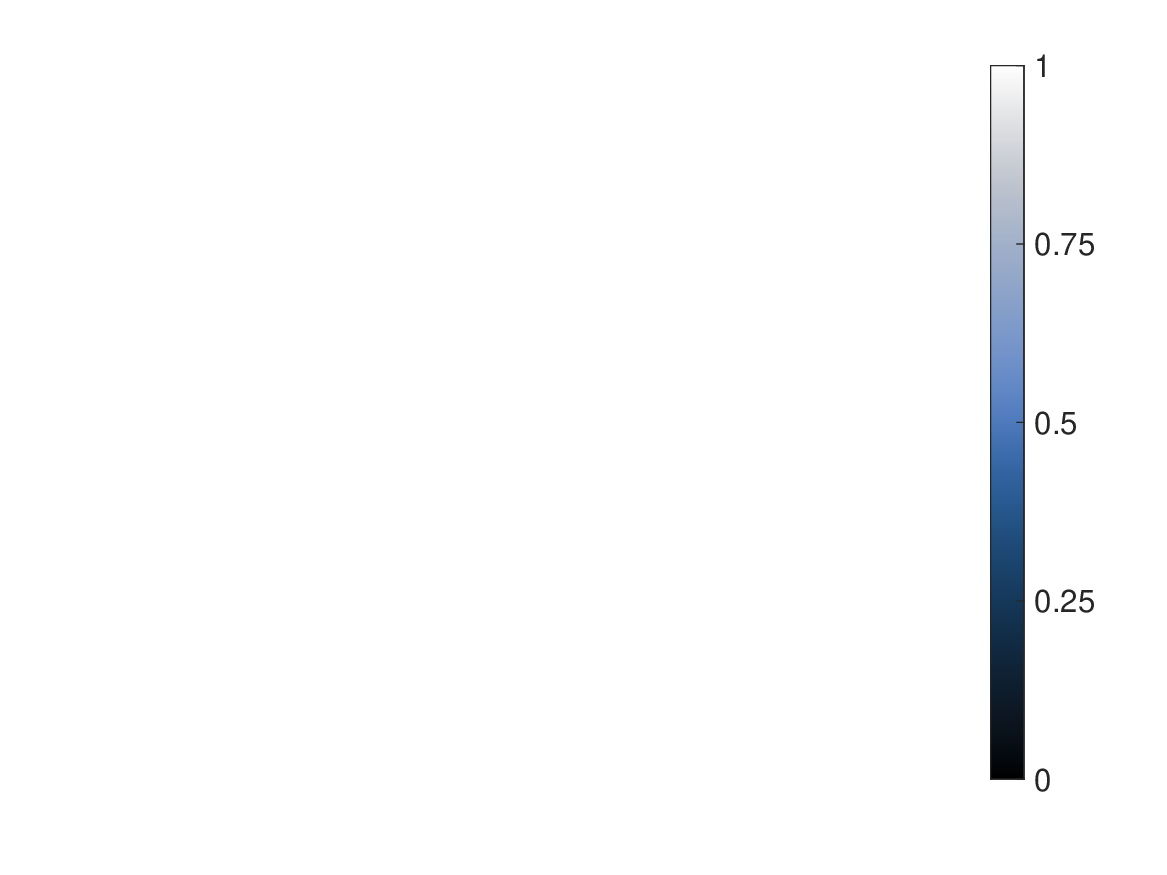}
               };
           }]{};
\end{scope}

\end{tikzpicture}

\caption{Ribbon/interface states for the hexagonal chiral pair. Panel (a) shows the ribbon Floquet--Bloch dispersion, derived from the eigenvalues of \eqref{HelmholtzHardScheme_1}; the shaded region is the projected bulk band gap from Figure~\ref{fig:HexValley} and the coloured curves are interfacial ZLM branches. Panels (b) and (c) show the normalised magnitude of the corresponding interface-localised eigenmodes (absolute value) at the marked $\kappa$ values at $\boldsymbol{\square}$ and $\boldsymbol{\diamond}$ respectively.}

\label{fig:HexRibbon}
\end{figure}

Across Figures \ref{fig:HexDirac}, \ref{fig:HexValley}, \ref{fig:SQDirac} and \ref{fig:SQValley}, the Floquet–Bloch spectra computed from the generalised eigenvalue problem \eqref{HelmholtzHardScheme_1}, derived via the method of matched asymptotic expansions, are seen to be in very good agreement with the corresponding FEM results (from full numerical simulations using

\begin{figure}[H]
\centering
\hspace*{-0cm}
\vspace{-1cm}
\begin{tikzpicture}[scale=0.3, transform shape]

\begin{scope}[xshift=14.5cm, yshift=22cm,scale=1.4]
               \node[rotate=0] at (-0.5,-0.25){
                   \includegraphics[scale=1.25]{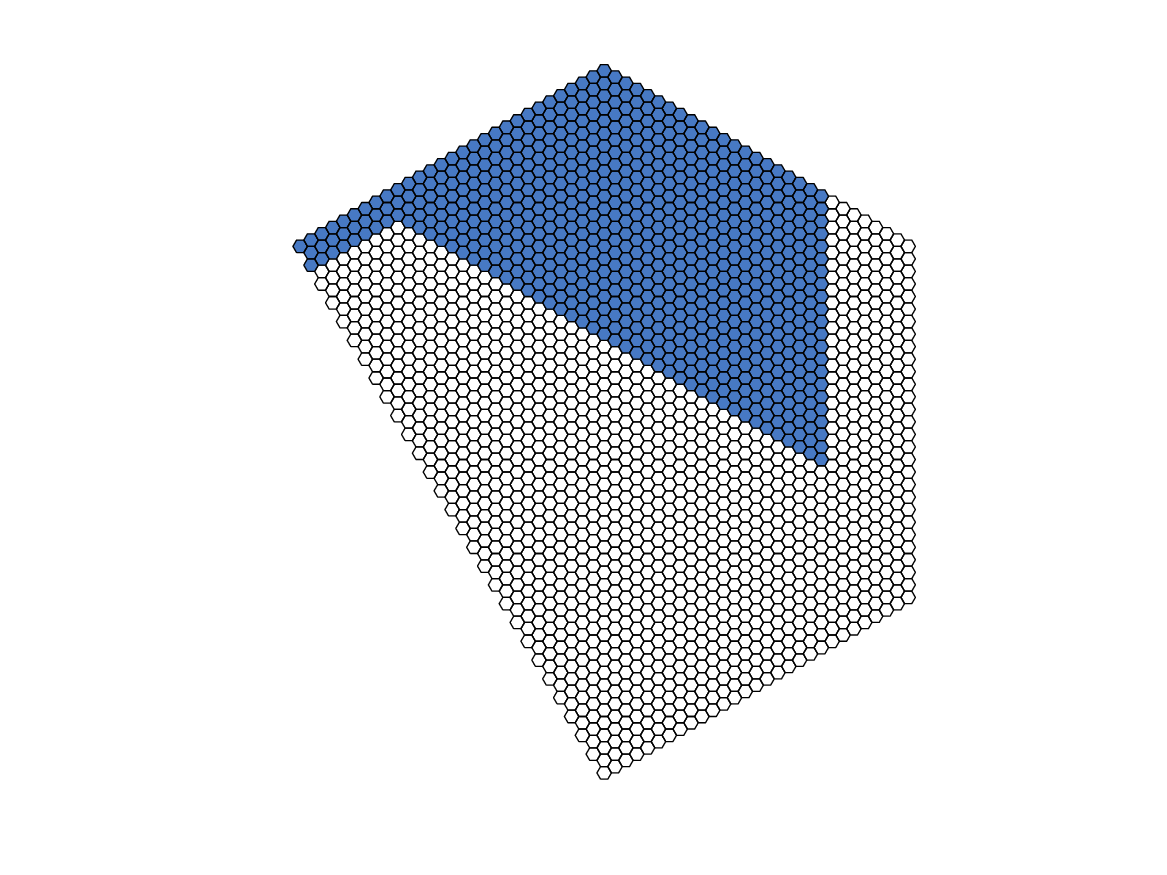}
               };    
		\node[below, scale=2.86, black] at (-8.5,8.75) {$\displaystyle (a)$};        
		
  \begin{scope}[xshift=-4cm, yshift= -2 cm,rotate=210,scale=4, transform shape]
 \draw[<-,line width=2.0] (0,0) -- (0.5*\xmax,0) node[right] {$\,$};
  \draw[-,line width=2.0] (0.3*\xmax,-0.15*\xmax) -- (0.3*\xmax,0.15*\xmax) node[right] {$\,$};
   \draw[-,line width=2.0] (0.35*\xmax,-0.15*\xmax) -- (0.35*\xmax,0.15*\xmax) node[right] {$\,$};
    \draw[-,line width=2.0] (0.4*\xmax,-0.15*\xmax) -- (0.4*\xmax,0.15*\xmax) node[right] {$\,$};
    \node[below, scale=0.65, black,rotate=-180] at (0.3*\xmax,-0.275*\xmax+0.425*\xmax) {Incident};
       \node[below, scale=0.65, black,rotate=-180] at (0.3*\xmax,-0.15*\xmax+0.4*\xmax)  {$\quad$ plane wave};
\end{scope}

\end{scope}

\begin{scope}[xshift=38cm, yshift=22cm,scale=1.4]
		\node[regular polygon, regular polygon sides=4,draw, inner sep=5.5cm,rotate=0,line width=0.0mm, white,
           path picture={
               \node[rotate=0] at (-0.5,-0.25){
                   \includegraphics[scale=1.25]{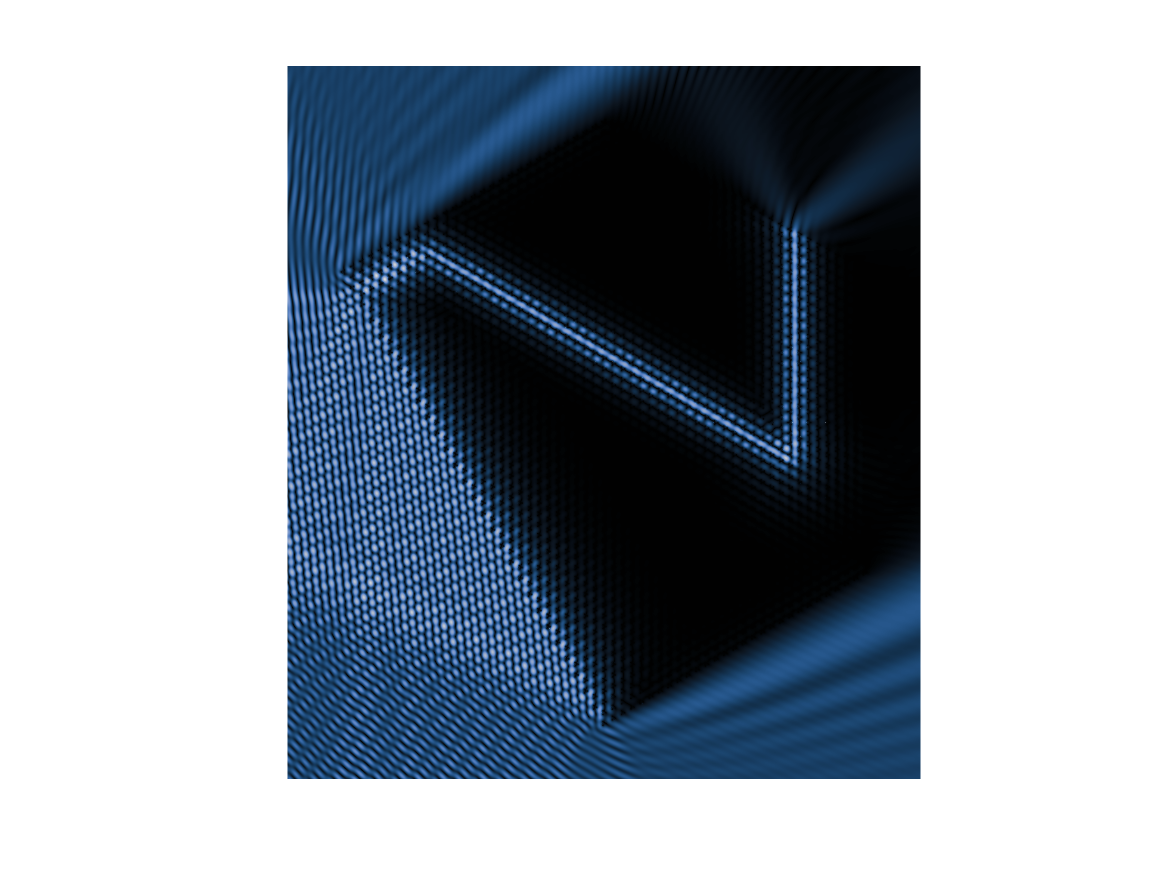}
               };
           }]{};
		\node[below, scale=2.86, black] at (-8.5,8.75) {$\displaystyle (b)$};           
\end{scope}

\begin{scope}[xshift=14.5cm, yshift=-2cm,scale=1.4]
               \node[rotate=0] at (-0.5,-0.25){
                   \includegraphics[scale=1.25]{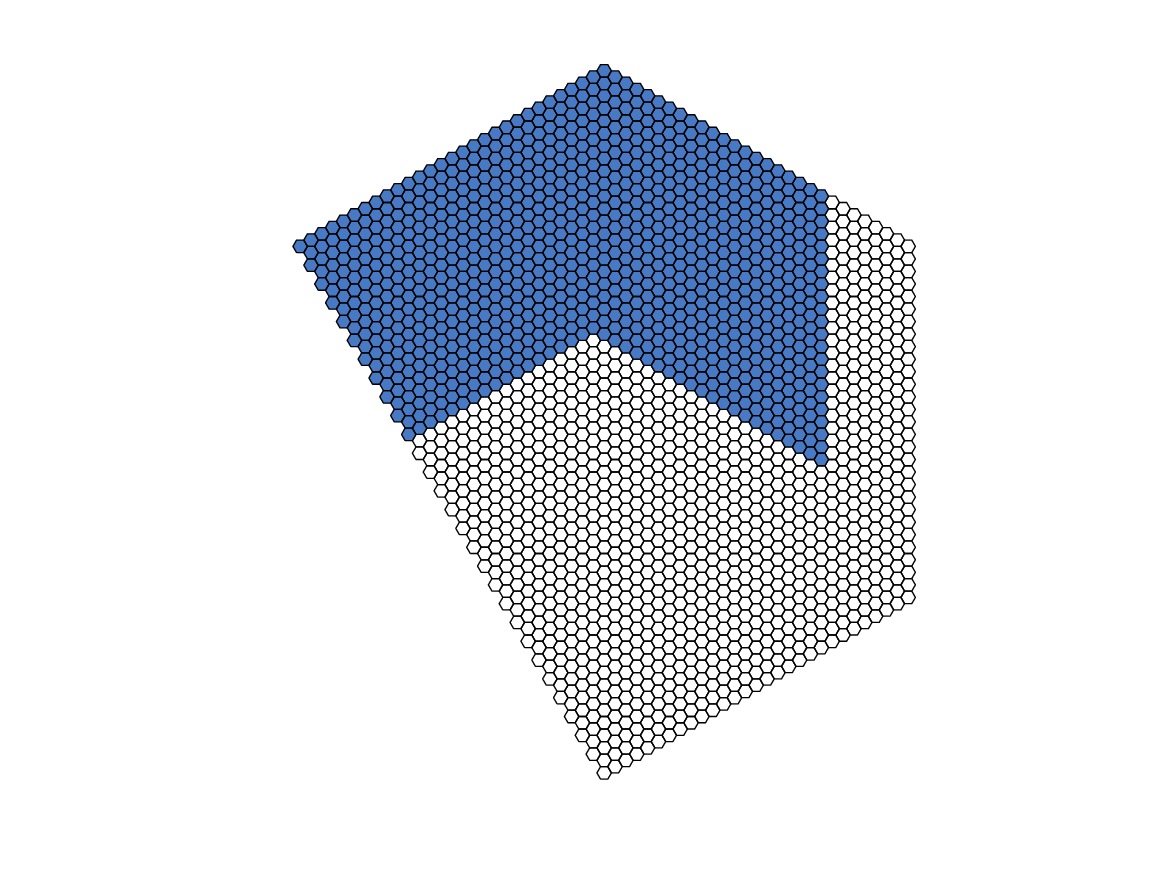}
               };
		\node[below, scale=2.86, black] at (-8.5,8.75) {$\displaystyle (c)$};           
 \begin{scope}[xshift=-4cm, yshift= -2 cm,rotate=210,scale=4, transform shape]
 \draw[<-,line width=2.0] (0,0) -- (0.5*\xmax,0) node[right] {$\,$};
  \draw[-,line width=2.0] (0.3*\xmax,-0.15*\xmax) -- (0.3*\xmax,0.15*\xmax) node[right] {$\,$};
   \draw[-,line width=2.0] (0.35*\xmax,-0.15*\xmax) -- (0.35*\xmax,0.15*\xmax) node[right] {$\,$};
    \draw[-,line width=2.0] (0.4*\xmax,-0.15*\xmax) -- (0.4*\xmax,0.15*\xmax) node[right] {$\,$};
\end{scope}
\end{scope}

\begin{scope}[xshift=38cm, yshift=-2cm,scale=1.4]
		\node[regular polygon, regular polygon sides=4,draw, inner sep=5.5cm,rotate=0,line width=0.0mm, white,
           path picture={
               \node[rotate=0] at (-0.5,-0.25){
                   \includegraphics[scale=1.25]{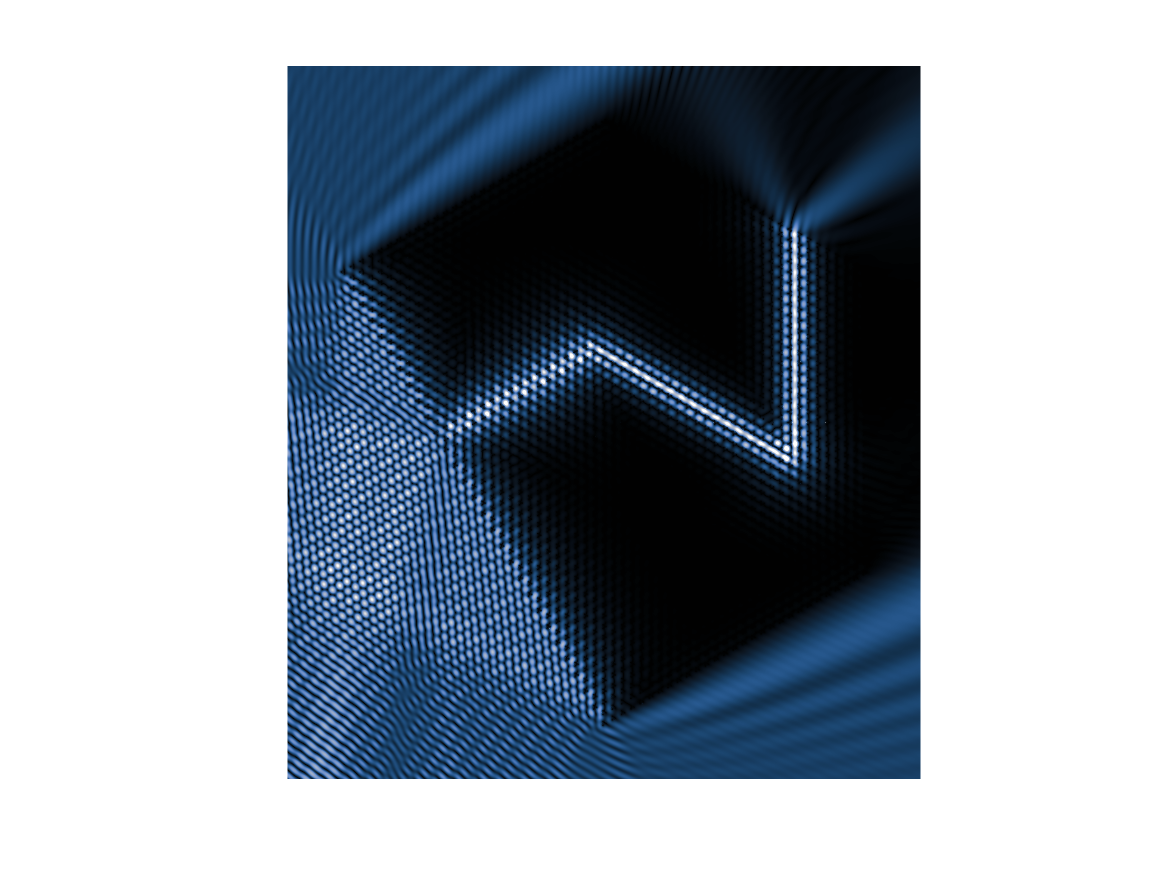}
               };
           }]{};
		\node[below, scale=2.86, black] at (-8.5,8.75) {$\displaystyle (d)$};           
\end{scope}

\begin{scope}[xshift=14.5cm, yshift=-26cm,scale=1.4]
               \node[rotate=0] at (-0.5,-0.25){
                   \includegraphics[scale=1.25]{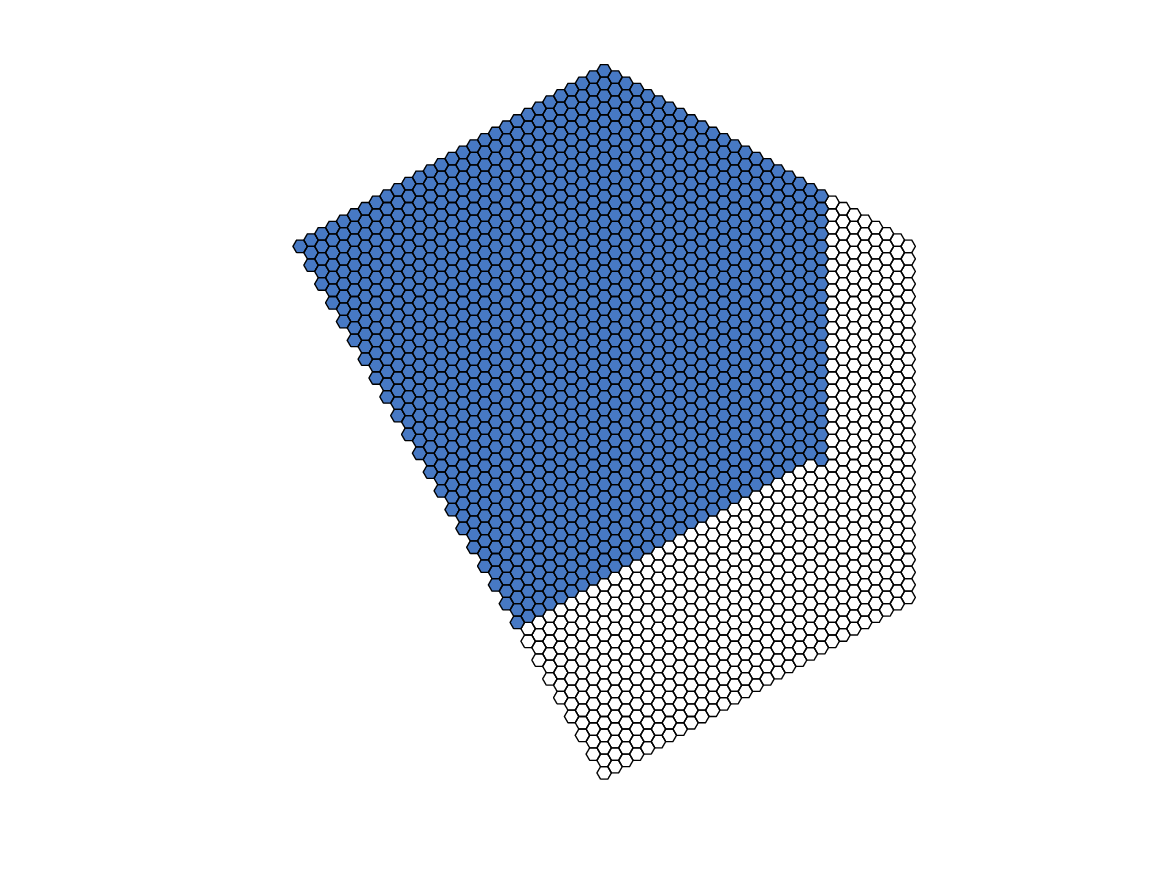}
               };
		\node[below, scale=2.86, black] at (-8.5,8.75) {$\displaystyle (e)$};           
		
		 \begin{scope}[xshift=-4cm, yshift= -2 cm,rotate=210,scale=4, transform shape]
 \draw[<-,line width=2.0] (0,0) -- (0.5*\xmax,0) node[right] {$\,$};
  \draw[-,line width=2.0] (0.3*\xmax,-0.15*\xmax) -- (0.3*\xmax,0.15*\xmax) node[right] {$\,$};
   \draw[-,line width=2.0] (0.35*\xmax,-0.15*\xmax) -- (0.35*\xmax,0.15*\xmax) node[right] {$\,$};
    \draw[-,line width=2.0] (0.4*\xmax,-0.15*\xmax) -- (0.4*\xmax,0.15*\xmax) node[right] {$\,$};
\end{scope}
\end{scope}

\begin{scope}[xshift=38cm, yshift=-26cm,scale=1.4]
		\node[regular polygon, regular polygon sides=4,draw, inner sep=5.5cm,rotate=0,line width=0.0mm, white,
           path picture={
               \node[rotate=0] at (-0.5,-0.25){
                   \includegraphics[scale=1.25]{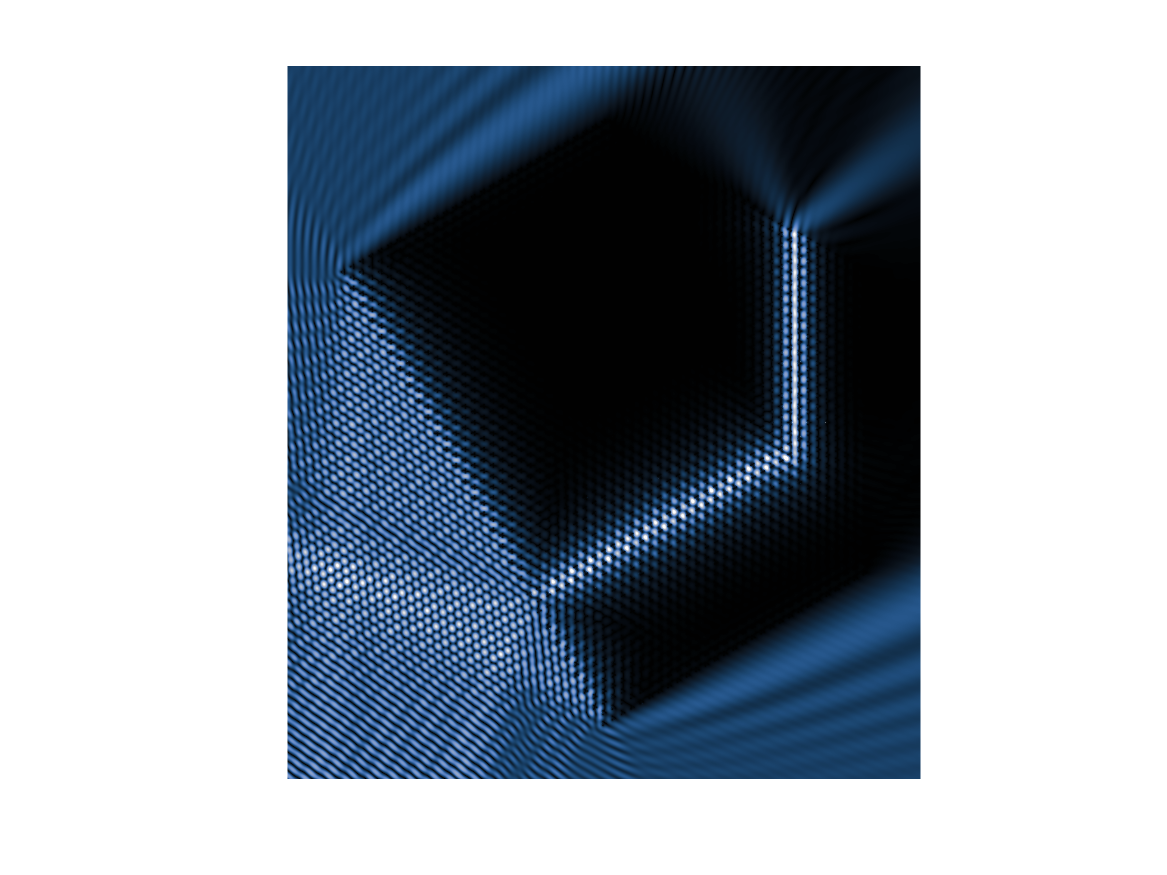}
               };
           }]{};
		\node[below, scale=2.86, black] at (-8.5,8.75) {$\displaystyle (f)$};           
\end{scope}

\begin{scope}[xshift=64.0cm, yshift=-3cm,scale=1.6]
		\node[regular polygon, regular polygon sides=4,draw, inner sep=7cm,rotate=0,line width=0.0mm, white,
           path picture={
               \node[rotate=0] at (-18.25,0){
                   \includegraphics[scale=1.25]{Figs/ColorBarAbsOslo.eps}
               };
           }]{};
\end{scope}  

\end{tikzpicture}

\caption{Plane-wave multiple-scattering simulation (the absolute value of the normalised total field is plotted, solving \eqref{scatteringSoln} to determine the unknown coefficients in \eqref{mscSOLNtotal}) for a finite collection of $2031$ hexagonal cells from Figure~\ref{fig:HexValley}(a),(b), with $\Omega = 4.8304$, $\theta_{\mathrm{inc}} = \frac{7 \pi}{6}$, $A_{\mathrm{inc}} = 1$ and $\textbf{X}_{\mathrm{inc}}$ the centre of the collection. The schematic configurations (a),(c),(e) differ only by the Dirichlet/Neumann assignments on inclusions (geometry fixed), thereby relocating the internal interface; the corresponding field magnitudes (b),(d),(f) show localisation along the moved interface. The frequency is chosen within the topologically non-trivial bulk band gap from fig. \ref{fig:HexValley}. The code required to compute panel (d) of this figure is given in the ancillary files, refer to section \ref{CodeAv}.}

\label{fig:HexScatt}
\end{figure}

\noindent the open source FEM package FreeFEM++ \cite{hecht2012new,laude2015phononic}). In both the hexagonal and square settings, and for both the unperturbed and symmetry-broken configurations, the asymptotic model accurately captures the location of the relevant degeneracies, the neighbouring dispersion branches, and the opening of the valley-type bulk band gaps. Although the derivation of \eqref{HelmholtzHardScheme_1} formally assumes asymptotically small inclusion radii, these comparisons indicate that the resulting simplified problem remains quantitatively reliable even beyond the strict classical asymptotic regime, namely in cases where the inclusion radii are not vanishingly small in the traditional sense. The asymptotically matched solutions are very accurate at low frequencies, with the observed error increasing (in comparison to FEM) at higher frequencies, when the wavelength becomes appreciable with respect to the radius of the inclusion. Moreover, the semi-analytical methods developed in Sections \ref{EigenSection} and \ref{MSTmannn} yield small matrix systems, enabling fast and efficient computation.

To obtain interface states, we form a ribbon super cell structure derived by stacking unit cells from Figure~\ref{fig:HexValley}(a) and (b) as shown in Figure~\ref{fig:HexRibbon}. The ribbon dispersion in Figure~\ref{fig:HexRibbon}(a) exhibits interfacial branches lying within the projected bulk gap, and Figure~\ref{fig:HexRibbon}(b) and (c) shows eigenmodes confined to the interface, i.e. the zero-line modes (ZLMs).

Crucially, the two bulk phases adjoining the interface are \emph{geometrically indistinguishable}: they differ only in which inclusions carry Dirichlet versus Neumann conditions. This makes the interface \emph{programmable}. Figure~\ref{fig:HexScatt} demonstrates this reconfigurability in a finite multiple-scattering simulation: panels (a), (c), and (e) differ only by their boundary-condition assignments, while the corresponding fields in panels (b), (d), and (f) show that the guided localisation follows the relocated interface. This is the central mechanism of the paper: changing the scatterer conditions alone changes where the topological interface exists, and hence repositions the ZLM within the same underlying lattice.

\subsection{Square arrays}

\begin{figure}[H]
\centering
\hspace*{-0.5cm}
\begin{tikzpicture}[scale=0.3, transform shape]

\begin{scope}[xshift=24.0cm, yshift=50.5cm]
		\node[regular polygon, regular polygon sides=4,draw, inner sep=7.0cm,rotate=0,line width=0.0mm, white,
           path picture={
               \node[rotate=0] at (-1,1){
                   \includegraphics[scale=1.25]{Figs/ColourBarEigenModes.eps}
               };
           }]{};
\end{scope}

\begin{scope}[xshift=2cm, yshift=50.5cm]
		\node[regular polygon, regular polygon sides=4,draw, inner sep=7.0cm,rotate=0,line width=0.0mm, white,
           path picture={
               \node[rotate=0] at (1,1){
                   \includegraphics[scale=1.25]{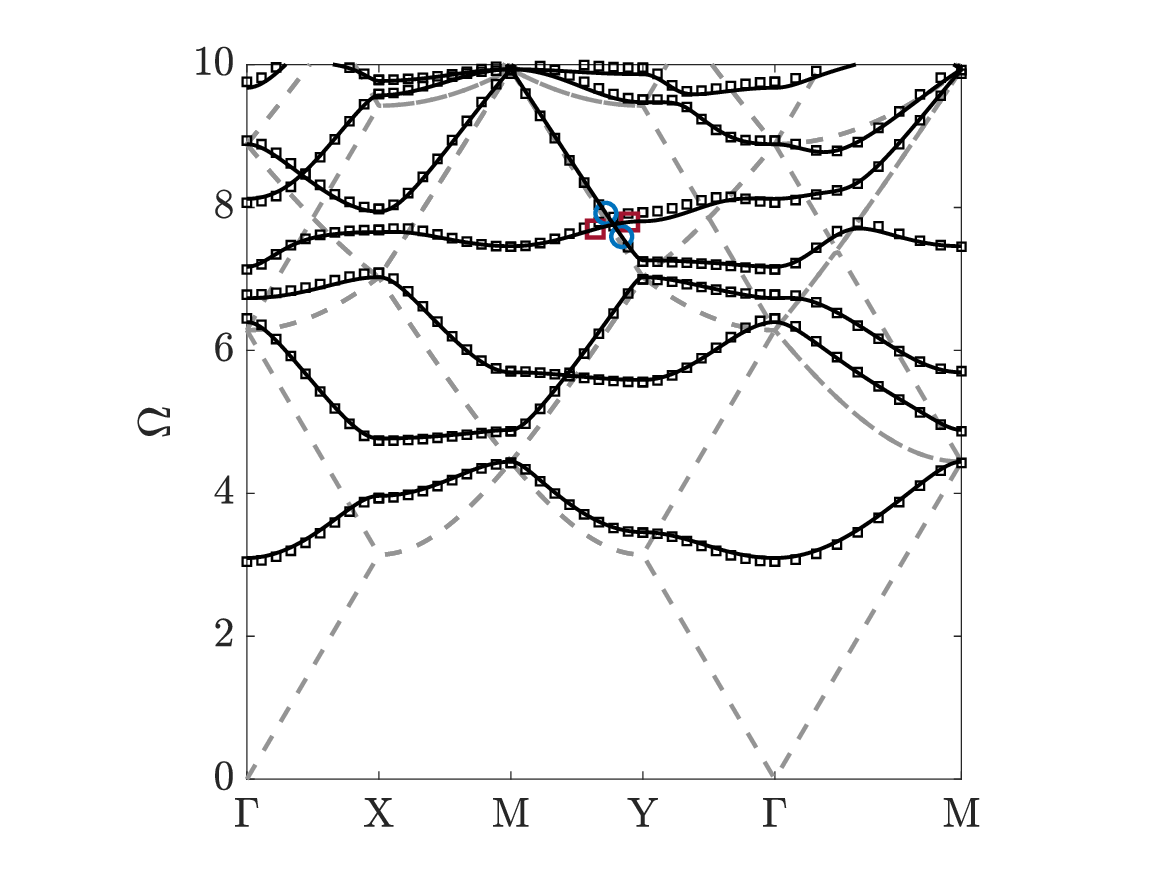}
               };
           }]{};
		\node[below, scale=2.5, black] at (1,-7.5) {$\displaystyle (a)$};           
\end{scope}

\begin{scope}[xshift=17cm, yshift=56cm,scale=0.45]
		\node[regular polygon, regular polygon sides=4,draw, inner sep=6.5cm,rotate=0,line width=0.0mm, white, opacity=0.0,
           path picture={
               \node[rotate=0,opacity=1.0] at (-0.475,-0.25){
                   \includegraphics[scale=1.4]{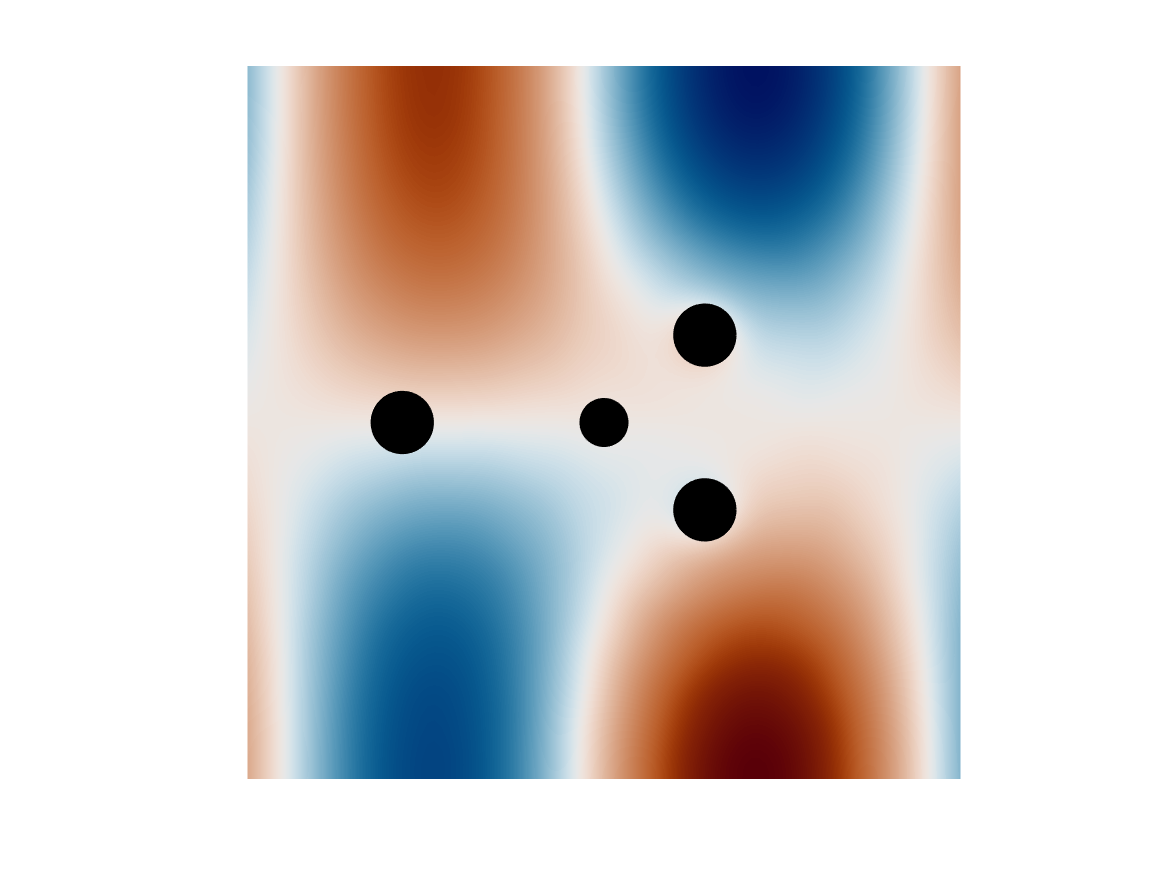}
               };
           }](polygon){};
		\node[below, scale=5.5, black] at (-10,8.75) {$\displaystyle (b)$};           
		\foreach \x in {2,4}{
        \draw [black,dashed, shorten <=-0.1cm,shorten >=-0.1cm,line width=0.5mm,](polygon.center) -- (polygon.side \x);}
         
\end{scope}

\begin{scope}[xshift=17cm, yshift=47.75cm,scale=0.45]
	\node[regular polygon, regular polygon sides=4,draw, inner sep=6.5cm,rotate=0,line width=0.0mm, white, opacity=0.0,
           path picture={
               \node[rotate=0,opacity=1.0] at (-0.475,-0.25){
                   \includegraphics[scale=1.4]{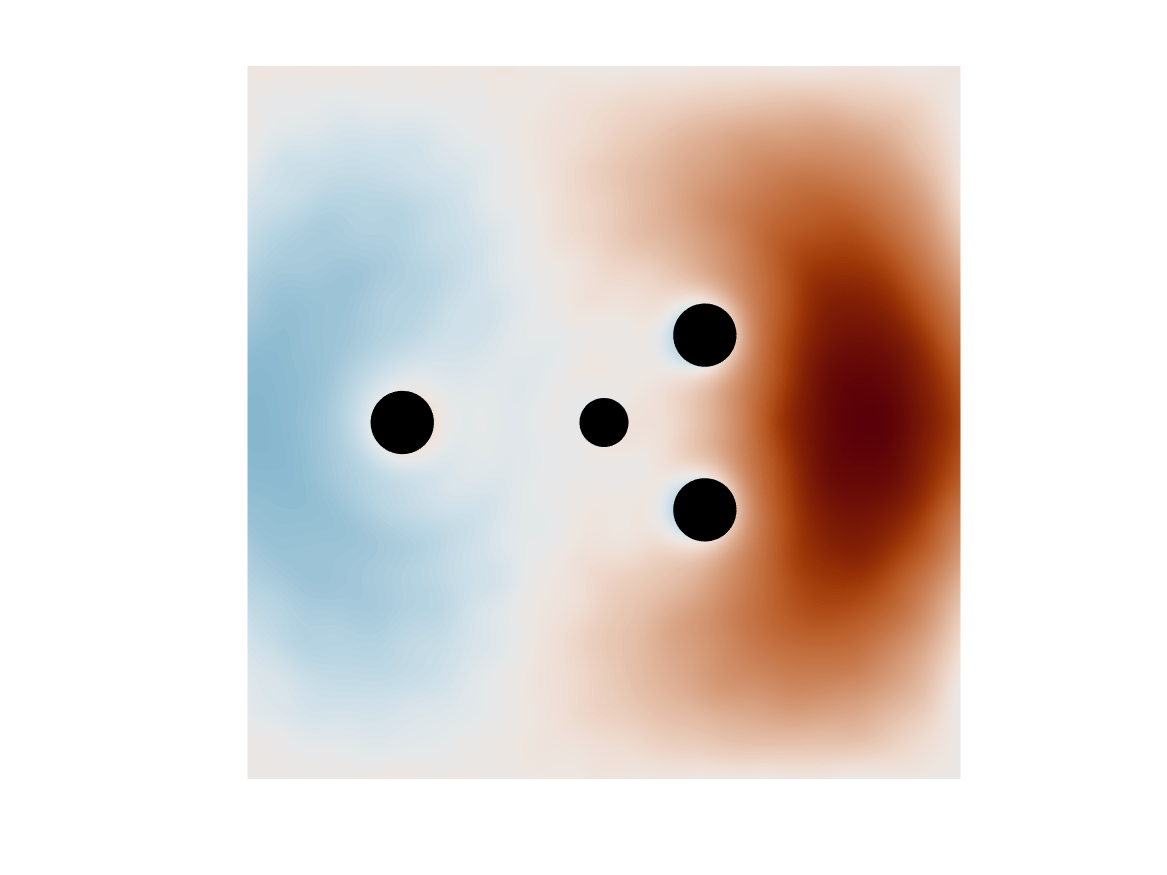}
               };
           }](polygon){};
		\node[below, scale=5.5, black] at (-10,8.75) {$\displaystyle (c)$};     
		\foreach \x in {2,4}{
        \draw [black,dashed, shorten <=-0.1cm,shorten >=-0.1cm,line width=0.5mm,](polygon.center) -- (polygon.side \x);}     
\end{scope}  

\begin{scope}[xshift=25.25cm, yshift=56cm,scale=0.45]
	\node[regular polygon, regular polygon sides=4,draw, inner sep=6.5cm,rotate=0,line width=0.0mm, white, opacity=0.0,
           path picture={
               \node[rotate=0,opacity=1.0] at (-0.475,-0.25){
                   \includegraphics[scale=1.4]{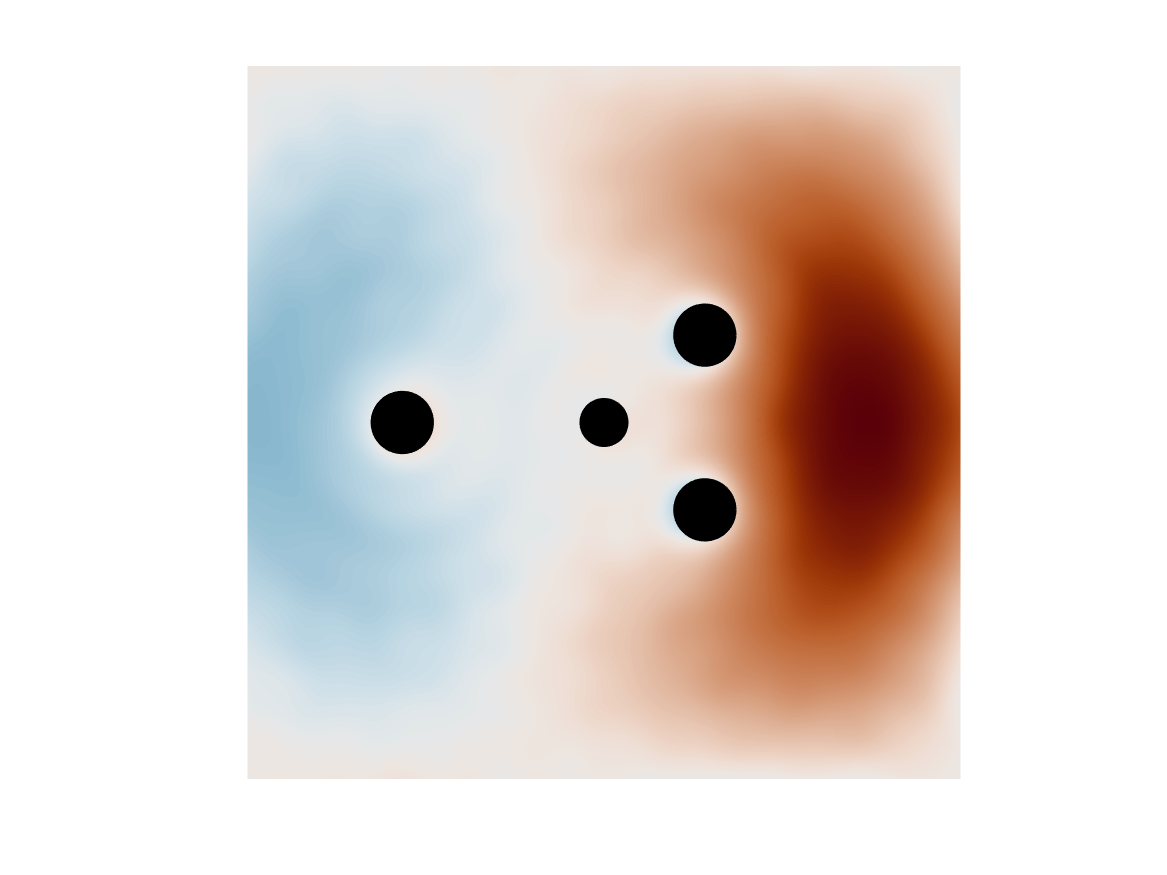}
               };
           }](polygon){};
		\node[below, scale=5.5, black] at (10,8.75) {$\displaystyle (d)$};       
		\foreach \x in {2,4}{
        \draw [black,dashed, shorten <=-0.1cm,shorten >=-0.1cm,line width=0.5mm,](polygon.center) -- (polygon.side \x);}      
\end{scope}

\begin{scope}[xshift=25.25cm, yshift=47.75cm,scale=0.45]
		\node[regular polygon, regular polygon sides=4,draw, inner sep=6.5cm,rotate=0,line width=0.0mm, black, opacity=0.0,
           path picture={
               \node[rotate=0,opacity=1.0] at (-0.475,-0.25){
                   \includegraphics[scale=1.4]{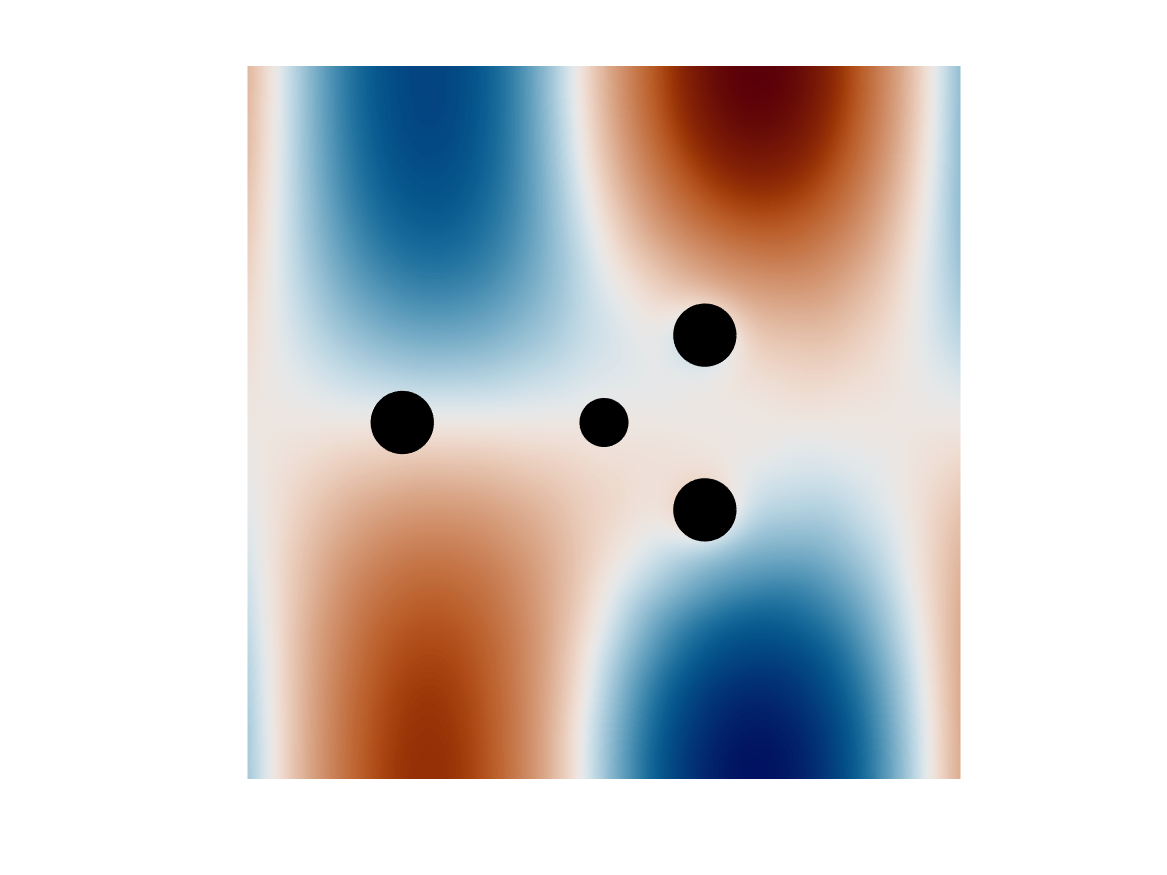}
               };
           }](polygon){};
		\node[below, scale=5.5, black] at (10,8.75) {$\displaystyle (e)$};       
		\foreach \x in {2,4}{
        \draw [black,dashed, shorten <=-0.1cm,shorten >=-0.1cm,line width=0.5mm,](polygon.center) -- (polygon.side \x);}           
\end{scope}  

\begin{scope}[xshift=-2.0cm, yshift=50.0cm,scale=1.0]
\node[regular polygon, regular polygon sides=4, draw, inner sep=0.5*6.28*10.0 pt,rotate=0,fill = white] at (0 pt,-6.28*10.0*1.5 pt) {};
\draw[line width=0.5mm,gray,-] (0pt,-6.28*10.0*1.5 pt) -- (0+ 0.5*6.28*10*1.41 pt,-6.28*10.0*1.5 pt);
\draw[line width=0.5mm,gray,-] (0+ 0.5*6.28*10*1.41 pt,-6.28*10.0*1.5 pt) -- (0+ 0.5*6.28*10*1.41  pt, 0.5*6.28*10*1.41 -6.28*10.0*1.5 pt);
\draw[line width=0.5mm,gray,-] (0+ 0.5*6.28*10*1.41  pt, 0.5*6.28*10*1.41 -6.28*10.0*1.5 pt) -- (0pt,0.5*6.28*10*1.41 -6.28*10.0*1.5 pt);
\draw[line width=0.5mm,gray,-] (0pt,0.5*6.28*10*1.41 -6.28*10.0*1.5 pt) -- (0pt,-6.28*10.0*1.5 pt);
\draw[line width=0.5mm,gray,-] (0pt,-6.28*10.0*1.5 pt) -- (0+ 0.5*6.28*10*1.41  pt, 0.5*6.28*10*1.41 -6.28*10.0*1.5 pt);
\node[below,left,scale=2.4] at (0pt,-6.28*10.0*1.5 pt) {$\displaystyle  \Gamma$}; 
\node[below,right,scale=2.4] at (0+ 0.5*6.28*10*1.41 pt,-6.28*10.0*1.5 pt) {$\displaystyle  X$};
\node[above,right,scale=2.4] at (0+ 0.5*6.28*10*1.41  pt, 0.5*6.28*10*1.41 -6.28*10.0*1.5+10 pt) {$\displaystyle M$};
\node[above,left,scale=2.4] at (5pt,0.5*6.28*10*1.41 -6.28*10.0*1.5+10 pt) {$\displaystyle  Y$};
\node[regular polygon, circle, draw, inner sep=1.25pt,rotate=0,line width=0.5mm,shading=fill,outer color=gray,gray] at (0pt,-6.28*10.0*1.5 pt)  {};
\node[regular polygon, circle, draw, inner sep=1.25pt,rotate=0,line width=0.5mm,shading=fill,outer color=gray,gray] at (0+ 0.5*6.28*10*1.41 pt,-6.28*10.0*1.5 pt)  {};
\node[regular polygon, circle, draw, inner sep=1.25pt,rotate=0,line width=0.5mm,shading=fill,outer color=gray,gray] at (0+ 0.5*6.28*10*1.41  pt, 0.5*6.28*10*1.41 -6.28*10.0*1.5 pt)  {};
\node[regular polygon, circle, draw, inner sep=1.25pt,rotate=0,line width=0.5mm,shading=fill,outer color=gray,gray] at (0pt,0.5*6.28*10*1.41 -6.28*10.0*1.5 pt)  {};
\end{scope}

\end{tikzpicture}

\caption{Square primitive cell with four Dirichlet inclusions (Dirichlet inclusions are shown as black filled inclusions). Parameters are $\boldsymbol{\alpha}_{1} =  \textbf{e}_{x}$, $\boldsymbol{\alpha}_{2} = \textbf{e}_{y}$, $\textbf{X}_{11} = 0.1415\textbf{e}_{x} - 0.1225\textbf{e}_{y}$, $\textbf{X}_{12} = -0.2829\textbf{e}_{x}$, $\textbf{X}_{13} = 0.1415\textbf{e}_{x} + 0.1225\textbf{e}_{y}$ and $\textbf{X}_{14}=\textbf{0}$, with $\eta_{1J}=0.04$ for $J=1,\ldots,3$ and $\eta_{14}=0.03$. Panel (a) shows the Floquet--Bloch dispersion, the black lines are eigenvalues from \eqref{HelmholtzHardScheme_1} and the black squares are FEM computed results. The four marked points  - $\color{myBLUE}\boldsymbol{\bigcirc}$   correspond to (b) \& (e) and $\color{myRED}\boldsymbol{\square}$ to (c) \& (d) - lie on the two branches adjacent to the degeneracy. Panels (b),(c) correspond to the two marked points on the left side of the degeneracy along the $MY$ path and panels (d),(e) to the two marked points on the right side; they show the real part of the corresponding normalised Bloch eigenmodes. The dotted line indicates the reflection symmetry line used for parity classification; the unperturbed configuration retains a single $\sigma_{v}$ reflection symmetry.}

\label{fig:SQDirac}
\end{figure}

Consider the square primitive cell in Fig.~\ref{fig:SQDirac}, which possesses a single vertical reflection symmetry $\sigma_{v}$. We denote by $\boldsymbol{\kappa}_{YM}$ any Bloch wavevector $\boldsymbol{\kappa}$ along $YM$, and note that $\sigma_{v}$ permits accidental degeneracies to occur for some $\boldsymbol{\kappa}_{YM}$ \cite{makwana19a}. In \cite{makwana2019topological}, it was shown that the irreducible representations along $\boldsymbol{\kappa}_{YM}$ are compatible with those at $G_{Y}$; hence, following \cite{sakoda2004optical}, the symmetry assignment for $G_{Y}$ can be deduced by inspecting any $G_{\boldsymbol{\kappa}_{YM}}$, and vice versa.

\begin{table}[H]
\centering
\begin{tabular}{c|c c c c|c}
\cline{2-5}
& \multicolumn{4}{c|}{Classes} \\
\hline
\multicolumn{1}{||c|}{IRs} & $E$ & $C_{2}$ & $\sigma_{v}$ & $\sigma_{h}$ & \multicolumn{1}{|c||}{Basis} \\
\hline\hline
\multicolumn{1}{||c|}{$A_{2}$} & $+1$ & $+1$ & $-1$ & $-1$ & \multicolumn{1}{|c||}{$xy$} \\
\multicolumn{1}{||c|}{$B_{1}$} & $+1$ & $-1$ & $+1$ & $-1$ & \multicolumn{1}{|c||}{$x$} \\
\hline
\end{tabular}
\caption{Excerpt from the $C_{2v}$ character table showing the basis functions relevant to the symmetry classification used below.}
\label{table:C2VCharacter}
\end{table}

Consider the eigenmodes in Fig.~\ref{fig:SQDirac}(b)--(e). The two branches adjacent to the crossing, marked by $\color{myBLUE}\boldsymbol{\bigcirc}$ and $\color{myRED}\boldsymbol{\square}$ in Fig.~\ref{fig:SQDirac}(a), match the relevant basis functions in Table~\ref{table:C2VCharacter} and have opposite parity under $\sigma_{v}$: one branch is purely even, whereas the other is purely odd. The presence of $\sigma_{v}$ symmetry therefore prevents these modes from hybridising, yielding a symmetry-protected degeneracy along $MY$. Applying the above compatibility condition, we deduce that $G_{Y}=C_{2v}$. Moreover, since $G_{Y}\leq G_{\Gamma}$, it follows immediately that $G_{\Gamma}=C_{2v}$.

\begin{figure}[H]
\centering
\hspace*{0.5cm}
\begin{tikzpicture}[scale=0.3, transform shape]

\begin{scope}[xshift=28.5cm, yshift=50.5cm]
		\node[regular polygon, regular polygon sides=4,draw, inner sep=7.0cm,rotate=0,line width=0.0mm, white,
           path picture={
               \node[rotate=0] at (-1,1){
                   \includegraphics[scale=1.25]{Figs/ColourBarEigenModes.eps}
               };
           }]{};
\end{scope}  

\begin{scope}[xshift=15cm, yshift=50.5cm]
		\node[regular polygon, regular polygon sides=4,draw, inner sep=7.0cm,rotate=0,line width=0.0mm, white,
           path picture={
               \node[rotate=0] at (1,1){
                   \includegraphics[scale=1.25]{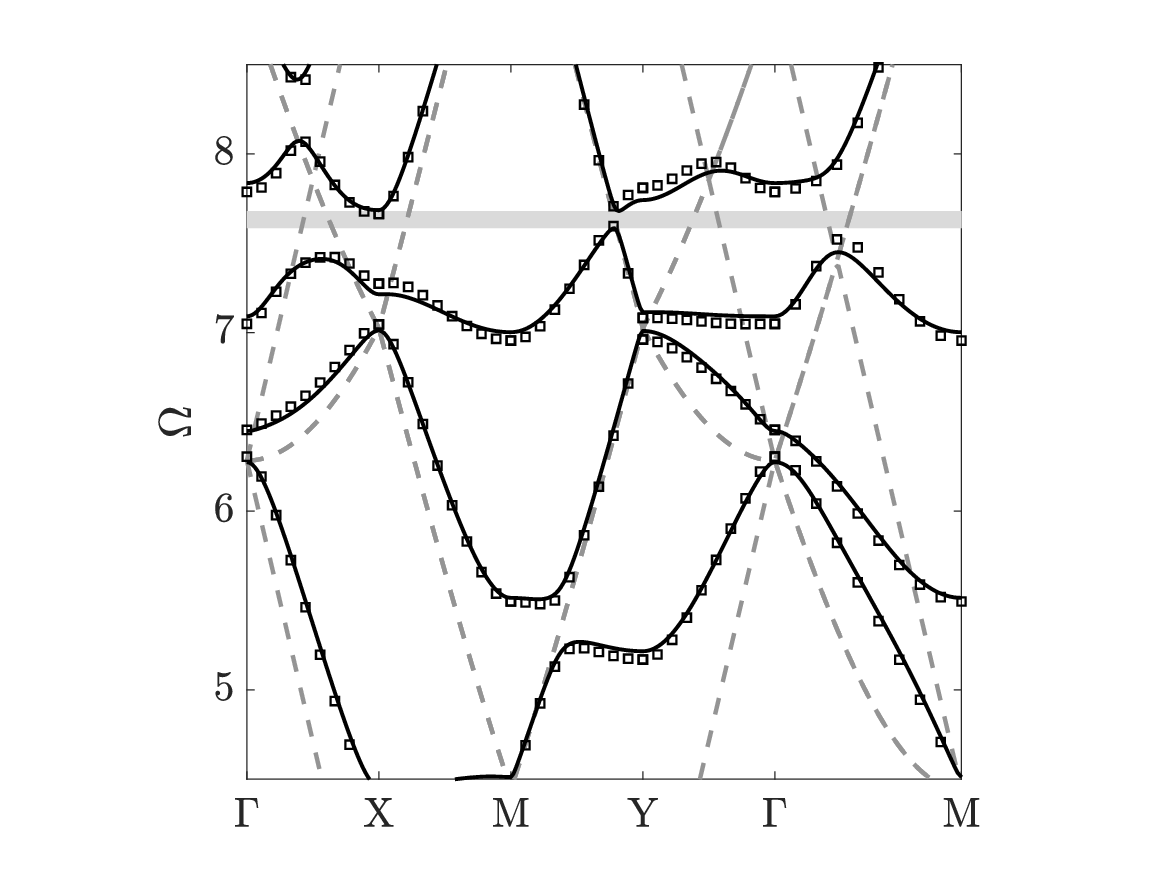}
               };
           }]{};
		\node[below, scale=2.5, black] at (1,-7.5) {$\displaystyle (c)$};           
\end{scope}  

\begin{scope}[xshift=4.0cm, yshift=56cm,scale=0.4]
		\node[regular polygon, regular polygon sides=4,draw, inner sep=6.5cm,rotate=0,line width=0.0mm, white,
           path picture={
               \node[rotate=0] at (1.45,-0.25){
                   \includegraphics[scale=1.4]{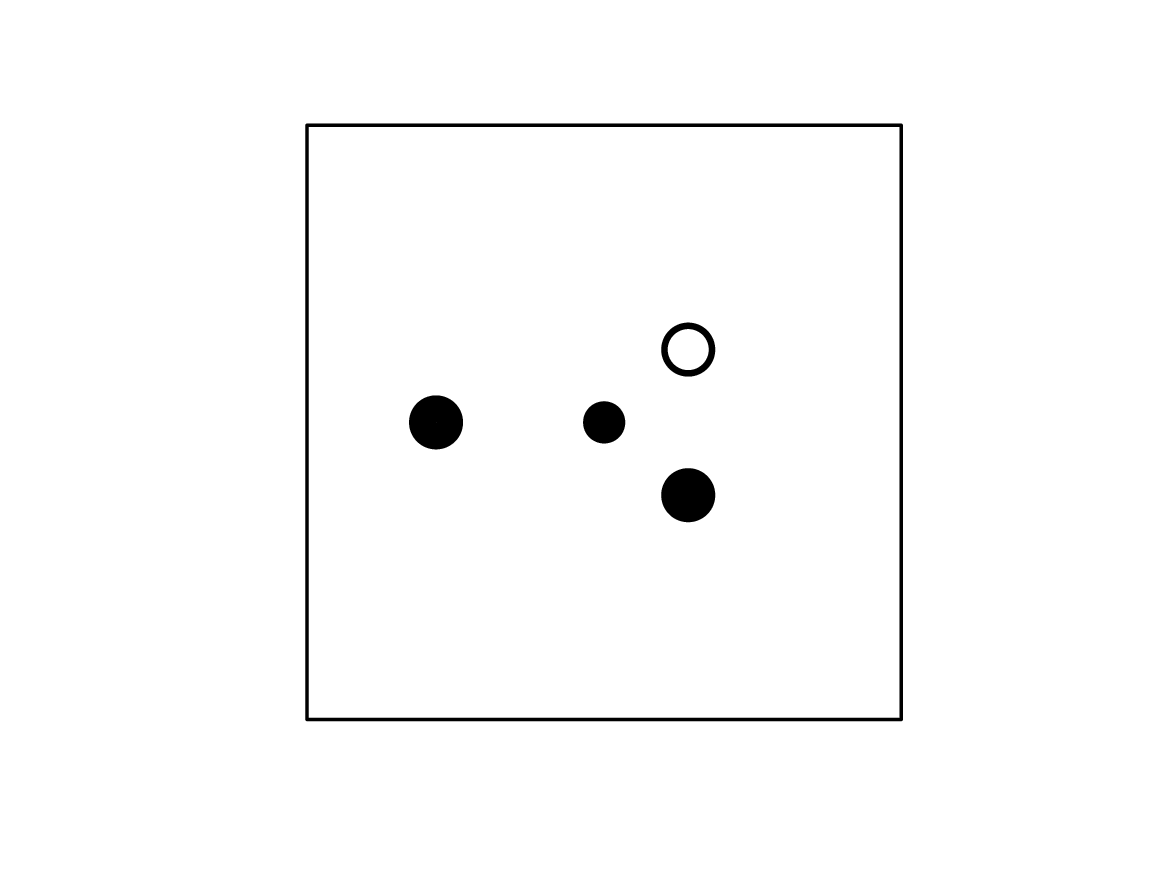}
               };
           }]{};
		\node[below, scale=6.3, black] at (1,-7.5) {$\quad \displaystyle (a)$};             
\end{scope}

\begin{scope}[xshift=4.0cm, yshift=47.25cm,scale=0.4]
	\node[regular polygon, regular polygon sides=4,draw, inner sep=6.5cm,rotate=0,line width=0.0mm, white,
           path picture={
               \node[rotate=0] at (1.45,-0.25){
                   \includegraphics[scale=1.4]{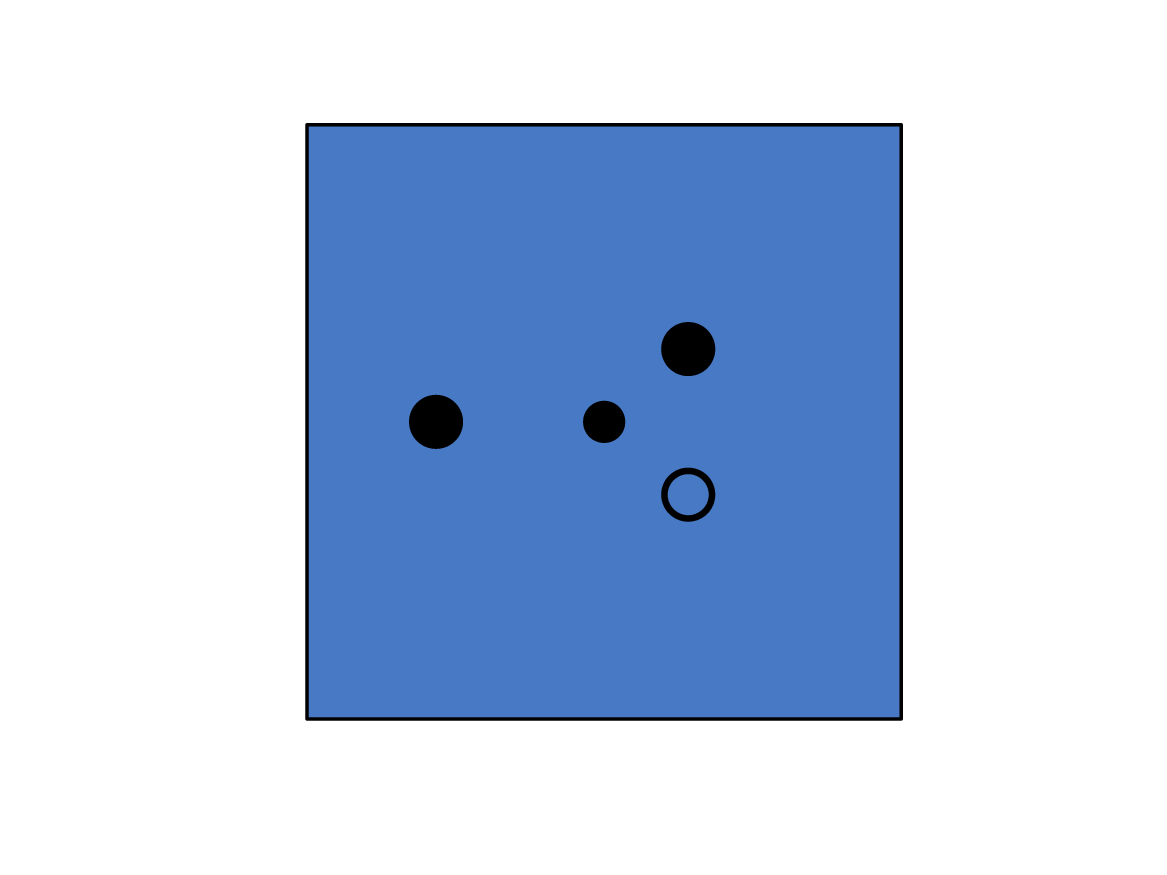}
               };
           }]{};
		\node[below, scale=6.3, black] at (1,-7.5) {$\quad \displaystyle (b)$};           
\end{scope}  

\begin{scope}[xshift=30cm, yshift=58cm,scale=0.6]
		\node[regular polygon, regular polygon sides=4,draw, inner sep=6.5cm,rotate=0,line width=0.0mm, white,
           path picture={
               \node[rotate=0] at (-0.35,-0.25){
                   \includegraphics[scale=1.4]{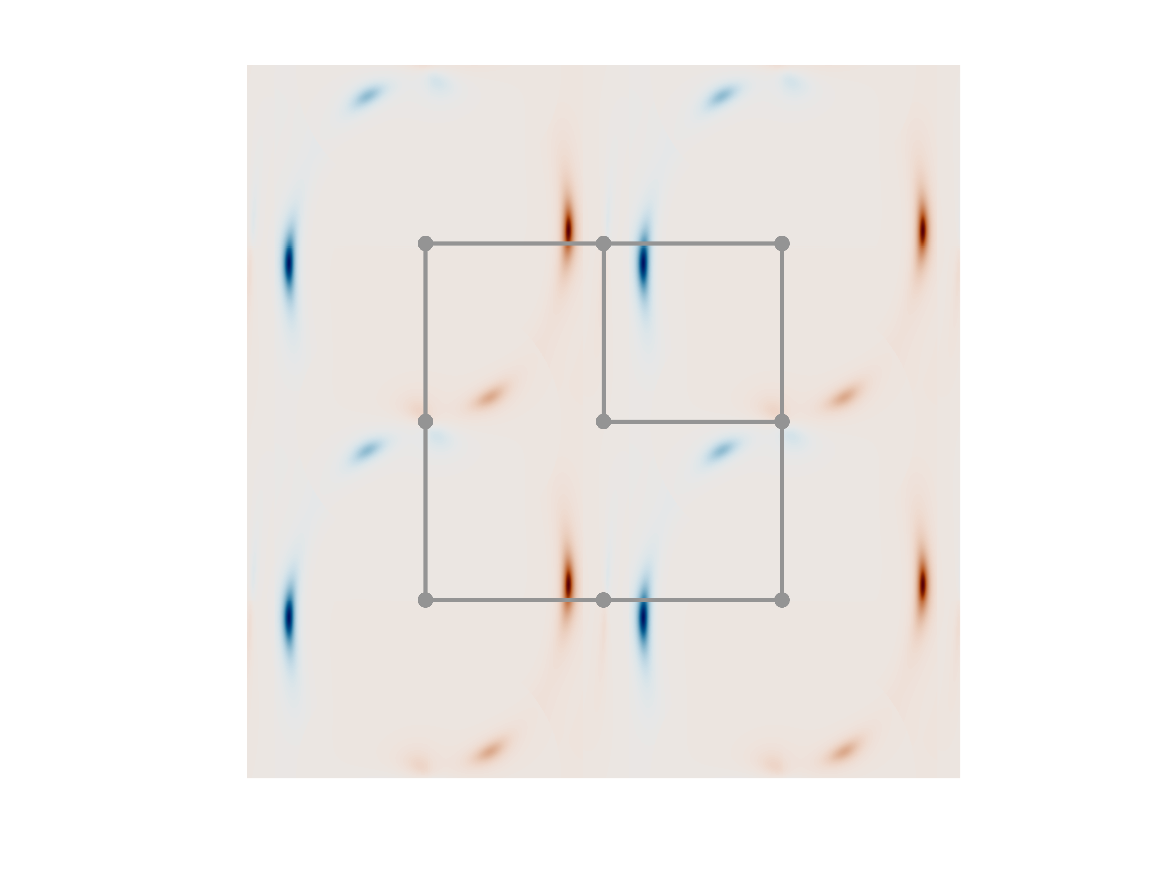}
               };
           }]{};
           \node[below,left,scale=4.5] at (0.5,-0.5) {$\displaystyle  \Gamma$}; 
			\node[below,right,scale=4.5] at (4.25,-0.5) {$\displaystyle  X$};
			\node[above,right,scale=4.5] at (4.25, 5.25) {$\displaystyle M$};
						\node[above,left,scale=4.5] at (1.5, 5.25) {$\displaystyle Y$};
		\node[below, scale=4.2, black] at (7,8.75) {$\displaystyle (d)$};              
\end{scope}

\begin{scope}[xshift=30cm, yshift=46.25cm,scale=0.6]
	\node[regular polygon, regular polygon sides=4,draw, inner sep=6.5cm,rotate=0,line width=0.0mm, white,
           path picture={
               \node[rotate=0] at (-0.35,-0.25){
                   \includegraphics[scale=1.4]{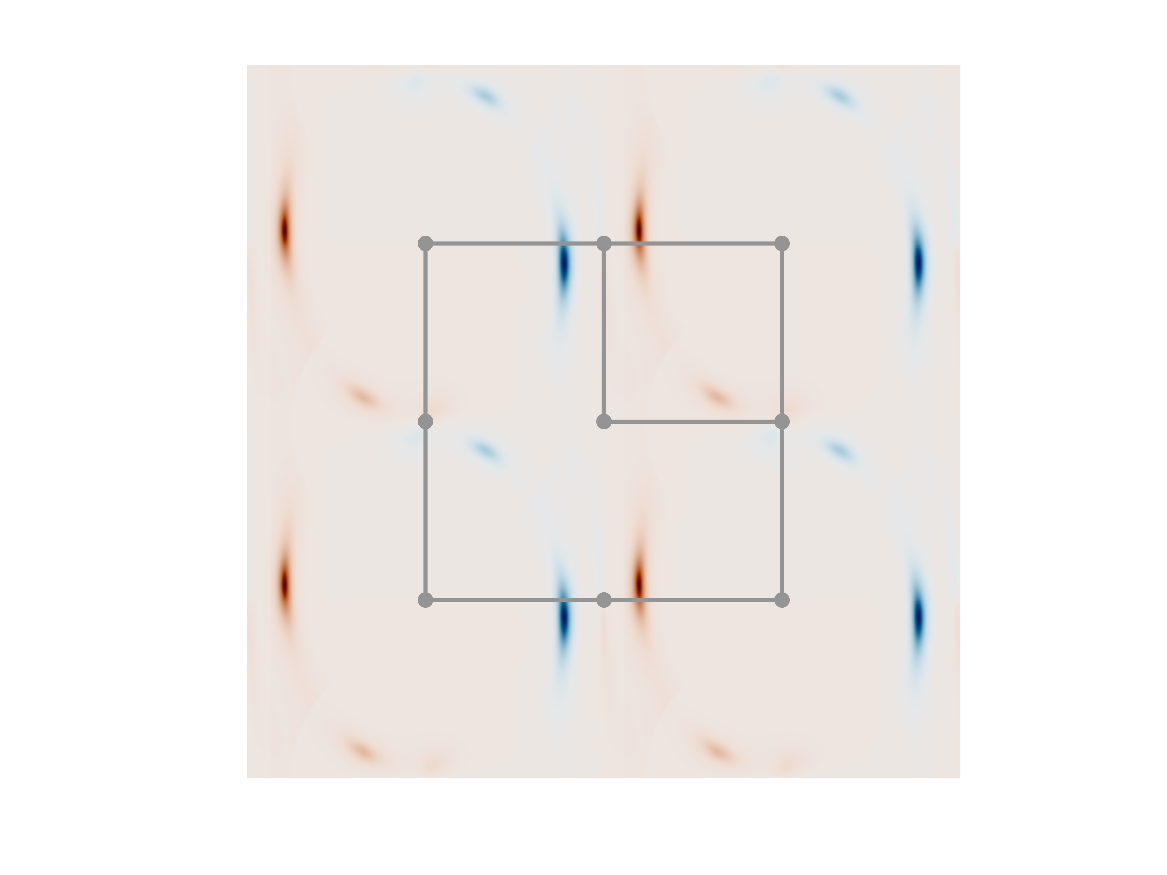}
               };
           }]{};
           \node[below,left,scale=4.5] at (0.5,-0.5) {$\displaystyle  \Gamma$}; 
			\node[below,right,scale=4.5] at (4.25,-0.5) {$\displaystyle  X$};
			\node[above,right,scale=4.5] at (4.25, 5.25) {$\displaystyle M$};
						\node[above,left,scale=4.5] at (1.5, 5.25) {$\displaystyle Y$};
		\node[below, scale=4.2, black] at (7,8.75) {$\displaystyle (e)$};            
\end{scope}

\end{tikzpicture}

\caption{Perturbing the structure in Figure~\ref{fig:SQDirac} by making the highest or lowest inclusion Neumann (Neumann inclusions are shown as white or blue filled inclusions) creates the mirror-related chiral unit cells in (a) and (b). Panel (c) shows the Floquet--Bloch dispersion relation for cell (a), with the opened bulk band gap shaded in grey - again the black lines are eigenvalues from \eqref{HelmholtzHardScheme_1} and the black squares are FEM computed results. Panels (d) and (e) show the Berry curvature over the Brillouin zone for the bands immediately below and above the gap, respectively.}

\label{fig:SQValley}
\end{figure}

We perturb the primitive cell in Fig.~\ref{fig:SQDirac} by switching either the uppermost or lowermost inclusion from Dirichlet to Neumann, thereby producing the mirror-related chiral cells in Fig.~\ref{fig:SQValley}(a),(b). Breaking $\sigma_{v}$ reduces the symmetry set from $\{G_{\Gamma},G_{Y}\}=\{C_{2v},C_{2v}\}$ for the unperturbed cell to $\{G_{\Gamma},G_{Y}\}=\{C_{2},C_{2}\}$ for the perturbed cells. Once the reflection symmetry is removed, the parity distinction between the two branches is lost, the modes can hybridise, and the accidental crossing is lifted. The resulting band repulsion opens the topologically non-trivial bulk band gap shown in Fig.~\ref{fig:SQValley}(c).

\begin{figure}[H]
\centering
\hspace*{2.5cm}
\begin{tikzpicture}[scale=0.3, transform shape]

\begin{scope}[xshift=15.5cm, yshift=-4cm,scale=1.4]
		\node[regular polygon, regular polygon sides=4,draw, inner sep=7cm,rotate=0,line width=0.0mm, white,
           path picture={
               \node[rotate=0] at (-9,0){
                   \includegraphics[scale=1.25]{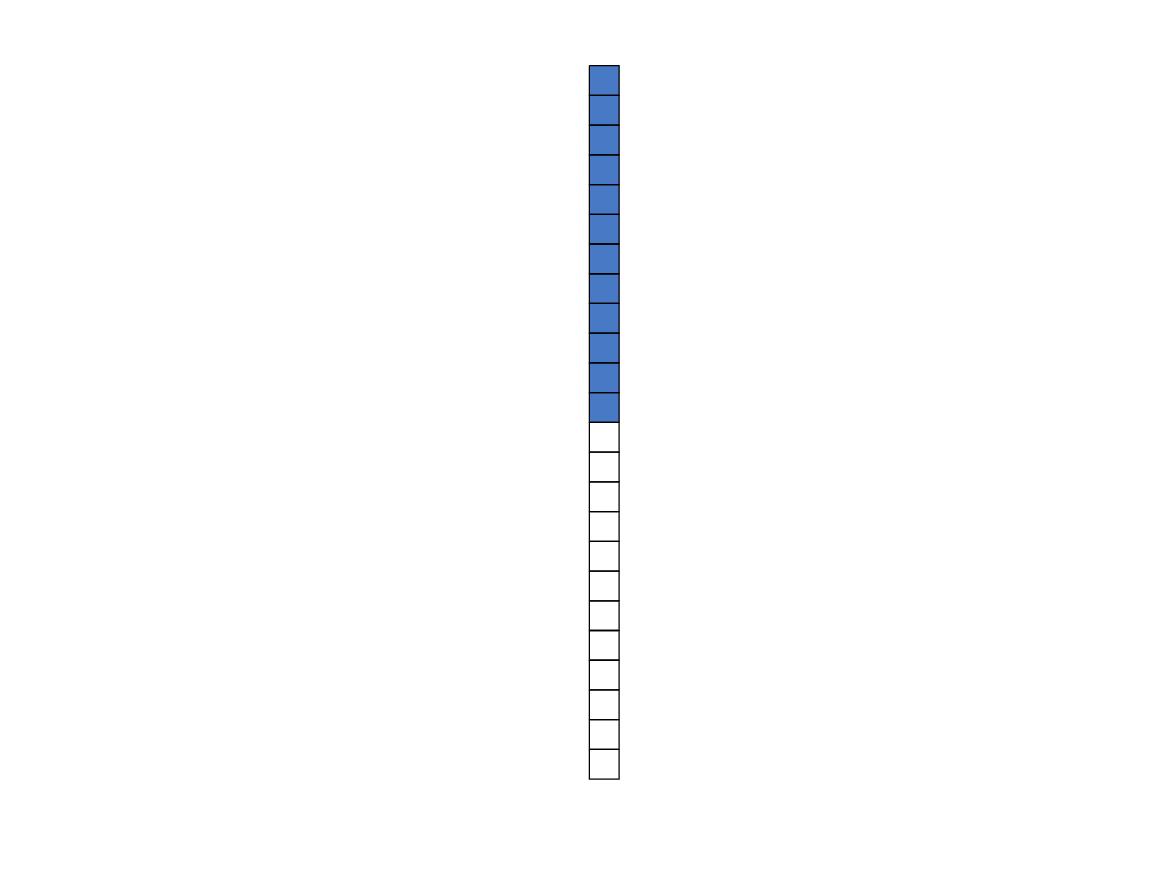}
               };
           }]{};
\end{scope}  

\begin{scope}[xshift=18.5cm, yshift=-4cm,scale=1.4]
		\node[regular polygon, regular polygon sides=4,draw, inner sep=7cm,rotate=0,line width=0.0mm, white,
           path picture={
               \node[rotate=0] at (-9.75,0){
                   \includegraphics[scale=1.25]{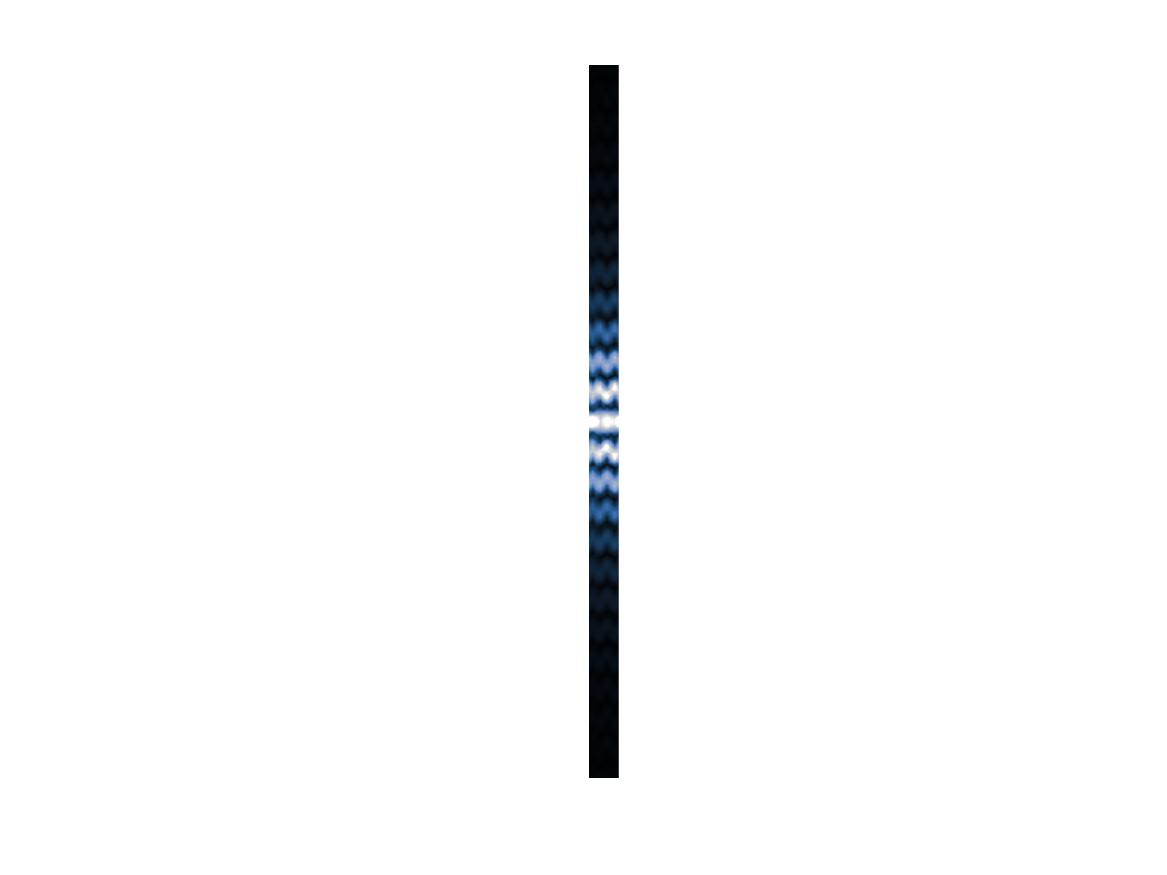}
               };
           }]{};
				\node[below, scale=2.86, black] at (-10,-7.5) {$\displaystyle (b)$};           
\end{scope}  

\begin{scope}[xshift=20.5cm, yshift=-4cm,scale=1.4]
		\node[regular polygon, regular polygon sides=4,draw, inner sep=7cm,rotate=0,line width=0.0mm, white,
           path picture={
               \node[rotate=0] at (-0.25,0){
                   \includegraphics[scale=1.25]{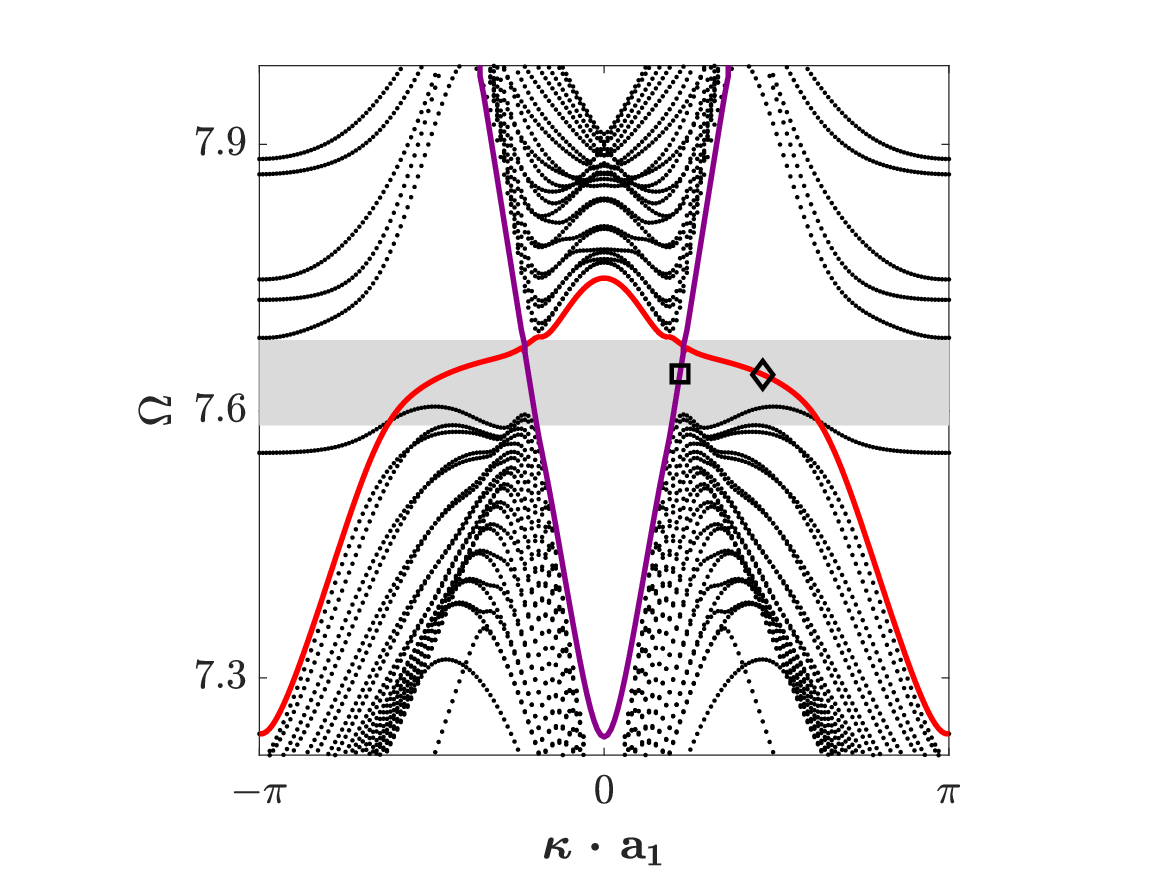}
               };
           }]{};
		\node[below, scale=2.86, black] at (-6.25,8.0) {$\displaystyle (a)$};           
\end{scope}  

\begin{scope}[xshift=45.5cm, yshift=-4cm,scale=1.4]
		\node[regular polygon, regular polygon sides=4,draw, inner sep=7cm,rotate=0,line width=0.0mm, white,
           path picture={
               \node[rotate=0] at (-9,0){
                   \includegraphics[scale=1.25]{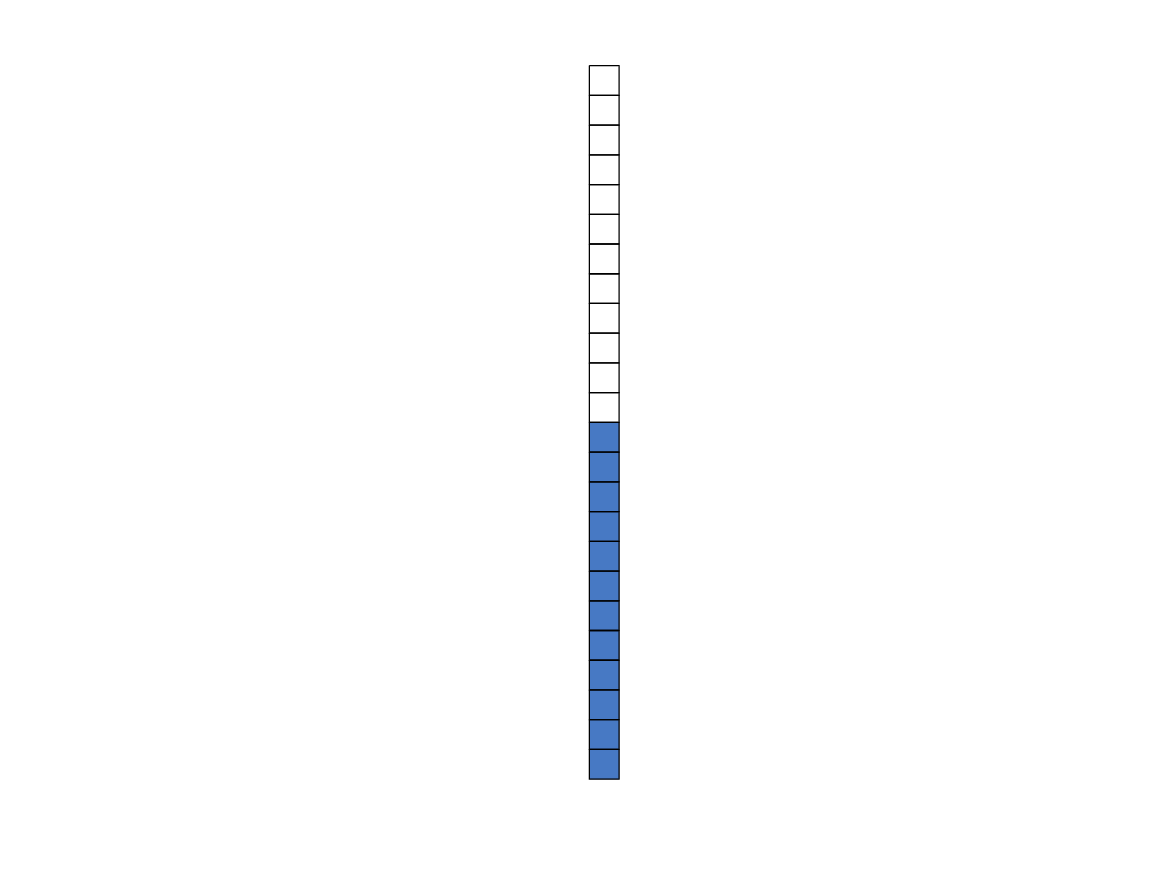}
               };
           }]{};
\end{scope}  

\begin{scope}[xshift=48.5cm, yshift=-4cm,scale=1.4]
		\node[regular polygon, regular polygon sides=4,draw, inner sep=7cm,rotate=0,line width=0.0mm, white,
           path picture={
               \node[rotate=0] at (-9.75,0){
                   \includegraphics[scale=1.25]{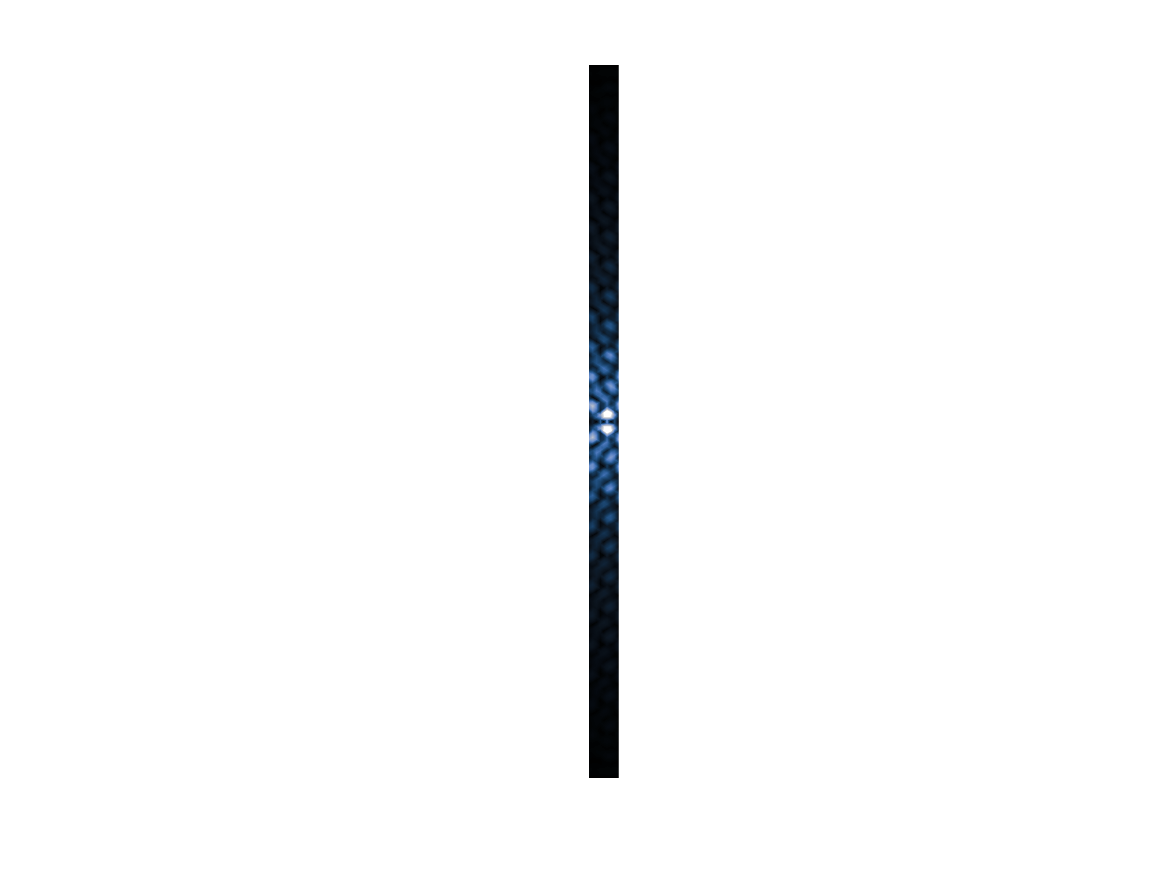}
               };
           }]{};
				\node[below, scale=2.86, black] at (-10,-7.5) {$\displaystyle (c)$};           
\end{scope}

\begin{scope}[xshift=51.5cm, yshift=-4cm,scale=1.4]
		\node[regular polygon, regular polygon sides=4,draw, inner sep=7cm,rotate=0,line width=0.0mm, white,
           path picture={
               \node[rotate=0] at (-18.25,0){
                   \includegraphics[scale=1.25]{Figs/ColorBarAbsOslo.eps}
               };
           }]{};
\end{scope}

\end{tikzpicture}

\caption{Ribbon/interface states for the square-lattice chiral pair. Panel (a) shows the ribbon Floquet--Bloch dispersion; the shaded region is the projected bulk band gap from \ref{fig:SQValley} and the coloured curves are interfacial ZLM branches. Panels (b) and (c) show the normalised magnitude of the corresponding interface-localised eigenmodes (absolute value) at the marked $\kappa$ values of $\boldsymbol{\square}$ and $\boldsymbol{\diamond}$, respectively.}

\label{fig:SQRibbon}
\end{figure}

The Berry curvature of the bands immediately below and above the gap, shown in Fig.~\ref{fig:SQValley}(d),(e), is localised near the $\boldsymbol{\kappa}$-points associated with the lifted degeneracy and changes sign between the two chiral cells. As in the hexagonal case, a ribbon formed by adjoining opposite chiral phases supports ZLM branches within the projected bulk gap; see Fig.~\ref{fig:SQRibbon}(a). The corresponding eigenmodes in Fig.~\ref{fig:SQRibbon}(b),(c) are strongly localised to the interface.

\begin{figure}[H]
\centering
\hspace*{-0.5cm}
\begin{tikzpicture}[scale=0.275, transform shape]

\begin{scope}[xshift=14.5cm, yshift=22cm,scale=1.4]
               \node[rotate=0] at (-0.5,-0.25){
                   \includegraphics[scale=1.25]{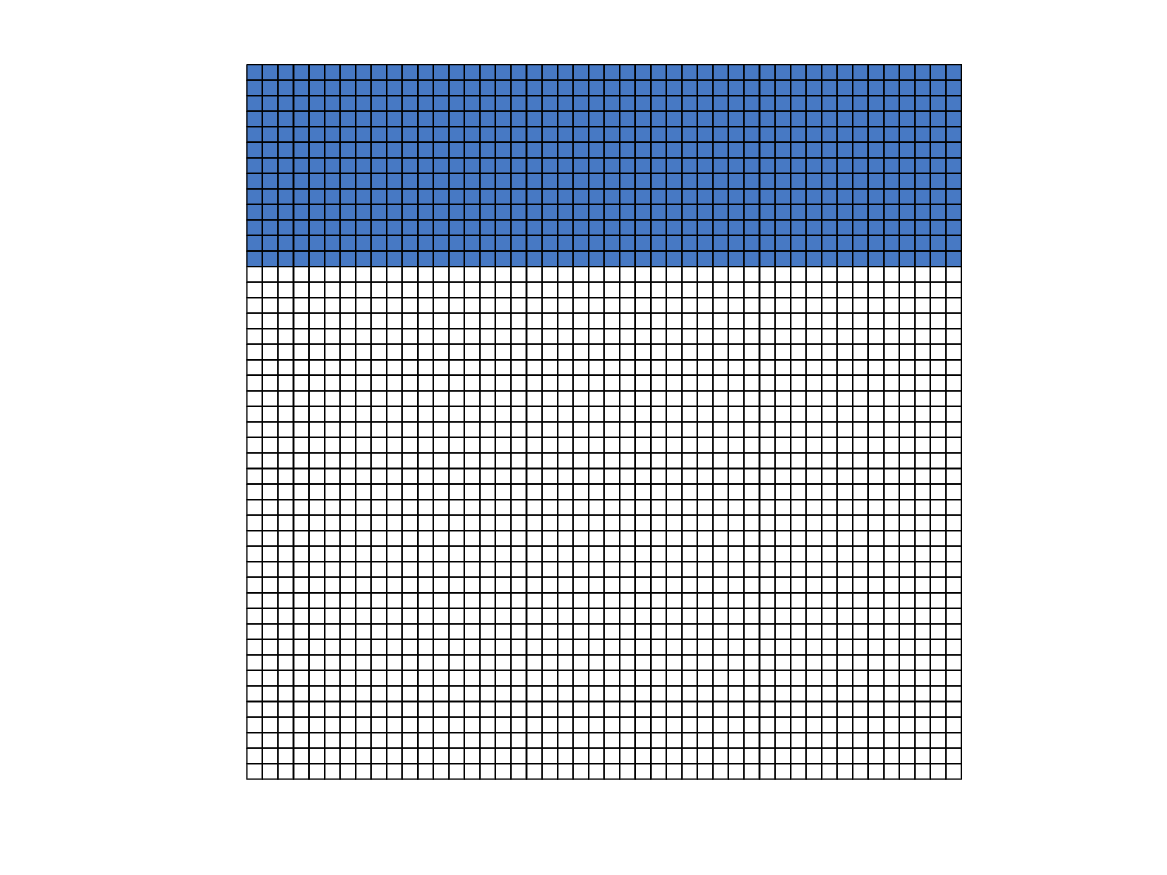}
               };    
		\node[below, scale=2.86, black] at (-8.5,8.75) {$\displaystyle (a)$};        
		
  \begin{scope}[xshift=-8cm, yshift= 0 cm,rotate=180,scale=4, transform shape]
 \draw[<-,line width=2.0] (0,0) -- (0.5*\xmax,0) node[right] {$\,$};
  \draw[-,line width=2.0] (0.3*\xmax,-0.15*\xmax) -- (0.3*\xmax,0.15*\xmax) node[right] {$\,$};
   \draw[-,line width=2.0] (0.35*\xmax,-0.15*\xmax) -- (0.35*\xmax,0.15*\xmax) node[right] {$\,$};
    \draw[-,line width=2.0] (0.4*\xmax,-0.15*\xmax) -- (0.4*\xmax,0.15*\xmax) node[right] {$\,$};
    \node[below, scale=0.65, black,rotate=-180] at (0.3*\xmax,-0.275*\xmax+0.425*\xmax) {Incident};
       \node[below, scale=0.65, black,rotate=-180] at (0.3*\xmax,-0.15*\xmax+0.4*\xmax)  {$\quad$ plane wave};
\end{scope}

\end{scope}

\begin{scope}[xshift=38cm, yshift=22cm,scale=1.4]
		\node[regular polygon, regular polygon sides=4,draw, inner sep=5.5cm,rotate=0,line width=0.0mm, white,
           path picture={
               \node[rotate=0] at (-0.5,-0.25){
                   \includegraphics[scale=1.25]{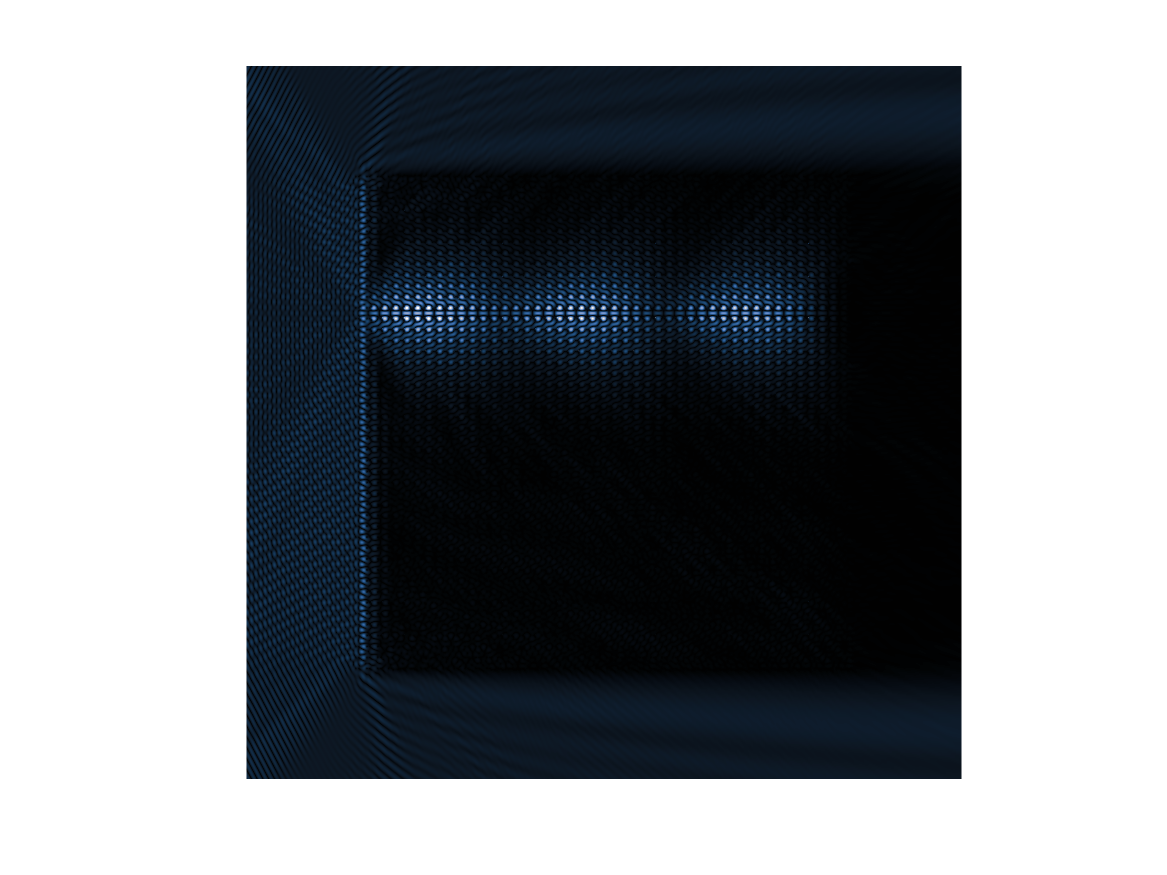}
               };
           }]{};
		\node[below, scale=2.86, black] at (-8.5,8.75) {$\displaystyle (b)$};           
\end{scope}

\begin{scope}[xshift=14.5cm, yshift=-2cm,scale=1.4]
               \node[rotate=0] at (-0.5,-0.25){
                   \includegraphics[scale=1.25]{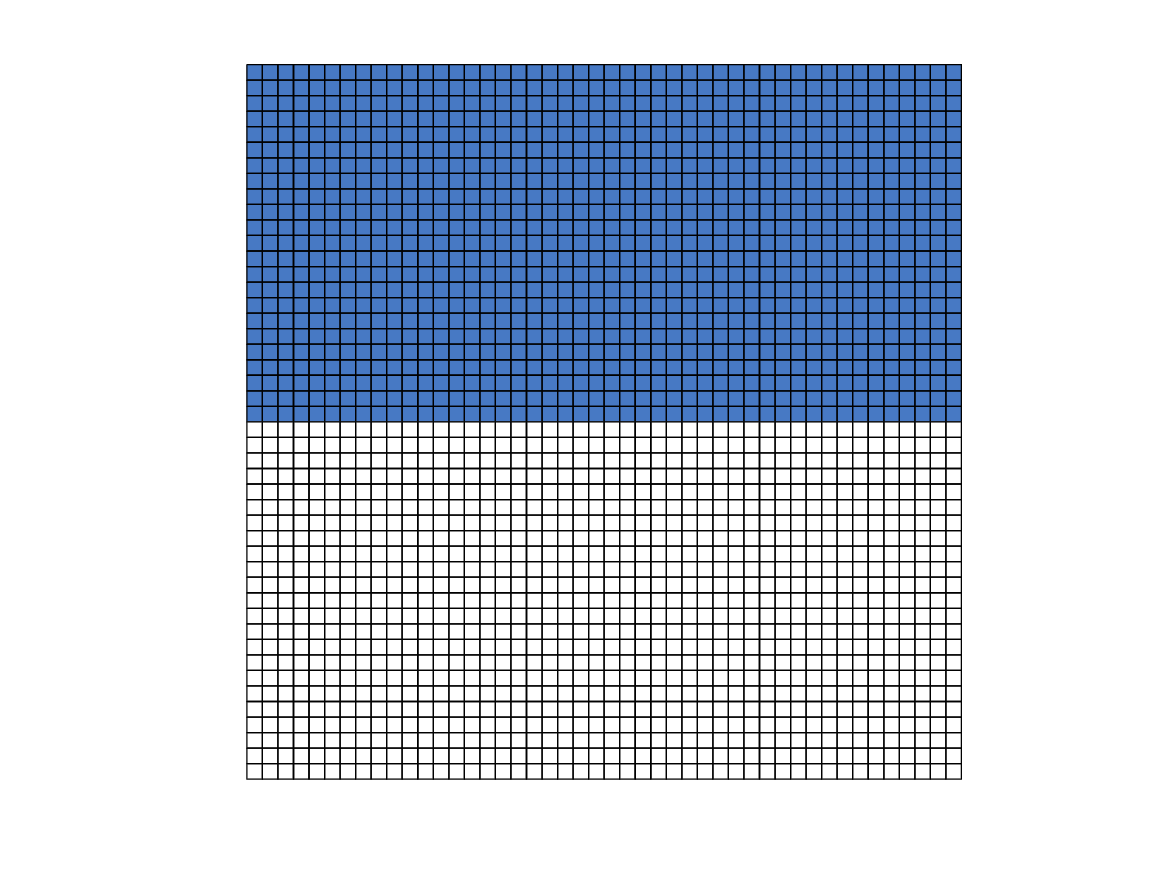}
               };
		\node[below, scale=2.86, black] at (-8.5,8.75) {$\displaystyle (c)$};           
 \begin{scope}[xshift=-8cm, yshift= 0 cm,rotate=180,scale=4, transform shape]
 \draw[<-,line width=2.0] (0,0) -- (0.5*\xmax,0) node[right] {$\,$};
  \draw[-,line width=2.0] (0.3*\xmax,-0.15*\xmax) -- (0.3*\xmax,0.15*\xmax) node[right] {$\,$};
   \draw[-,line width=2.0] (0.35*\xmax,-0.15*\xmax) -- (0.35*\xmax,0.15*\xmax) node[right] {$\,$};
    \draw[-,line width=2.0] (0.4*\xmax,-0.15*\xmax) -- (0.4*\xmax,0.15*\xmax) node[right] {$\,$};
\end{scope}
\end{scope}

\begin{scope}[xshift=38cm, yshift=-2cm,scale=1.4]
		\node[regular polygon, regular polygon sides=4,draw, inner sep=5.5cm,rotate=0,line width=0.0mm, white,
           path picture={
               \node[rotate=0] at (-0.5,-0.25){
                   \includegraphics[scale=1.25]{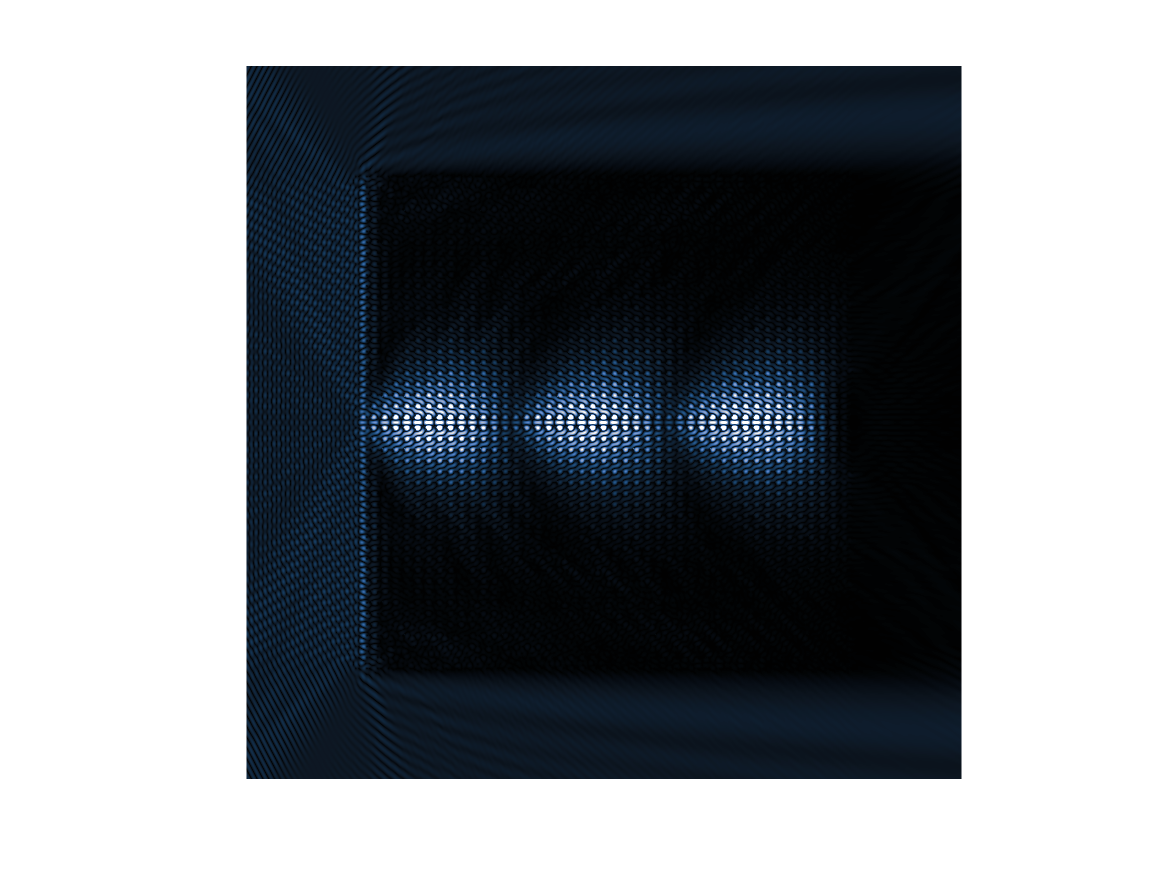}
               };
           }]{};
		\node[below, scale=2.86, black] at (-8.5,8.75) {$\displaystyle (d)$};           
\end{scope}

\begin{scope}[xshift=14.5cm, yshift=-26cm,scale=1.4]
               \node[rotate=0] at (-0.5,-0.25){
                   \includegraphics[scale=1.25]{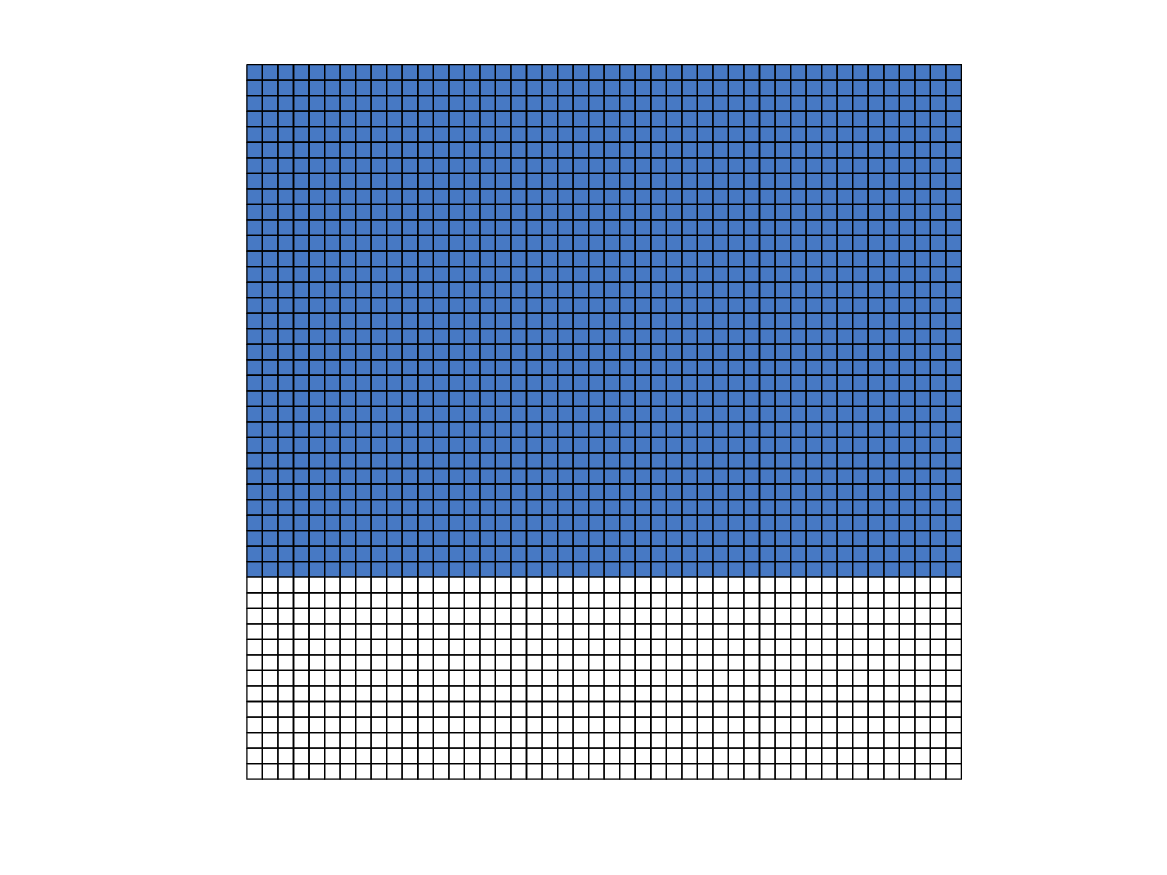}
               };
		\node[below, scale=2.86, black] at (-8.5,8.75) {$\displaystyle (e)$};           
		
		 \begin{scope}[xshift=-8cm, yshift= 0 cm,rotate=180,scale=4, transform shape]
 \draw[<-,line width=2.0] (0,0) -- (0.5*\xmax,0) node[right] {$\,$};
  \draw[-,line width=2.0] (0.3*\xmax,-0.15*\xmax) -- (0.3*\xmax,0.15*\xmax) node[right] {$\,$};
   \draw[-,line width=2.0] (0.35*\xmax,-0.15*\xmax) -- (0.35*\xmax,0.15*\xmax) node[right] {$\,$};
    \draw[-,line width=2.0] (0.4*\xmax,-0.15*\xmax) -- (0.4*\xmax,0.15*\xmax) node[right] {$\,$};
\end{scope}
\end{scope}

\begin{scope}[xshift=38cm, yshift=-26cm,scale=1.4]
		\node[regular polygon, regular polygon sides=4,draw, inner sep=5.5cm,rotate=0,line width=0.0mm, white,
           path picture={
               \node[rotate=0] at (-0.5,-0.25){
                   \includegraphics[scale=1.25]{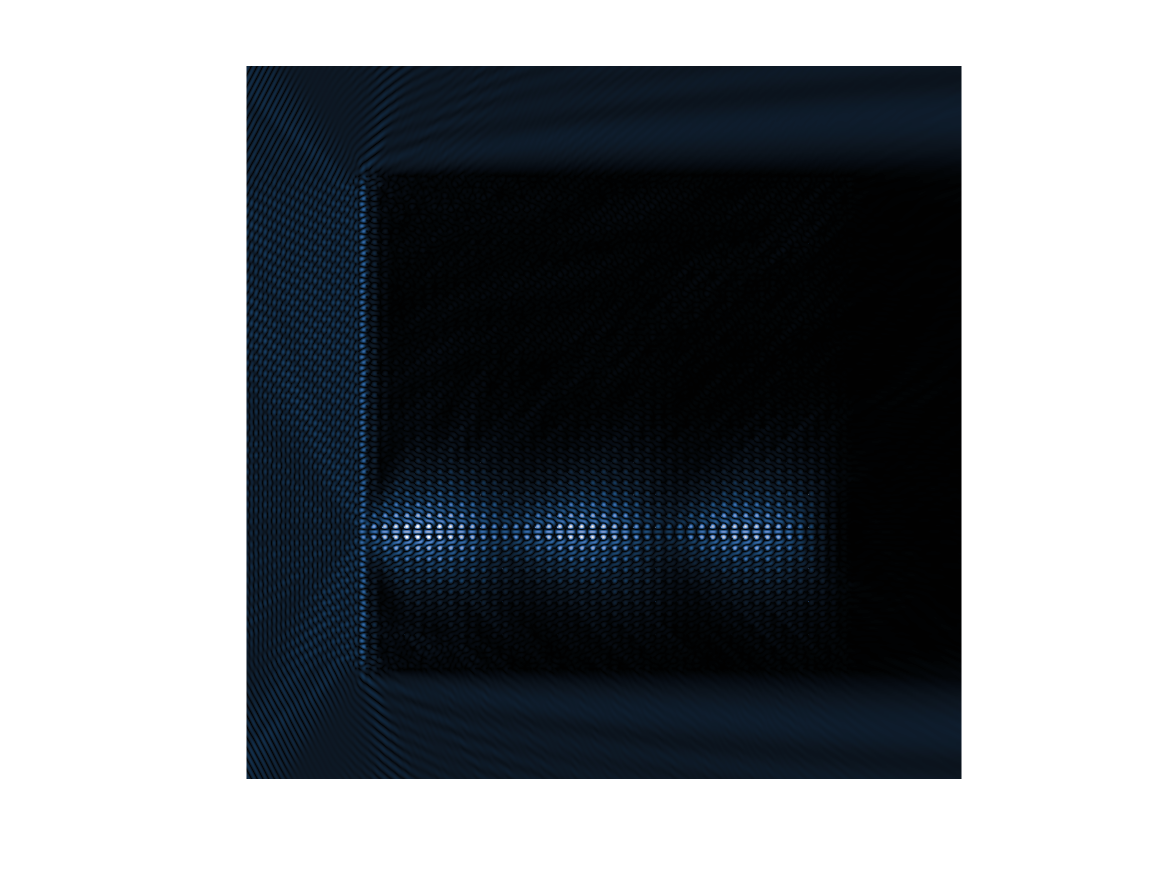}
               };
           }]{};
		\node[below, scale=2.86, black] at (-8.5,8.75) {$\displaystyle (f)$};           
\end{scope}

\begin{scope}[xshift=64.5cm, yshift=-3cm,scale=1.6]
		\node[regular polygon, regular polygon sides=4,draw, inner sep=7cm,rotate=0,line width=0.0mm, white,
           path picture={
               \node[rotate=0] at (-17.0,0){
                   \includegraphics[scale=1.25]{Figs/ColorBarAbsOslo.eps}
               };
           }]{};
\end{scope}  

\end{tikzpicture}

\caption{Plane-wave multiple-scattering simulation (the absolute value of the normalised total field is plotted, solving \eqref{scatteringSoln} to determine the unknown coefficients in \eqref{mscSOLNtotal})  for a finite collection of $2025$ square cells from Figure~\ref{fig:SQValley}(a),(b), with $\Omega = 7.548$, $\theta_{\mathrm{inc}} = \pi$, $A_{\mathrm{inc}} = 1$ and $\textbf{X}_{\mathrm{inc}}$ the centre of the collection. The schematic configurations (a),(c),(e) differ only by the Dirichlet/Neumann assignments on inclusions (geometry fixed), thereby translating the internal interface; the corresponding field magnitudes (b),(d),(f) show localisation along the moved interface. The frequency is chosen within the topologically non-trivial bulk band gap from fig. \ref{fig:SQValley}.}
\label{fig:SQscattering}
\end{figure}

Finally, Fig.~\ref{fig:SQscattering} demonstrates reconfigurable interface relocation in a finite square array under plane-wave excitation. The three schematic configurations differ only by the Dirichlet/Neumann assignments imposed on the inclusions, while the geometry remains fixed; the resulting localised response follows the translated interface. This confirms, within the present point-scatterer framework, that switching inclusion boundary conditions provides a practical means of moving valley-Hall interfaces and hence repositioning ZLM trajectories in square-lattice settings.

\section{Conclusions}
We have presented a unified asymptotic framework for analysing and designing reconfigurable valley-Hall guiding in two-dimensional lattices whose inclusions are modelled by idealised boundary conditions. Starting from the scalar Helmholtz formulation \eqref{Helmholtz} with the high-contrast limits \eqref{Neumann}--\eqref{Dirichlet}, we employed matched asymptotic expansions to derive the point-scatterer approximation \eqref{PFSmyDrude} for a lattice composed of distinct Dirichlet and Neumann inclusions. This simplification yields, for infinite doubly periodic arrays, a finite-dimensional generalised eigenproblem \eqref{HelmholtzHardScheme_1} whose solutions provide Floquet--Bloch dispersion relations and eigenmodes; and, for finite collections, a generalised Foldy \cite{foldy1945multiple,martin2006multiple} multiple-scattering system \eqref{scatteringSoln} for computing driven responses. The resulting semi-analytical method was validated against full FEM simulations and found to be accurate, while also producing small matrix systems that enable fast and efficient computation.

Within this framework we have established, by explicit band-structure and Berry-curvature computations, that both hexagonal and square lattices admit symmetry-broken bulk band gaps of valley type when an appropriate reflection symmetry of the primitive cell is broken by switching a subset of inclusions between Dirichlet and Neumann conditions. In each lattice, mirror-related boundary-condition assignments produce chiral bulk phases with opposite valley signatures while retaining an overlapping gap. When these phases are adjoined, ribbon calculations exhibit interfacial ZLM branches lying within the projected bulk gap, and the associated eigenmodes are localised to the interface between phases.

The principal design implication is that the interface is not tied to a geometric partition of two distinct crystals. Instead, it is determined by the choice of scatterer conditions. By altering that choice--without moving inclusions--one creates, removes, and translates internal interfaces, and hence repositions the ZLM spatially within the same underlying crystal. Finite multiple-scattering simulations confirm that the localised guided response follows the moved interface in both hexagonal and square lattices. This provides a convenient approach to actively reconfigure topological insulators in which waveguiding pathways are programmed by switching scatterer conditions, with the spectral/topological character of the guiding inherited from the valley-Hall band gap structure computed from~\eqref{HelmholtzHardScheme_1}.

Possible extensions of the theory include the active re-routing of photonic or phononic circuitry and the creation of switchable interface networks through time modulation \cite{touboul2024high}, for example by allowing $\epsilon_i=\epsilon_i(t)$. In the present work, however, we have focused on deriving the long-time-limit scattering formulations and on demonstrating, within this framework, boundary-condition-controlled switching and the resulting repositioning of ZLMs in representative hexagonal and square lattices.

\section{Code availability} \label{CodeAv}
The semi-analytical solvers developed in this study are provided in the ancillary files. In particular, the code used to solve the generalised eigenvalue problem in \eqref{HelmholtzHardScheme_1}, in order to determine $\Omega = \Omega(\boldsymbol{\kappa})$ for Figs.~\ref{fig:HexDirac} and \ref{fig:HexValley}, is included. In addition, the multiple-scattering solver used to solve \eqref{scatteringSoln} for the unknown coefficients in \eqref{mscSOLNtotal} is provided to reproduce the example shown in Fig.~\ref{fig:HexScatt}(c,d). Further details are given in the README file included with the ancillary material. These codes can also be easily altered to investigate the other problems considered in this paper.

\section*{Funding}

R. W. acknowledges funding from the EU, H2020 projects MetaVEH (grant agreement number 952039) and DYNAMO (grant agreement number 101046489). H.J.P. acknowledges funding from EPSRC Grant No. EP\textbackslash X012689\textbackslash1.

\section*{Acknowledgements}
R.W. wishes to thank Richard Craster for continued support, guidance, advice, and encouragement throughout R.W.’s academic development. R.W. is also grateful for the hospitality of the Quantum Measurement and Instrumentation Laboratory at the University of Osaka. The welcoming and supportive environment of the group made the research visit both productive and personally meaningful. In particular, R.W. would like to thank Hirotsugu Ogi, Kichitaro Nakajima, Natsumi Fujiwara, Hiroki Okita, and Ambuj Kumar for their kindness and generosity, and for the warm welcome they extended throughout the visit. Particular thanks are due to Miwa Yamamoto for her support throughout the visit, especially for the practical assistance she provided. Lastly, R.W. gratefully acknowledges the guidance, encouragement, and generosity of Oliver B. Wright, whose support has extended beyond formal academic mentorship. 

\vspace{0.1cm}
\begin{center}
\includegraphics[width=.9\linewidth]{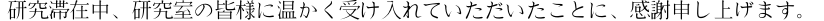}
\end{center}

\noindent The scientific colour maps Oslo and Vik \cite{crameri2018scientific} are employed in this study to minimise visual distortion and to ensure accessibility for readers with variations in colour vision \cite{crameri2020misuse}.


\end{document}